\documentclass[preprintnumbers,superscriptaddress,nofootinbib,aps,prd,floatfix,notitlepage]{revtex4-1}

\usepackage{hyperref}
\usepackage{amsmath,slashed}
\usepackage{graphicx}
\usepackage{caption}
\usepackage{subcaption}
\usepackage{relsize}
\usepackage{mathtools}

\usepackage{empheq}
\usepackage{amssymb}

\newcommand{\be}{\begin{eqnarray*}}
\newcommand{\ee}{\end{eqnarray*}}

\newcommand{\bee}{\begin{eqnarray}}
\newcommand{\eee}{\end{eqnarray}}
\newcommand{\beeq}{\begin{equation}}
\newcommand{\eeeq}{\end{equation}}
\newcommand{\gev}{{\rm{GeV}}}

\newcommand{\obs}[1]{\mathrm{O}_{#1}}

\newcommand{\one}[2]{\mathds{1}^{#1}_{#2}}

\newcommand{\dire}{\textsc{Dire}\;}
\newcommand{\pythia}{\textsc{Pythia}\;}

\newcommand{\nnnlo}{\textnormal{\smaller N3LO}}
\newcommand{\nnlo}{\textnormal{\smaller NNLO}}
\newcommand{\nlo}{\textnormal{\smaller NLO}}
\newcommand{\qcd}{\textnormal{\smaller QCD}}
\newcommand{\pdf}{\textnormal{\smaller PDF}}

\newcommand{\nlopps}{\textnormal{{\smaller NLO}+{\smaller PS}}}
\newcommand{\nnlopps}{\textnormal{{\smaller NNLO}+{\smaller PS}}}
\newcommand{\nnnlopps}{\textnormal{{\smaller N3LO}+{\smaller PS}}}

\let\ss\undefined
\let\i\undefined
\newcommand{\ff}[5]{ {#1}_{#2}^{(#3) \left[\textnormal{\smaller$#4$}\right]}(#5)}
\newcommand{\ss}[4]{ d\sigma_{#1}^{(#2) \left[\textnormal{\smaller$#3$}\right]}(#4)}
\newcommand{\f}[4]{ {#1}_{#2}^{(#3)}(#4)}
\newcommand{\s}[3]{ d\sigma_{#1}^{(#2)}(#3)}
\newcommand{\w}[1]{ w_{#1}(\Phi_{#1})}

\newcommand{\i}[1]{ \int \mathrm{d}\Phi_{#1}}

\newcommand{\unlops}{\textnormal{\smaller UNLOPS}}
\newcommand{\unnlops}{\textnormal{\smaller UN$^2$LOPS}}

\newcommand{\nnnlops}{\textnormal{\textsc{Tomte}}}

\DeclareGraphicsExtensions{.pdf,.png,.jpg}

\usepackage{tikz}
\usetikzlibrary{tikzmark}
\usetikzlibrary{calc} 
\usetikzlibrary{decorations.pathmorphing}

\usepackage{tikz-feynman}
\tikzfeynmanset{compat=1.0.0}

\tikzstyle{description} = [rectangle, minimum width=1cm, minimum height=0.5cm, draw=black, fill=green!05, inner sep=2pt, inner ysep=2pt, inner xsep=2pt]
\tikzstyle{nobox} = [rectangle, minimum width=0cm, minimum height=0.0cm, text centered, draw=black!00, fill=black!00, inner sep=1pt, inner ysep=1pt]
\tikzstyle{box} = [rectangle, minimum width=0cm, minimum height=0.0cm, text centered, draw=black, fill=black!00, inner sep=1pt, inner ysep=1pt]
\tikzstyle{circ} = [circle, minimum width=0cm, minimum height=0.0cm, text centered, draw=black, fill=black!00, inner sep=1pt, inner ysep=1pt]
\tikzstyle{arrow} = [thick,->,>=stealth]
\tikzstyle{fixedbox15} = [rectangle, minimum width=1.5cm, minimum height=1.5cm, text centered, draw=black, fill=black!00, inner sep=1pt, inner ysep=1pt]
\tikzstyle{fixedbox11} = [rectangle, minimum width=1.1cm, minimum height=1.1cm, text centered, draw=black, fill=black!00, inner sep=1pt, inner ysep=1pt]
\tikzstyle{fixedbox10} = [rectangle, minimum width=1.0cm, minimum height=1.0cm, text centered, draw=black, fill=black!00, inner sep=1pt, inner ysep=1pt]
\tikzstyle{fixedbox09} = [rectangle, minimum width=0.9cm, minimum height=0.9cm, text centered, draw=black, fill=black!00, inner sep=1pt, inner ysep=1pt]
\tikzstyle{fixedbox08} = [rectangle, minimum width=0.8cm, minimum height=0.8cm, text centered, draw=black, fill=black!00, inner sep=1pt, inner ysep=1pt]
\tikzstyle{fixedbox07} = [rectangle, minimum width=0.7cm, minimum height=0.7cm, text centered, draw=black, fill=black!00, inner sep=1pt, inner ysep=1pt]
\tikzstyle{fixedbox06} = [rectangle, minimum width=0.6cm, minimum height=0.6cm, text centered, draw=black, fill=black!00, inner sep=1pt, inner ysep=1pt]
\tikzstyle{fixedbox05} = [rectangle, minimum width=0.5cm, minimum height=0.5cm, text centered, draw=black, fill=black!00, inner sep=1pt, inner ysep=1pt]

\usepackage{tcolorbox}
\tcbuselibrary{theorems}

\tcbsetforeverylayer{arc=0pt,outer arc=0pt,top=0pt,bottom=0pt,left=0pt,right=0pt,colback=black!10,colframe=black, boxrule=0.1pt, boxsep=0.1pt}
\tcbuselibrary{skins}
\tcbsetforeverylayer{skin=enhanced}

\usepackage{amsfonts}

\usepackage{color}

\usepackage{cleveref}
\usepackage{dsfont}

\definecolor{mypurple}{rgb}{0.8,0.1,0.7}

\definecolor{myolive}{rgb}{0.2,0.45,0.25}

\definecolor{mybrown}{rgb}{0.54,0.27,0.07}

\definecolor{mymidgrey}{rgb}{0.5,0.5,0.5}

\makeatletter
\newenvironment{nonfloatfig}{\par\noindent\minipage{\textwidth}
\def\@captype{table}%
\centering\medskip}{\medskip\endminipage}
\makeatother

\allowdisplaybreaks

\makeatletter
\def\upintkern@{\mkern-7mu\mathchoice{\mkern-3.5mu}{}{}{}}
\def\upintdots@{\mathchoice{\mkern-4mu\@cdots\mkern-4mu}%
 {{\cdotp}\mkern1.5mu{\cdotp}\mkern1.5mu{\cdotp}}%
 {{\cdotp}\mkern1mu{\cdotp}\mkern1mu{\cdotp}}%
 {{\cdotp}\mkern1mu{\cdotp}\mkern1mu{\cdotp}}}
\newcommand{\upiint}{\DOTSI\protect\UpMultiIntegral{2}}

\newcommand{\UpMultiIntegral}[1]{%
  \edef\ints@c{\noexpand\upintop
    \ifnum#1=\z@\noexpand\upintdots@\else\noexpand\upintkern@\fi
    \ifnum#1>\tw@\noexpand\upintop\noexpand\upintkern@\fi
    \ifnum#1>\thr@@\noexpand\upintop\noexpand\upintkern@\fi
    \noexpand\upintop
    \noexpand\ilimits@
  }%
  \futurelet\@let@token\ints@a
}
\makeatother

\DeclareFontFamily{OMX}{mdbch}{}
\DeclareFontShape{OMX}{mdbch}{m}{n}{ <->s * [0.8]  mdbchr7v }{}
\DeclareFontShape{OMX}{mdbch}{b}{n}{ <->s * [0.8]  mdbchb7v }{}
\DeclareFontShape{OMX}{mdbch}{bx}{n}{<->ssub * mdbch/b/n}{}

\DeclareSymbolFont{uplargesymbols}{OMX}{mdbch}{m}{n}
\SetSymbolFont{uplargesymbols}{bold}{OMX}{mdbch}{b}{n}
\DeclareMathSymbol{\upintop}{\mathop}{uplargesymbols}{82}
\DeclareMathSymbol{\upointop}{\mathop}{uplargesymbols}{"48}

\DeclareFontEncoding{MDB}{}{}
\DeclareFontFamily{MDB}{mdbch}{}
\DeclareFontShape{MDB}{mdbch}{m}{n}{ <->s * [0.8]  mdbchrmb }{}
\DeclareFontShape{MDB}{mdbch}{b}{n}{ <->s * [0.8]  mdbchbmb }{}
\DeclareFontShape{MDB}{mdbch}{bx}{n}{<->ssub * mdbch/b/n}{}
\DeclareFontSubstitution{MDB}{cmr}{m}{n}
\DeclareSymbolFont{mathdesignB}{MDB}{mdbch}{m}{n}%
\SetSymbolFont{mathdesignB}{bold}{MDB}{mdbch}{b}{n}%
\DeclareMathSymbol{\upintclockwise}{\mathop}{mathdesignB}{128}
\DeclareMathSymbol{\upointclockwise}{\mathop}{mathdesignB}{130}
\DeclareMathSymbol{\upointctrclockwise}{\mathop}{mathdesignB}{132}
\DeclareMathSymbol{\upoiint}{\mathop}{mathdesignB}{134}
\DeclareMathSymbol{\upoiiint}{\mathop}{mathdesignB}{136}

\makeatletter
\newcommand{\upint}{\DOTSI\upintop\ilimits@}
\newcommand{\upoint}{\DOTSI\upointop\ilimits@}

\newcommand{\intOne}{\mathop{\mathlarger{\upoint\displaylimits^{t^+}_{t^-}}}}
\newcommand{\intTwo}{\mathop{\mathlarger{\upoiint\displaylimits^{t^+}_{t^-}}}}
\newcommand{\intThree}{\mathop{\mathlarger{\upoiiint\displaylimits^{t^+}_{t^-}}}}
\renewcommand{\int}{\mathop{\mathlarger{\upint}}}

\begin{document}

\title{Matching N3LO QCD calculations to parton showers}

\begin{abstract}
The search for new interactions and particles in high-energy collider physics relies on precise background predictions. This has led to many advances in combining precise fixed-order cross-section calculations with detailed event generator simulations.
In recent years, fixed-order \qcd{} calculations of inclusive cross sections at \nnnlo{} precision have emerged, followed by an impressive progress at producing differential results. Once differential results become publicly available, it would be prudent to embed these into event generators to allow the community to leverage these advances.  
This note offers some concrete thoughts on {\smaller ME}+{\smaller PS} matching at third order in \qcd{}. As a method for testing these thoughts, a toy calculation of $e^+e^-\rightarrow u \bar u$ at $\mathcal{O}(\alpha_s^3)$ is constructed, and combined with an event generator through unitary matching.
The toy implementation may serve also as blueprint for high-precision \qcd{} predictions at future lepton colliders. As a byproduct of the \nnnlo{} matching formula, a new \nnlopps{} formula for processes with ``additional" jets is obtained.
\end{abstract}

\author{Stefan Prestel}
\affiliation{Department of Astronomy and Theoretical Physics,\\Lund University, S-223 62 Lund, Sweden}

%comment out for jhep

\preprint{LU-TP-21-21, MCNET-21-10}
\vspace*{4ex}

%\maketitle

% avoid new page after title page
\maketitle

\tableofcontents

%%%%%%%%%%%%%%%%%%%%%%%%%%%%%%%%%%%%%%%%%%%%%%%%%%%
\section{General Introduction}
\label{sec:intro}

\noindent
High-energy collider physics tries to provide insights into a consistent quantum field theory of nature by accumulating immense amounts of experimental data, and confronting it with precision calculations -- the expectation being that enough data on as many scattering processes as possible might expose flaws in our current understanding. Unfortunately, since calculations in the Standard Model of particle physics quickly become tedious, hints of new interactions are often explained by phenomena that were omitted in the background calculation. This has resulted a large and vibrant sub-community producing precise and detailed calculations of background processes -- especially for experiments at the Large Hadron Collider. The typical model for calculations used by experimentalists is a combination of higher-order \qcd{} (or electroweak) fixed-order calculations (to obtain the best available prediction of the highest-energy scattering process), combined with an event generator handling the evolution of the high-energy particles into composite hadrons, energetic jets and other remnants of the colliding beams. 

The first \emph{general} methods to improve the precision and/or accuracy of such a computational model emerged at the turn of the century, when the matching of \nlo{} \qcd{} calculations with event generators~\cite{Frixione:2002ik,*Nason:2004rx,Frixione:2007vw} and the merging of multiple leading-order multi-jet calculations~\cite{Catani:2001cc,Lonnblad:2001iq,Mangano:2001xp,*Mrenna:2003if,*Alwall:2007fs} were formulated. This kick-started steady progress of incorporating higher \qcd{} orders and/or a more accurate treatment of the other SM interactions~\cite{Frixione:2010ra,*Torrielli:2010aw,*Alioli:2010xd,*Hoeche:2010pf,*Hoeche:2011fd, *Platzer:2011bc,*Alwall:2014hca,*Jadach:2015mza,*Czakon:2015cla,*Hamilton:2009ne,*Hamilton:2010wh,*Hoche:2010kg,*Lonnblad:2012ng,*Lavesson:2005xu,*Lonnblad:2011xx,*Platzer:2012bs,*Gehrmann:2012yg,*Hoeche:2012yf,*Lonnblad:2012ix,*Frederix:2012ps,*Alioli:2012fc,*Bellm:2017ktr,Lavesson:2008ah}. Apart from a few impressive examples~\cite{Alioli:2020qrd,*Mazzitelli:2020jio}, this progress has slowed in recent years, and is being complemented with improvements of all-order parton showers. Parton showers form the backbone of matching fixed-order calculations to event generators, by e.g. furnishing differential subtractions to make unweighted (or unweightable) fixed-order \emph{event} generation by Monte-Carlo methods possible. Fully-differential matching procedures are presently only available in \nlo{} \qcd{}, since the (fully differential) singularity-structure of \qcd{} is not yet captured by any available parton shower. This has however not prevented several successful combinations of \nnlo{} predictions and event generators~\cite{Lavesson:2008ah,Hoeche:2014aia,Hoche:2014dla, Hamilton:2013fea, *Karlberg:2014qua,*Hamilton:2015nsa,*Alioli:2015toa,*Astill:2018ivh,*Hoche:2018gti,*Re:2018vac,*Monni:2019whf,*Monni:2020nks,*Lombardi:2020wju,*Hu:2021rkt}. Phase-space slicing inspired unitarized merging methods offer a convenient stepping stone towards high accuracy. 

Recent years have seen impressive progress in calculating \qcd{} corrections at \nnnlo, both to inclusive cross sections~\cite{Anastasiou:2015vya,*Duhr:2019kwi,*Duhr:2020seh,*Chen:2019lzz,Duhr:2020sdp} and even at differential level~\cite{Dulat:2017prg,Currie:2018fgr,Dreyer:2018qbw,Cieri:2018oms,Mondini:2019gid,Chen:2021isd,Billis:2021ecs,Camarda:2021ict,Re:2021con}. Most of the latter results (\cite{Currie:2018fgr,Dreyer:2018qbw,Cieri:2018oms,Mondini:2019gid,Billis:2021ecs,Camarda:2021ict,Re:2021con}) rely on the subdivision (``slicing") of phase space into individually manageable sub-calculations. Such a strategy has already proven successful in combining \nnlo{} calculations with parton showering~\cite{Hoeche:2014aia,Hoche:2014dla}, suggesting that a similar strategy might be successful also at \nnnlo{} precision. This hope is further compounded by\cite{Banfi:2015pju}, which presented a method to match \nnnlo{} calculations to (analytic) jet-veto resummation, and by the very recent progress \cite{Billis:2021ecs,Camarda:2021ict,Re:2021con} combining \nnnlo inclusive cross sections with (analytic) transverse-momentum resummation.

This note aims to present ideas towards matching \nnnlo{} \qcd{} calculations to event generators. The overall philosophy of the approach is straight-forward and based on enforcing the desired target precision through all-order subtractions inspired by the unitarity of parton showers. This allows to combine several calculations executed with minimal phase-space cuts (to avoid the most singular regions) into a precise matched calculation, similar in spirit to e.g.\ $q_\perp$~\cite{Catani:2007vq,*Catani:2008me,*Catani:2009sm,*Catani:2011qz,*Bonciani:2015sha,*Grazzini:2017mhc} or Njettiness-sliced~\cite{Boughezal:2011jf,Gaunt:2015pea,Boughezal:2015dva,*Boughezal:2015aha,Campbell:2017hsw,*Campbell:2019dru} fixed-order results. Section \ref{sec:match_intro} will be used to set the scene for the third-order\footnote{The somewhat ill-defined term "third-order" will be used to avoid the impression that \nnnlo-accurate results are presented in this note.} method from a pedagogic angle, while sec.\ \ref{sec:fixed_order} explains the construction of a \emph{toy fixed-order} calculation. In the absence of differential \nnnlo{} calculations that produce finite-weight \emph{events}, it is convenient to work with a toy example to validate the third-order matching in a very controlled environment. Details about third-order matching are set out in sec.\ \ref{sec:matching}, and the suggestions are studied with the help of the toy calculation in sec.\ \ref{sec:results}. The note also contains detailed appendices with background information on parton showers and the implementation of matching factors. 

%%%%%%%%%%%%%%%%%%%%%%%%%%%%%%%%%%%%%%%%%%%%%%%%%%%
\section{Introduction to Matching up to NNLO precision}
\label{sec:match_intro}

\begin{figure}[ht!]
\begin{tikzpicture}[remember picture]
\node (function) [nobox] {\Large{$f$}};
\node (subscript) [nobox, right of=function, xshift=-4.5ex,yshift=-1ex]{\footnotesize{$n$}};
\node (superscript) [nobox, right of=function, xshift=-4.5ex, yshift=1ex] {\footnotesize{$(i)$}};
\node (superscript2) [nobox, right of=function, xshift=-1.8ex, yshift=1ex] {\scriptsize{$[M]$}};
\node (argument) [nobox, right of=function, xshift=2.5ex] {($\Phi_{n}$)};
\node (functiondescription) [description, left of=function, text width=0.33\textwidth,xshift=-4cm] {\footnotesize function of interest; typically, the differential cross section $\sigma$, or the Sudakov factor $\Delta$};
\node (subscriptdescription) [description, below of=subscript, text width=0.65\textwidth,yshift=-1cm] {\footnotesize parton multiplicity; the number of additional partons is defined relative to the lowest-multiplicity process; e.g.\ $e^+e^-\rightarrow u \bar u$ entails $n=0$, $e^+e^-\rightarrow u \bar u g$ means $n=1$\dots};
\node (superscriptdescription) [description, above of=superscript, text width=0.9\textwidth,yshift=0.2cm, xshift=0cm] {\footnotesize perturbative order; the order is defined relative to the lowest order necessary to define the function $f$; combinations may be used; e.g.\ $(0+1)$ signifies the Born contributions and first-order corrections; a value $\infty$ signifies an all-order factor};
\node (superscript2description) [description, right of=superscript2, text width=0.33\textwidth,yshift=-0.1cm, xshift=4cm] {\footnotesize additional tag to further specify the function; e.g.\ $[\mathrm{inc}]$ for "inclusive" or $[t_c]$ for "exclusive, with veto on real emissions above $t_c$"; tags will be defined in the text.};
\node (argumentdescription) [description, below of=superscript2description, text width=0.33\textwidth,yshift=-0.25cm, xshift=0cm] {\footnotesize phase space point $\Phi$ at which the function is evaluated.};
\draw[red,thick,dotted] ($(function.north west)+(-0.3,0.4)$)  rectangle ($(argument.south east)+(0.1,-0.4)$);
\path[every node/.style={font=\sffamily\small},->,>=stealth]
    (subscriptdescription) edge [out=north, in=south]  (subscript);
\path[every node/.style={font=\sffamily\small},->,>=stealth]
    (functiondescription) edge [out=east, in=west]  (function);
\path[every node/.style={font=\sffamily\small},->,>=stealth]
    (superscriptdescription) edge [out=south, in=90]  (superscript);
\path[every node/.style={font=\sffamily\small},->,>=stealth]
    (superscript2description) edge [out=west, in=90]  (superscript2);
\path[every node/.style={font=\sffamily\small},->,>=stealth]
    (argumentdescription) edge [out=west, in=-90]  (argument);
\end{tikzpicture}
\caption{\label{fig:notation}Reference for the sub- and superscript notation used in this note.}
\end{figure}
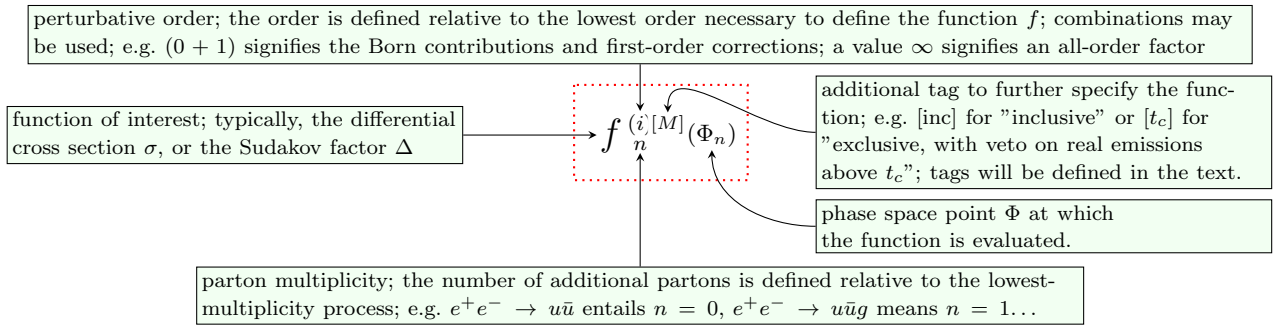

\noindent
Integrating fixed-order calculations into event generators almost always requires removing the overlap between the parton shower approximation embedded in the event generator and the fixed-order result. This overlap can be removed by subtraction or by diving phase-space into disjoint fixed-order and shower regions. Both cases yield the benefit of allowing for event generation, i.e.\ the production and storage of finite-weight phase-space points allowing an analysis independent of the computational details. For the purposes of this note, it is convenient to take the perspective of a (conventional) parton-shower calculation. This entails that
\begin{itemize}
\item $n$-parton phase-space points $\Phi_n$ can be obtained from parton-shower evolution; the parton shower sums logarithmic enhancements in perturbation theory
\item the "separation" of partons is determined by the parton-shower ordering variable $t$ calculated from their four-momenta; a sequence of ordering variables (as is necessary to assign parton-shower factors) is defined through a parton-shower history from the lowest- to the highest-multiplicity state; if necessary to determine an ordering sequence, one amongst all possible histories is chosen probabilistically (cf. Appendix \ref{app:histories}); the entries in a sequence of ordering variables are called $t(\Phi_0), t(\Phi_1)\dots t(\Phi_n)$ or abbreviated $t_0,t_1\dots t_n$
\item unitarity of the evolution holds, i.e.\ the action of the parton shower does not change inclusive cross sections; due to unitarity, parton-shower resummation assigns finite (or vanishing) cross sections even to phase-space points with collinear or soft partons; parton-shower all-order factors can be used to regularize fixed-order cross sections evaluated at phase-space points containing collinear or soft partons.  
\end{itemize}
The aim of matching a fixed-order calculation to a parton shower is to combine the strengths of either approximation, while ensuring that the resulting combination is fixed-order accurate and does not impair the accuracy of the parton-shower resummation. The term ``accuracy of the parton shower" is not immediately obvious. The heavily constraining definition applied in this note is given in Appendix \ref{app:shower-accuracy}. Before diving into specifics of matching, fig.\ \ref{fig:notation} introduces the features of the notation used throughout this note.

\noindent
Some examples of this notation that will appear repeatedly are given in Table \ref{tab:symbols}. Using this notation, the action $\mathcal{F}$ of the parton-shower on an ensemble of particles $\Phi_n$ with distribution {\smaller $\s{n}{0}{\Phi_n}$} either leads to no change in any observable $\obs{}$ (i.e.\ $\obs{}$ will be still evaluated at the phase-space point $\Phi_n$, i.e.\ $\obs{}=\obs{}(\Phi_n)\equiv \obs{n}$), or to the decay of one or more particles:
\begin{eqnarray}
\label{eq:ps1}
\f{\mathcal{F}}{n}{\infty}{\Phi_n, t_+,t_-}
 &\coloneqq& \s{n}{0}{\Phi_n}\Delta_n (t_+,t_-) \obs{n} \\
&+& \s{n}{0}{\Phi_n}  \int\limits^{t_+}_{t_-} \frac{dt}{t} dz d\phi \frac{\alpha_s(\mu)}{2\pi} P(t,z,\phi) \w{n+1} \Delta_n (t_+,t) 
\f{\mathcal{F}}{n+1}{\infty}{\Phi_n \cap \{t,z,\phi\}, t,t_-}\nonumber\\
\label{eq:ps2}
&=& \left( \s{n}{0}{\Phi_n}  - \s{n}{0}{\Phi_n}\int\limits^{t_+}_{t_-} \frac{dt}{t} dz d\phi \frac{\alpha_s(\mu)}{2\pi} P(t,z,\phi) \w{n+1} \Delta_n (t_+,t)\right) \obs{n}\\
&+& \s{n}{0}{\Phi_n}\int\limits^{t_+}_{t_-} \frac{dt}{t} dz d\phi \frac{\alpha_s(\mu)}{2\pi} P(t,z,\phi) \w{n+1} \Delta_n (t_+,t) 
\f{\mathcal{F}}{n+1}{\infty}{\Phi_n \cap \{t,z,\phi\}, t,t_-}\nonumber~,
\end{eqnarray}
where $P(t,z,\phi)$ is the sum of naive parton-shower branching kernels\footnote{For the sake of not obfuscating the arguments and derivations, the summation over different decay channels in the parton shower is, though relevant for any implementation, suppressed in the notation of the main text. In appendix \ref{app:details}, the sums are being made explicit when necessary.
Furthermore, the fact that limits of the $z$-integral may depend on the specific branching is also suppressed.}, and $\Delta_n (t_+,t_-)$ is the Sudakov factor encapsulating the no-branching rate. The evolution of incoming partons typically introduces ratios of parton-distribution functions~\cite{Sjostrand:1985xi}. Such ratios are suppressed for the sake for readability in this note.
The second equality \ref{eq:ps2} is due to the exponential form of the Sudakov factor,
\begin{eqnarray}
\label{eq:pssudakov}
\!\!\!\!\!\!\!\!
\Delta_n (t_+,t_-) = \exp\left[\! -\! \int\limits^{t_+}_{t_-}\! \frac{dt}{t} dz d\phi \frac{\alpha_s(\mu)}{2\pi} P(t,z,\phi) \w{n+1} \right] = 1 \!-\! \int\limits^{t_+}_{t_-}\! \frac{dt}{t} dz d\phi \frac{\alpha_s(\mu)}{2\pi} P(t,z,\phi) \w{n+1} \Delta_n (t_+,t)\,.
\end{eqnarray}
The notation $\w{n+1} \Delta_n (t_+,t)$ is a symbolic representation of the actual parton-shower result, which includes a sum over all possible ``paths" to arrive at the real-emission phase space point. Details of parton-shower factors are given in Appendix \ref{app:weights}. The main text will use the shorter, symbolic notation, to avoid over-crowding. 

Equation \ref{eq:pssudakov} shows that by construction, no interval $[t_+,t_-]$ will add to the overall cross section -- simply because the integrated radiation pattern between any two scales is \emph{subtracted} from the semi-exclusive lowest-multiplicity contribution, such that integrating over the last line will lead to the original distribution {\smaller $\s{n}{0}{\Phi_n}$}. Viewed slightly differently, this "parton-shower unitarity" can also be interpreted as calculating a radiation pattern in the collinear approximation, regularizing the radiation pattern through all-order Sudakov factors (i.e.\ the third line), and subtracting the regularized radiation pattern from the next-lower multiplicity (i.e.\ the second line). This reasoning can be exploited to obtain a more precise calculation, by following the steps

\begin{nonfloatfig}
\renewcommand{\arraystretch}{1.5}
\begin{tabular}{ l p{0.96\textwidth}}
1.\quad & Choose a precise target lowest-multiplicity prediction\\
2.\quad & Correct the radiation pattern to the desired accuracy, regularize with Sudakov factors and remove undesired higher orders introduced by this reweighting\\
3.\quad & Subtract the integrated form of the improved radiation pattern from the lowest-multiplicity contribution
\end{tabular}
\caption{\label{tab:steps} The three main steps of unitarized matching/merging}
\end{nonfloatfig}
This ''unitarized merging" approach~\cite{Lonnblad:2012ng,Platzer:2012bs,Lonnblad:2012ix} can be performed for an arbitrary $t_-$ value.
The details of this procedure can become rather intricate -- but are manageable, as elaborated on in Appendix \ref{app:details}. The unitarized merging paradigm rests on the assumption that the parton shower does not change the inclusive cross section. Certain types of threshold enhancements are known to violate this condition~\cite{Nagy:2016pwq}, and are thus typically not included. However, an appropriate redefinition of the ``target prediction" may circumvent this concern.
 
\begin{table}[t!]
\renewcommand{\arraystretch}{1.5}
\begin{tabular}{|c c p{0.6\textwidth} |}
\hline
$\ss{n}{0}{Q(\Phi_{n})>Q_c}{\Phi_{n}}$ & : & the leading-order fully differential cross section to produce the phase-space point $\Phi_n$, under the constraint that the phase-space point $\Phi_{n}$ should pass the $Q$-cut with minimal value $Q_c$ \\
$\Delta_n^{(\infty) [\textnormal{\smaller PS}]} (t_n,t_c)$ & : & the parton-shower Sudakov factor enforcing that the particle ensemble $\Phi_n$ did not change (via decay) between the scales $t_n\equiv t(\Phi_n)$ and $t_c$. 
For example, if $\Phi_n$ consists of the three particles $\{q,\bar q, g\}$, then 
\qquad\qquad\qquad
$\Delta_n^{(\infty) [\textnormal{\smaller PS}]} (t_n,t_c) = \exp\Big[ - \int\limits^{t_n}_{t_c} \frac{dt}{t} dz d\phi \frac{\alpha_s(t)}{2\pi} \Big(P_{qq} +P_{\bar q\bar q} + P_{gg} + P_{gq}\Big) \Big]$, with the splitting kernels $P_{ij}$
defined in~\cite{Hoche:2015sya}. Unless expressly necessary, the abbreviated symbol $\Delta_n (t_n,t_c)$ will be used synonymously.\\
$\Delta_n^{(m)} (t_n,t_c)$ & : & the $m$th term in the expansion of the parton-shower Sudakov factor pertaining to no emission off the particle ensemble $\Phi_n$ between the scales $t_n$ and $t_c$.\\
$\f{w}{n+1}{\infty}{\Phi_{n+1}}$ & : & the combination of all parton-shower all-order factors \emph{except} Sudakov factors that the shower would have applied to the ensemble $\Phi_{n+1}$. For $e^+e^-\rightarrow$ jets annihilation, $\f{w}{n+1}{\infty}{\Phi_{n+1}} = \alpha_s(t_{n+1}) / \alpha_s(\mu)$, where $\mu$ is a fixed value of the renormalization scale. If necessary in the interest of brevity, the notation ${w}_{n+1} (\Phi_{n+1})$ will be used synonymously.\\
$\f{w}{n+1}{m}{\Phi_{n+1}}$ & : & the $m$th term in the expansion of the expansion of the combination above.\\
$\ss{n}{0+1}{\mathrm{INC}}{\Phi_{n}}$ & : & the inclusive \nlo{} cross section to produce $n$ additional partons, differential only in the variables $\Phi_n$. The real-emission correction is integrated over the single-parton phase space. The 
{\smaller$\overline{\mathrm{B}}$}-cross sections in the \textsc{Powheg} method~\cite{Frixione:2007vw} fall under this category.
Symbolically,
$\ss{n}{0+1}{\mathrm{INC}}{\Phi_{n}}$ = $\s{n}{0}{\Phi_{n}} +$ $\left(\s{n}{1}{\Phi_{n}} + \mathcal{I}_{n}(\Phi_n) \right) +$ $\left[\sum\limits_{i\in\mathbb{I}}\upint d\Phi_{1}^{[i]} \left(d\sigma_{n+1}^{(0)[i]}(\Phi_{n+1}) - \mathcal{S}_{n+1}^{[i]}(\widetilde{\Phi}_{n},\Phi_{1}^{[i]})\right)\right]_{\Phi_n}$ where $\mathbb{I}$ is the set of all singular limits of the real corrections. Since various treatments of singular limits are possible, this may lead to slightly different definitions of the inclusive cross section. The matching method developed in this note should be flexible enough to handle differing definitions.
\\
$Q(\Phi_{n})>Q_c$  & : & constraint on phase space points $\Phi_n$: $\Phi_n$ should lead to $Q$-values above $Q_c$ (cf.\ Appendix \ref{app:details}). The notation $Q_n\equiv Q(\Phi_{n})$ will also be used, for brevity.\\
$Q(\Phi_{n})<Q_c$  & : & constraint that $\Phi_n$ should lead to $Q$-values below $Q_c$ (cf.\ Appendix \ref{app:details})\\
$\ss{n}{1}{Q_{n}>Q_c \land Q_{n+1}<Q_c }{\Phi_{n}}$ & : & the exclusive \nlo{} correction, i.e.\ the \nlo{} rate excluding hard real-emission corrections with $Q_{n+1}>Q_c$. The cross section is again only differential in the variables $\Phi_n$.\\
\hline
\end{tabular}
\caption{\label{tab:symbols}Some symbols that will feature heavily in this note. More details on (inclusive and exclusive) fixed-order cross sections can be found in Appendix \ref{app:fixed-order-xsections}. }
\end{table}

As a first example of unitarized matching, it is possible to obtain an \nlo-correct calculation from the previous result, by performing the operations\footnote{These rules are symbolic. In particular, the second replacement rule is only correct up to constants and Jacobian factors. The complete replacement rule can be found in the dipole factorization formulae in~\cite{Catani:1996vz}.}
\begin{eqnarray*}
\s{n}{0}{\Phi_n} &\rightarrow& \ss{n}{0+1}{\mathrm{INC}}{\Phi_n}\\
\s{n}{0}{\Phi_n} \alpha_s(\mu) P(t,z,\phi)/t &\rightarrow& \ss{n+1}{0}{Q_{n+1}>Q_c}{\Phi_{n+1}}
\end{eqnarray*}
where $\ss{n}{0+1}{\mathrm{INC}}{\Phi_{n+1}}$ is the inclusive \nlo{} cross section differential (only) in the variables $\Phi_n$. The cut $Q_{n+1}>Q_c$ is necessary to regularize a tree-level calculation of the real-emission configurations. The physical interpretation of the ``inclusive cross section" should not be over-stretched, since it only provides a suitable model for observables that cannot resolve any effect of an additional real-emission parton, irrespective of its hardness. Only very few measurements in a realistic environment have this trait. However, inclusive cross sections serve, together with differential real-emission cross sections, as building blocks for more realistic predictions. The steps in Table \ref{tab:steps} for example suggest that an \nlo{} matched rate is given by
\begin{eqnarray}
\label{eq:unlops}
\ff{\mathcal{F}}{n}{\infty}{\mathrm{\unlops}}{\Phi_n, t_+,t_-}
&\coloneqq& \left( \ss{n}{0+1}{\mathrm{INC}}{\Phi_n}  - \int\limits^{t_+}_{t_-} d\Phi_1 \ss{n+1}{0}{Q_{n+1}>Q_c}{\Phi_{n+1}} \w{n+1}\Delta_n (t_+,t_{n+1})\right) \obs{n}\\
&+& \int\limits^{t_+}_{t_-} d\Phi_{1} \ss{n+1}{0}{Q_{n+1}>Q_c}{\Phi_{n+1}} \w{n+1} \Delta_{n} (t_+,t_{n+1}) 
\f{\mathcal{F}}{n+1}{\infty}{\Phi_{n+1}, t_{n+1},t_-}\nonumber
\end{eqnarray}
where $\Phi_1=\{t_{n+1},z,\phi\}$ and $\Phi_{n+1}=\{\Phi_n \cap \Phi_1\}$. From this form, it becomes clear that the most natural functional definition of the cut is $Q=($parton shower evolution variable$)$, making for the most natural value $Q_c=t_-$. In this case, the division into an $\obs{n}$-component and a $\f{\mathcal{F}}{n+1}{\infty}{\Phi_{n+1}, t_{n+1},t_-}$ is -- by virtue of eq.\ \ref{eq:ps2} -- completely equivalent to the parton shower result.  However, eq.\ \ref{eq:unlops} is still appropriate if $Q_{n+1}\neq t_{n+1}$ and the parton shower has (effectively) vanishing support for $Q_{n+1}<Q_c$. This means that the calculation is sub-divided into an $n$-parton contribution and an $n+\geq 1$ part. The latter are only included fully differentially above $Q_c$, so that $Q_c$ should be chosen as small as possible. The presence of many high-energy $n$-parton contributions might lead to spuriously large hadronization effects, since these parton ensembles are directly hadronized, without dressing the states with further soft or collinear radiation.
Similar concerns may be raised in any event generator prediction that employs a parton-shower cut-off, since the no-emission contribution in eq.\ \ref{eq:ps1} also consists of high-energy few-parton states. In that case, though, no-emission events are typically rare, such that their hadronization
only produces marginal effects. In eq.\ \ref{eq:unlops}, two contributions that are numerically sampled with many events cancel to produce a small effective no-emission event count. This procedure may lead to a higher sensitivity to statistical outliers in the hadronization procedure, i.e.\ concerns about hadronization are of technical, not theoretical nature. Nevertheless, for \emph{jet} observables that guarantee that states with no emissions or emissions with $Q_{n+1}<Q_c$ are treated inclusively, it is permissible to replace $\obs{n}$ with the action of a parton shower $\f{\mathcal{F}}{n}{\infty,Q_{n+1}<Q_c}{\Phi_{n}, t_+,t_-}$ that omits emissions with $Q_{n+1}>Q_c$. This might improve the transition to the non-perturbative regime. This note will continue to use the notation ``$\obs{n}$", leaving the possibility of such a replacement implicit. Further comments on this ``sliced" or ``binned" approach are given in Appendix \ref{app:giggles}.

The extension to a unitarized \nnlo{} matching scheme is possible once an \nlo-accurate rate beyond lowest multiplicity is available. This can be combined with an inclusive \nnlo{} cross section {\smaller $\ss{n}{0+1+2}{\mathrm{INC}}{\Phi_n}$} that is differential (only) in the lowest-multiplicity variables $\Phi_n$. It is useful to introduce the short notation\footnote{The method to numerically generate such $\upoint$ integrals is based on the parton shower, and is described in Appendix \ref{app:integrals}.}
\begin{eqnarray*}
\intOne \sigma_{i+1}(\Phi_{i+1}) &=& \int\!\! d\Phi_1 ~\Theta[t(\Phi_{i+1}) - t^-] ~\Theta[t^+ - t(\Phi_{i+1})]~\sigma_{i+1}(\Phi_{i+1})  \quad,~ \Phi_1=\{t_{i+1},z_{i+1},\phi_{i+1}\}
\quad,~ \Phi_{i+1}=\{\Phi_{i} \cap \Phi_1\}
\end{eqnarray*}
for the integrations in the all-order subtraction terms. Assuming that $\obs{i}$ is a measurement of \emph{all degrees of freedom} of the phase-space point $\Phi_i$ (so that the second integral in eq.\ \ref{eq:unlops} can be omitted for lighter notation), the \unnlops\ matching formula then reads
\begin{eqnarray}
&&\ff{\mathcal{F}}{n}{\infty}{\mathrm{\unnlops}}{\Phi_n, t_+,t_-} \coloneqq \Bigg( \ss{n}{0+1+2}{\mathrm{INC}}{\Phi_n} \nonumber\\
&&\qquad-\intOne\ss{n+1}{0}{Q_{n+1}>Q_c}{\Phi_{n+1}} ~\tcboxmath[colback=green!10]{~\Big(1 - \f{w}{n+1}{1}{\Phi_{n+1}} - \f{\Delta}{n}{1}{t_+,t_{n+1}}\Big)\Delta_n (t_+,t_{n+1})}~\f{w}{n+1}{\infty}{\Phi_{n+1}}
\nonumber\\
&&\qquad-\intOne\ss{n+1}{1}{Q_{n+1}>Q_c}{\Phi_{n+1}} ~\tcboxmath[colback=red!10]{~\f{w}{n+1}{\infty}{\Phi_{n+1}}\Delta_n (t_+,t_{n+1})~}~\Bigg)\obs{n}\ \nonumber\\
&&\,+~
\Bigg( ~\ss{n+1}{0}{Q_{n+1}>Q_c}{\Phi_{n+1}} ~\tcboxmath[colback=green!10]{~\Big( 1 - \f{w}{n+1}{1}{\Phi_{n+1}} - \f{\Delta}{n}{1}{t_+,t_{n+1}}\Big)\Delta_n (t_+,t_{n+1})} \f{w}{n+1}{\infty}{\Phi_{n+1}} \nonumber\\
&&\qquad+ \ss{n+1}{1}{Q_{n+1}>Q_c}{\Phi_{n+1}} ~\tcboxmath[colback=red!10]{~\f{w}{n+1}{\infty}{\Phi_{n+1}}\Delta_n (t_+,t_{n+1})~} \nonumber\\
&&\qquad -\intOne \ss{n+2}{0}{Q_{n+2}>Q_c}{\Phi_{n+2}} ~\tcboxmath[colback=blue!10]{~\f{w}{n+1}{\infty}{\Phi_{n+1}} \f{w}{n+2}{\infty}{\Phi_{n+2}}\Delta_{n} (t_+,t_{n+1}) \Delta_{n+1} (t_{n+1},t_{n+2})~}~ \Bigg) \obs{n+1}\nonumber\\
&&\quad +~ \ss{n+2}{0}{Q_{n+2}>Q_c}{\Phi_{n+2}} ~\tcboxmath[colback=blue!10]{~\f{w}{n+1}{\infty}{\Phi_{n+1}} \f{w}{n+2}{\infty}{\Phi_{n+2}} \Delta_{n} (t_+,t_{n+1}) \Delta_{n+1} (t_{n+1},t_{n+2})~} \otimes\f{\mathcal{F}}{n+2}{\infty}{\Phi_{n+2}, t_{n+2},t_-}~.
\label{eq:un2lops-inc}
\end{eqnarray}
Upon integration over both one- and two-parton states, the terms in (and multiplying) the boxes \raisebox{1pt}{\tcboxmath[colback=red!10,boxsep=1.5pt]{\cdot}}, \raisebox{1pt}{\tcboxmath[colback=green!10,boxsep=1.5pt]{\cdot}} and \raisebox{1pt}{\tcboxmath[colback=blue!10,boxsep=1.5pt]{\cdot}} cancel pairwise, leaving only the desired \nnlo{} inclusive cross section. Integrating only over two-parton states, the terms in \raisebox{1pt}{\tcboxmath[colback=blue!10,boxsep=1.5pt]{\cdot}} cancel, so that for observables that require $(n+1)$-parton ensembles, \nlo{} precision is guaranteed. Note that for these $(n+1)$-parton observables, the impact of (parton-shower) resummation at higher orders remains if $t_+\gg t_{n+1}$, as desired. The accuracy of the resummation procedure is not jeopardized by the matching, as the all-order factors are only shifted by finite, multiplicative, fixed-order factors (e.g.\ {\smaller $\ss{n+1}{1}{Q_{n+1}>Q_c}{\Phi_{n+1}}$}). Also, no assumption on the logarithmic accuracy of the parton shower needs to be made. 

To avoid over-complicating the discussion at this point, eq.\ \ref{eq:un2lops-inc} assumes that all two-parton states admit an interpretation as ordered sequence of parton-shower transitions at scales $t_{n+1}$ and $t_{n+2}<t_{n+1}$. This assumption heavily depends on the details of the parton-shower, as well as the hard-scattering process, and does not necessarily hold in all regions of phase space. The bulk of the cross section is captured by configurations with parton-shower interpretation, but additional ``non-shower" configurations (called unordered in~\cite{Lonnblad:2011xx,Fischer:2017yja}, and exceptional in~\cite{Hoeche:2014aia}) need to be considered for an accurate prediction of sub-dominant phase-space regions. This complication (and related complications when matching three-parton states) are neglected in the main text, as well as the for the closure test in sec.\ \ref{sec:results}. A detailed discussion based on parton-shower ideas is given in Appendix \ref{app:unordered}. In a full-fledged matching implementation, the treatment of non-shower configurations will also depend on the details of the necessary fixed-order (input) calculations. 

The use of inclusive cross sections allows for particularly simple pairwise cancellation between configurations differing always by one parton. However, cross sections like {\smaller $\ss{n}{0+1}{\mathrm{INC}}{\Phi_n}$} or {\smaller $\ss{n}{0+1+2}{\mathrm{INC}}{\Phi_n}$} are often difficult to obtain due to the required integrations over the real-emission phase space. While inclusive cross sections are not available, it is easy to adjust the unitarized prescription to rely on ``jet-vetoed" (``exclusive") cross sections instead\footnote{As was the case for ``inclusive cross sections",
the physical meaning of ``exclusive cross section" should not be over-stated. Exclusive cross sections are employed as \emph{one component} of a realistic prediction.}. If the veto scale is sufficiently small, these can be obtained by expanding resummed analytic calculations. A sufficiently small veto scale further allows to combine these jet-vetoed cross sections with higher-multiplicity cross sections to form precise results that differ from the exact fixed-order results only by very small power corrections. This strategy is e.g.\ employed by $q_\perp$- or Njettiness-sliced fixed-order calculations.
When using only jet-vetoed cross sections, and introducing another short-hand,
\begin{eqnarray*}
&&\intTwo \sigma_{i+2}(\Phi_{i+2}) = \int\!\! d\bar\Phi_1 ~\Theta[t(\Phi_{i+1}) - t^-] ~\Theta[t^+ - t(\Phi_{i+1})]\int\!\! d\Phi_1 ~\Theta[t(\Phi_{i+2}) - t^-] ~\Theta[t^+ - t(\Phi_{i+2})]~\sigma_{i+2}(\Phi_{+2})\\
&&\textnormal{where}\quad \bar\Phi_1=\{t_{i+1},z_{i+1},\phi_{i+1}\},~ \Phi_1=\{t_{i+2},z_{i+2},\phi_{i+2}\},~ \Phi_{i+1}=\{ \Phi_{i} \cap \bar\Phi_1\},~ \Phi_{i+2}=\{ \Phi_{i+1} \cap \Phi_1\}~,
\end{eqnarray*}
the unitarized \nnlo{} matching prescription becomes
\begin{eqnarray}
&&\ff{\mathcal{F}}{n}{\infty}{\mathrm{\unnlops}}{\Phi_n, t_+,t_-} \coloneqq \Bigg( \ss{n}{0+1+2}{Q_{n+1}<Q_c \land Q_{n+2}<Q_c}{\Phi_n} \nonumber\\
&&
\qquad
+\intOne\ss{n+1}{0}{Q_{n+1}>Q_c}{\Phi_{n+1}} \left[\one{n+1}{n} - ~\tcboxmath[colback=green!10]{~\Big(1 - \f{w}{n+1}{1}{\Phi_{n+1}} - \f{\Delta}{n}{1}{t_+,t_{n+1}}\Big)\Delta_n (t_+,t_{n+1})\f{w}{n+1}{\infty}{\Phi_{n+1}}~}~\right]
\nonumber\\
&&\qquad
+\intOne\ss{n+1}{1}{Q_{n+1}>Q_c \land Q_{n+2}<Q_c }{\Phi_{n+1}} \left[\one{n+1}{n} - ~\tcboxmath[colback=red!10]{~\f{w}{n+1}{\infty}{\Phi_{n+1}}\Delta_n (t_+,t_{n+1})~}~\right]\nonumber\\
&&\qquad
+ \intTwo \ss{n+2}{0}{Q_{n+2}>Q_c}{\Phi_{n+2}} \left[\one{n+2}{n} - ~\tcboxmath[colback=orange!10]{~\f{w}{n+1}{\infty}{\Phi_{n+1}}\Delta_n (t_+,t_{n+1}) \one{n+2}{n+1} ~}~\right]
~\Bigg)\obs{n}\ \nonumber\\
&&\,+
\Bigg(~\ss{n+1}{0}{Q_{n+1}>Q_c}{\Phi_{n+1}} ~\tcboxmath[colback=green!10]{~\Big( 1 - \f{w}{n+1}{1}{\Phi_{n+1}} - \f{\Delta}{n}{1}{t_+,t_{n+1}}\Big) \Delta_n (t_+,t_{n+1}) \f{w}{n+1}{\infty}{\Phi_{n+1}}~}\nonumber\\
&&\qquad~+ \ss{n+1}{1}{Q_{n+1}>Q_c \land Q_{n+2}<Q_c}{\Phi_{n+1}} ~\tcboxmath[colback=red!10]{~\f{w}{n+1}{\infty}{\Phi_{n+1}} \Delta_n (t_+,t_{n+1})~} \nonumber\\
&&\qquad~+ \intOne \ss{n+2}{0}{Q_{n+2}>Q_c}{\Phi_{n+2}} \left[ ~\tcboxmath[colback=orange!10]{~\f{w}{n+1}{\infty}{\Phi_{n+1}}\Delta_{n} (t_+,t_{n+1}) \one{n+2}{n+1} ~}
\right.\nonumber\\
&&\left.
\quad\quad\quad\qquad
- ~\tcboxmath[colback=blue!10]{~\f{w}{n+1}{\infty}{\Phi_{n+1}}\f{w}{n+2}{\infty}{\Phi_{n+2}} \Delta_{n} (t_+,t_{n+1}) \Delta_{n+1} (t_{n+1},t_{n+2})~}~\right]~ \Bigg) \obs{n+1}\nonumber\\
&&\quad +~ \ss{n+2}{0}{Q_{n+2}>Q_c}{\Phi_{n+2}} ~\tcboxmath[colback=blue!10]{~\f{w}{n+1}{\infty}{\Phi_{n+1}}\f{w}{n+2}{\infty}{\Phi_{n+2}} \Delta_{n} (t_+,t_{n+1}) \Delta_{n+1} (t_{n+1},t_{n+2}) ~}~\otimes\f{\mathcal{F}}{n+2}{\infty}{\Phi_{n+2}, t_{n+2},t_-}~.
\label{eq:un2lops}
\end{eqnarray}
The main difference to the previous result is that the improved radiation patterns are only partially subtracted from the lower-multiplicity configurations, so that the full inclusive cross section is recovered upon integration. The coupling-independent kinematic factors $\one{i+j}{i}\approx 1$ ensure that the method to generate complementary $i$-parton contributions from $i+j$-parton contributions does not introduce $\Phi_i$-dependent biases that are not present in an inclusive fixed-order calculation. These factors need to be included to reconcile the requirement that the parton-shower accuracy is preserved (see Appendix \ref{app:shower-accuracy} for the definition of ``shower accuracy"), and the requirement that the result is unbiased when expanded to fixed order. It is sufficient to think of $\one{i+j}{i} = 1$ on first reading\footnote{The reason for non-unity $\one{i+j}{i}$ factors is explained in Appendix \ref{app:integrals}. While inclusive predictions using eq.\ \ref{eq:un2lops} are prone to methodological bias (since exact complements need to be generated), exclusive predictions using eq.\ \ref{eq:un2lops-inc} can exhibit related issues, since the first term in the expansion of the unitarity subtraction should remove integrated hard jet configurations -- which again has should avoid unwanted bias. Any unitary matching or merging method is prone to these issues. The similarity of using unitarity subtractions to the subtractions in the recent Projection-to-Born method~\cite{Cacciari:2015jma} suggests that, if a selection between several Born-level states were required in that method, similar considerations may also become relevant. An assessment of $\one{i+j}{i}$ factors required for the present note is given in Appendix \ref{app:toycalc}.}. 

The cancellation mechanism between matched contributions in eq.\ \ref{eq:un2lops} is now slightly more involved than in eq.\ \ref{eq:un2lops-inc}, since the two-jet contributions \raisebox{1pt}{\tcboxmath[colback=orange!10,boxsep=1.5pt]{\cdot}} now also cancel between one- and zero-jet observables. Note that eq.\ \ref{eq:un2lops} deviates slightly from the matching formula presented in~\cite{Hoeche:2014aia}, since the original proposition builds on an MC@NLO-matched one-jet radiation pattern, and handles the running \qcd{} coupling differently. The matching formula \ref{eq:un2lops} is closely related to the {\smaller UNLOPS-PC} prescription in~\cite{Gellersen:2020tdj}. Either method has benefits and down-sides, as discussed Appendix \ref{app:giggles}.

Generalizations of the unitarized matching procedure to higher precision will also follow the logic suggested in Table \ref{tab:steps}. Section \ref{sec:matching} employs this reasoning to introduce a third-order prescription, reusing eq.\ \ref{eq:un2lops} in the process.

%%%%%%%%%%%%%%%%%%%%%%%%%%%%%%%%%%%%%%%%%%%%%%%%%%%
\section{Constructing a toy fixed-order implementation for closure testing}
\label{sec:fixed_order}

\noindent
To extend the precision of an event generator calculation to higher order, precise fixed-order calculations are required. This serves both as testing ground for new developments as well as physics deliverable. This note will only be concerned with the testing aspect of the method, since no differential \nnnlo{} calculations are available at present. A very controlled testing environment is in itself quite useful when developing a concrete matching implementation. To this end, a toy third-order calculation is constructed by combining and rescaling readily available tree-level results. This allows maximal control, and enables tests that would not be possible in a theoretically more rigorous calculation. 

Gluonic corrections to $e^+e^-\rightarrow u \bar{u}$ form the simplest laboratory for third-order matching ideas, due to the uncolored, non-composite initial state. To be applicable to hadron collider predictions, the matching prescription presented in section \ref{sec:matching} would require a detailed -- and currently absent -- understanding of the interplay between the factorization of parton distribution functions (\pdf{}s) and renormalization within parton shower resummation\footnote{Also, current \nnnlo{} calculations for hadron collider processes employ \nnlo{} \pdf{} fits (as \nnnlo{} \pdf{} fits are not yet available), leading to further \pdf{}-related ambiguities, see e.g.~\cite{Duhr:2020sdp}. Lepton colliders provide a more solid environment for developing \nnnlopps{} methods.}. The toy third-order calculation for  $e^+e^-\rightarrow u \bar{u}$ is constructed by 
\begin{enumerate}
\item Producing tree-level event samples $\s{0}{0}{\Phi_{0}}, \ss{1}{0}{S(\Phi_{1})>S_c}{\Phi_{1}}, \ss{2}{0}{S(\Phi_{2})>S_c}{\Phi_{2}}$ and $\ss{3}{0}{S(\Phi_{3})>S_c}{\Phi_{3}}$ for $e^+e^-\rightarrow u \bar{u} + n g$ where $0\leq n \leq 3$. For $n>0$, very minimal regularization cuts $S(\Phi_{n})>S_c$ are applied on the projection of the (sum of) gluon four-momenta onto the four-momenta of the other partons, see Appendix \ref{app:toycalc}.
\item Constructing a sequence of toy fixed-order calculations. A toy exclusive \nlo{} calculation for four-parton states is constructed from {\smaller$\ss{2}{0}{S(\Phi_{2})>S_c}{\Phi_{2}}$} and {\smaller$\ss{3}{0}{S(\Phi_{3})>S_c}{\Phi_{3}}$} by
\begin{eqnarray}
\label{eq:toynlo}
&&\ss{2}{0+1}{\mathrm{TOY}}{\Phi_{2}}\\
&=& \Bigg\{\i{2}\ss{2}{0}{S(\Phi_{2})>S_c\land Q(\Phi_{2})>Q_c}{\Phi_{2}} \cdot \left[ 1 + \frac{\alpha_s}{2\pi}\textnormal{\smaller$\left(a_2^{q\bar q}y_{q\bar q} + a_2^{q g_1}y_{q g_1} + a_2^{q g_2}y_{q g_2} + a_2^{\bar q g_1}y_{\bar q g_1} + a_2^{\bar q g_2}y_{\bar q g_2} + a_2^{gg}y_{gg} \right)$} \right]\nonumber\\
&& ~-~\tcboxmath[colback=red!10]{~\i{3}\ss{3}{0}{S(\Phi_{3})>S_c\land Q(\Phi_{2})>Q_c}{\Phi_{3}}~}\Bigg\}~\obs{2} 
%\nonumber\\
%&&
~+~
\tcboxmath[colback=green!10]{~\i{3}\ss{3}{0}{S(\Phi_{3})>S_c \land Q(\Phi_{2})>Q_c \land Q(\Phi_{3})<Q_c}{\Phi_{3}}}~ \obs{2}\nonumber\\
&=& 
\label{eq:toynlo2}
%\i{2}\Bigg\{
%~\ss{2}{0}{S(\Phi_{2})>S_c\land Q(\Phi_{2})>Q_c}{\Phi_{2}}~ \cdot \left[ 1 + \frac{\alpha_s}{2\pi} \textnormal{\smaller$\left(a_2^{q\bar q}y_{q\bar q} + a_2^{q g_1}y_{q g_1} + a_2^{q g_2}y_{q g_2} + a_2^{\bar q g_1}y_{\bar q g_1} + a_2^{\bar q g_2}y_{\bar q g_2} + a_2^{gg}y_{gg} \right)$}  \right] \nonumber\\
%&&- ~\tcboxmath[colback=black!00,colframe=mypurple,boxrule=0.5pt,boxsep=1pt]{~\i{\mathrm{R}}\ss{3}{0}{S(\Phi_{3})>S_c\land Q(\Phi_{2})>Q_c \land Q(\Phi_{3})>Q_c }{\Phi_{3}}~}~\Bigg\}~\obs{2}\\
\i{2}\Bigg\{
~\ss{2}{0+1}{\mathrm{TOY\, INC}}{\Phi_{2}}~ - ~\tcboxmath[colback=black!00,colframe=mypurple,boxrule=0.5pt,boxsep=1pt]{~\i{\mathrm{R}}\ss{3}{0}{S(\Phi_{3})>S_c\land Q(\Phi_{2})>Q_c \land Q(\Phi_{3})>Q_c }{\Phi_{3}}~}~\Bigg\}~\obs{2}\\
&\approx&
\ss{2}{0+1}{Q(\Phi_{2})>Q_c \land Q(\Phi_{3})<Q_c }{\Phi_{2}} 
\nonumber~,
\end{eqnarray}
where the ``{\smaller$\mathrm{TOY\, INC}$}" contribution is a short-hand for term $\propto\frac{\alpha_s}{2\pi}\left[\cdots\right]$,
$\mathrm{d}\Phi_\mathrm{R}$ are the degrees of freedom for one additional parton, and $y_{ij}=2p_ip_j/M_Z^2$. The method to produce $\mathrm{d}\Phi_\mathrm{R}$ integrals and the application of $Q$-constraints are explained in Appendix \ref{app:toycalc}. The term in \raisebox{1pt}{\tcboxmath[colback=red!10,boxsep=1.5pt]{\cdot}} serves as proxy of logarithmic contributions of loop integrals, the \raisebox{1pt}{\tcboxmath[colback=green!10,boxsep=1.5pt]{\cdot}} term as real contribution below the jet veto scale, and the rescaling $\frac{\alpha_s}{2\pi}\left[\cdots\right]$ mimics the deformation of the spectra at \nlo, after cancellation of logarithms between real and virtual corrections. The \raisebox{1pt}{\tcboxmath[colback=black!00,colframe=mypurple,boxrule=0.5pt,boxsep=1.5pt]{\cdot}} term in eq.\ \ref{eq:toynlo2} only depends on five-parton states away from the phase space boundaries (since $Q(\Phi_{2})>Q_c \land Q(\Phi_{3})>Q_c$) -- the cancellation of real and virtual corrections close to the boundary is explicit, as is the miscancellation away from the boundary due to the veto condition. This also means that a dependence on the regularization criterion $S(\Phi_{n})$ is effectively absent. Overall, this toy prediction has features typical for fixed-order jet-vetoed cross sections, e.g.\ a strongly negative contribution.

A toy \nnlo{} calculation for three-parton states is constructed from {\smaller$\ss{1}{0}{S(\Phi_{1})>S_c}{\Phi_{1}}$} and {\smaller$\ss{2}{0+1}{\mathrm{TOY\, INC}}{\Phi_{2}}$} by
\begin{eqnarray}
\label{eq:toynnlo}
&&
\ss{1}{0+1+2}{\mathrm{TOY}}{\Phi_{1}}\\
&=&
\Bigg\{\i{1}\ss{1}{0}{S(\Phi_{1})>S_c\land Q(\Phi_{1})>Q_c}{\Phi_{1}} \cdot \left[ 1 + \frac{\alpha_s}{2\pi}\textnormal{\smaller$\left(a_1^{q\bar q}y_{q\bar q} + a_1^{q g}y_{q g} + a_1^{\bar q g}y_{\bar q g} \right)$} + \left(\frac{\alpha_s}{2\pi}\right)^2 \textnormal{\smaller$\left(b_1^{q\bar q}y_{q\bar q} + b_1^{q g}y_{q g} + b_1^{\bar q g}y_{\bar q g}\right)$}  \right]\nonumber\\
&& \tcboxmath[colback=red!10]{~-~\i{2} \ss{2}{0+1}{\mathrm{TOY\, INC}, Q(\Phi_{1})>Q_c }{\Phi_{2}} ~}\Bigg\}\obs{1} ~~\tcboxmath[colback=green!10]{~+~\i{2} \ss{2}{0+1}{\mathrm{TOY\, INC}, Q(\Phi_{1})>Q_c \land Q(\Phi_{2})<Q_c }{\Phi_{2}} ~}~\obs{1}\nonumber\\
&=&
\label{eq:toynnlo2}
%\i{1}\Bigg\{~\ss{1}{0}{S(\Phi_{1})>S_c\land Q(\Phi_{1})>Q_c}{\Phi_{1}}~ \cdot \left[ 1 + \frac{\alpha_s}{2\pi}\textnormal{\smaller$\left(a_1^{q\bar q}y_{q\bar q} + a_1^{q g}y_{q g} + a_1^{\bar q g}y_{\bar q g} \right)$} + \left(\frac{\alpha_s}{2\pi}\right)^2 \textnormal{\smaller$\left(b_1^{q\bar q}y_{q\bar q} + b_1^{q g}y_{q g} + b_1^{\bar q g}y_{\bar q g}\right)$}  \right]\nonumber\\
%&& -~\tcboxmath[colback=black!00,colframe=mypurple,boxrule=0.5pt,boxsep=1pt]{~\i{\mathrm{R}}\ss{2}{0+1}{\mathrm{TOY}, Q(\Phi_{1})>Q_c \land Q(\Phi_{2})>Q_c }{\Phi_{2}} ~}~ \Bigg\}~\obs{1} \\
\i{1}\Bigg\{~\ss{1}{0+1+2}{\mathrm{TOY\, INC}}{\Phi_{1}}~ -~\tcboxmath[colback=black!00,colframe=mypurple,boxrule=0.5pt,boxsep=1pt]{~\i{\mathrm{R}}\ss{2}{0+1}{\mathrm{TOY\, INC}, Q(\Phi_{1})>Q_c \land Q(\Phi_{2})>Q_c }{\Phi_{2}} ~}~ \Bigg\}~\obs{1} \\
&\approx&
\ss{1}{0+1+2}{Q(\Phi_{1})>Q_c \land Q(\Phi_{2})<Q_c \land Q(\Phi_{3})<Q_c}{\Phi_{1}} ~.
\nonumber
\end{eqnarray}
Again, the term in \raisebox{1pt}{\tcboxmath[colback=red!10,boxsep=1.5pt]{\cdot}} approximates loop integrals, the \raisebox{1pt}{\tcboxmath[colback=green!10,boxsep=1.5pt]{\cdot}} 
term implements jet-vetoed real-contributions, and a rescaling that mimics finite \nnlo{} corrections is included. Again, the miscancellation due to the veto condition is again explicit.
Finally, the toy {\nnnlo{}} calculation for two-parton states is assembled from {\smaller$\s{0}{0}{\Phi_{0}}$} and {\smaller$\ss{1}{0+1+2}{\mathrm{TOY\, INC}}{\Phi_{1}}$} by
\begin{eqnarray}
\label{eq:toynnnlo}
&&
\ss{0}{0+1+2+3}{\mathrm{TOY}}{\Phi_{0}}\\
&=&
\Bigg\{\i{0}\s{0}{0}{\Phi_{0}} \cdot \left[ 1
 + \frac{\alpha_s}{2\pi} \left(a_0^{qe}y_{qe} + a_0^{\bar qe}y_{\bar qe}\right)
 + \left(\frac{\alpha_s}{2\pi}\right)^2 \left( b_0^{qe} \left(1-y_{qe}\right)\ln{y_{qe}} + b_0^{\bar q e}\left(1-y_{\bar qe}\right)\ln{y_{\bar qe}} \right)\right.\nonumber\\
&&\left.
\qquad\qquad\qquad\qquad\quad~
 + \left(\frac{\alpha_s}{2\pi}\right)^3 \left(c_0^{qe}y_{qe}\cos{2\pi y_{qe}} + c_0^{\bar q e}y_{\bar qe}\sin{2\pi y_{\bar qe}} \right)   \right]\nonumber\\
&& \tcboxmath[colback=red!10]{~-~\i{1} \ss{1}{0+1+2}{\mathrm{TOY\, INC}}{\Phi_{1}} ~}\Bigg\}\obs{0} ~~\tcboxmath[colback=green!10]{~+~\i{1} \ss{2}{0+1+2}{\mathrm{TOY\, INC} , Q(\Phi_{1})<Q_c }{\Phi_{1}} ~}~\obs{0}\nonumber\\
&=&
\label{eq:toynnnlo2}
%\i{0}\Bigg\{ \s{0}{0}{\Phi_{0}} \cdot 
%\left[ 1
% + \frac{\alpha_s}{2\pi} \left(a_0^{qe}y_{qe} + a_0^{\bar qe}y_{\bar qe}\right)
% + \left(\frac{\alpha_s}{2\pi}\right)^2 \left( b_0^{qe} \left(1-y_{qe}\right)\ln{y_{qe}} + b_0^{\bar q e}\left(1-y_{\bar qe}\right)\ln{y_{\bar qe}} \right)\right.\nonumber\\
%&&\left.
%\qquad\qquad\qquad\qquad\quad~
% + \left(\frac{\alpha_s}{2\pi}\right)^3 \left(c_0^{qe}y_{qe}\cos{2\pi y_{qe}} + c_0^{\bar q e}y_{\bar qe}\sin{2\pi y_{\bar qe}} \right)   \right]\nonumber\\
%&&
% -~\tcboxmath[colback=black!00,colframe=mypurple,boxrule=0.5pt,boxsep=1pt]{~\i{\mathrm{R}}\ss{1}{0+1+3}{\mathrm{TOY}, Q(\Phi_{1})>Q_c}{\Phi_{1}} ~}\Bigg\}~\obs{0}\\
\i{0}\Bigg\{ \ss{0}{0+1+2+3}{\mathrm{TOY\, INC}}{\Phi_{0}}
 ~-~\tcboxmath[colback=black!00,colframe=mypurple,boxrule=0.5pt,boxsep=1pt]{~\i{\mathrm{R}}\ss{1}{0+1+3}{\mathrm{TOY\, INC}, Q(\Phi_{1})>Q_c}{\Phi_{1}} ~}\Bigg\}~\obs{0}\\
&\approx&
\ss{0}{0+1+2+3}{Q(\Phi_{1})<Q_c \land Q(\Phi_{2})<Q_c \land Q(\Phi_{3})<Q_c}{\Phi_{0}}  ~.
\nonumber
\end{eqnarray}
The argument for including the terms in colored boxes is identical to the previous cross sections. The peculiar functional dependences multiplying the coefficients $a_0$, $b_0$ or $c_0$ would never arise from \qcd{} corrections, but are chosen to allow for very clean closure testing.
\item Choosing the coefficients $a_n^X,b_n^X$ and $c_0^X$ in the toy calculation to produce exaggerated higher-order effects. The reproduction of these effects in the full third-order matched calculation will serve as affirmation of the correctness of the matching implementation. Explicit values are given in Table \ref{tab:toycalc-coefficients}.
\end{enumerate}
The resulting toy calculations exhibit the typical features of jet-vetoed cross sections, and thus serve as a realistic laboratory to test third-order matching ideas.

%%%%%%%%%%%%%%%%%%%%%%%%%%%%%%%%%%%%%%%%%%%%%%%%%%%
\section{Third-order matching}
\label{sec:matching}

\noindent
Having constructed a toy third-order calculation, one step towards concrete third-order event generators has been taken. This calculation should now be combined with parton shower evolution in a manner that guarantees that neither the accuracy of the fixed-order inputs nor of the parton shower are impaired. A unitarized matching ansatz offers a convenient route, since it does not put stringent (and not yet obtainable) restrictions on the parton shower implementation. The only relevant constraint is that on-shell intermediate states should exist at each step in the parton shower evolution, and that the parton-shower rate of any (pre-generated) phase-space point can be calculated numerically. The outlook in sec.\ \ref{sec:outlook} will theorize on other potential third-order matching methods. A third-order matching method should fulfill the criteria
\begin{nonfloatfig}
\renewcommand{\arraystretch}{1.5}
\begin{tabular}{ |l p{0.96\textwidth}|}
\hline
1.\quad & 3rd-order precision for inclusive zero-jet observables\\
2.\quad & 2nd-order precision for one-jet observables, combined with resummation when the jet becomes unresolved\\
3.\quad & 1st-order precision for two-jet observables, combined with resummed effects when either of the two jets becomes unresolved individually\\
4.\quad & 0th-order precision for three-jet observables, combined with resummed effects when any of the three jets become unresolved individually\\
5.\quad & parton-shower resummation of any observable sensitive to unresolved partons should not be impaired\\
\hline
\end{tabular}
\caption{\label{tab:criteria} The criteria for a consistent \nnnlo{} matching method}
\end{nonfloatfig}

The starting point for a simple third-order matching is the availability of an \nnlopps{}-matched $+1$-parton calculation {\smaller $\ff{\mathcal{F}}{n+1}{\infty}{\mathrm{\unnlops}, Q(\Phi_{1})>Q_c}{\Phi_n, t_+,t_-}$} performed using eq.~\ref{eq:un2lops}, and the availability of an \nnnlo{} jet-vetoed $+0$-jet cross section {\smaller$\ss{n}{0+1+2+3}{Q(\Phi_{1})<Q_c \land Q(\Phi_{2})<Q_c \land Q(\Phi_{3})<Q_c}{\Phi_{0}}$} ($\equiv$ {\smaller$\ss{n}{0+1+2+3}{\mathrm{EXC}}{\Phi_{0}}$} for brevity). The recent emergence of \nnnlo{} differential cross sections in $q_\perp$ subtraction
and Projection-to-Born methods suggests that this is not a particularly far-fetched requirement. The construction of the simple third-order matching method proceeds in the same spirit as the $\unnlops$ matching prescription. Concretely, the method is constructed in the following steps:

\begin{nonfloatfig}
\renewcommand{\arraystretch}{1.5}
\begin{tabular}{ |l p{0.96\textwidth}|}
\hline
{\bf A.}\quad & Regularize a one-jet ${\unnlops}$ calculation with $\Delta_{n} (t_+,t_{n+1})$ factors so that the hardest jet can become unresolved. \\
{\bf B.}\quad & Remove unwanted \nnlo{} terms from the regularized the one-jet spectrum.\\
{\bf C.}\quad & Unitarize, i.e. subtract one-jet spectrum from zero-jet terms.\\
{\bf D.}\quad & Include and \nnnlo{} jet-vetoed zero-jet cross section.\\
\hline
\end{tabular}
\end{nonfloatfig}
The result is a valid third-order matching method, and will be referred to as \nnnlops{} (for third-order matched transition events).
It should be noted that the prescription does not depend on the toy fixed-order calculation introduced before -- the \nnnlops{} method would yield an \nnnlopps{}  prediction given appropriate inputs. Furthermore, the scheme does not depend on the details of the parton shower, meaning that improved showers with reduced uncertainty will directly improve the \nnnlops{} method and decrease its uncertainty in regions dominated by partons becoming unresolved.

\subsection{Constructing the improved radiation pattern}

\noindent
The baseline for producing a precise radiation pattern is eq.~\ref{eq:un2lops} after performing the shift $n\rightarrow n+1$. In this case, \nnlo{} accuracy means that no unwanted terms are introduced at $\mathcal{O}(\alpha_s^3)$. Conversely, this means that reweighting most contributions in {\smaller$\ff{\mathcal{F}}{n+1}{\infty}{\mathrm{\unnlops}}{\Phi_{n+1}, t_+,t_-}$} with Sudakov- and running-coupling factors will only result in $\mathcal{O}(\alpha_s^4)$ shifts. There are two terms term that, when reweighted, threaten to introduce undesirable contributions. First first is the is the jet-vetoed 1-parton cross section {\smaller$\ss{n+1}{0+1+2}{Q_{n+2}<Q_c \land Q_{n+3}<Q_c}{\Phi_{n+1}}$}. The potentially problematic term can be isolated by
\begin{eqnarray}
\label{eq:replacement}
\ss{n+1}{0+1+2}{Q_{n+2}<Q_c \land Q_{n+3}<Q_c}{\Phi_{n+1}}
 &=& \ss{n+1}{2}{Q_{n+2}<Q_c \land Q_{n+3}<Q_c}{\Phi_{n+1}} + \ss{n+1}{1}{Q_{n+2}<Q_c}{\Phi_{n+1}} \nonumber\\
&&+ 
\s{n+1}{0}{\Phi_{n+1}}
\end{eqnarray}
Any reweighting of the first line will only introduce terms of $\mathcal{O}(\alpha_s^4)$ or higher, while the second line requires careful consideration. Further unsafe terms are highlighted through \raisebox{1pt}{\tcboxmath[colback=blue!10,boxsep=1.5pt]{\cdot}} in eq.~\ref{eq:un2lops}. An appropriate weighting strategy is outlined in sec. \ref{sec:fixingweights}. 
Performing the replacement in eq.\ \ref{eq:replacement}, and introducing the placeholders $u_1,u_2, u_3$ and $u_4$ for the correct weights of dangerous terms, the one-jet contribution for \nnnlops{} can be written as
\begin{eqnarray}
&&\ff{\mathcal{F}}{n+1}{\infty}{\mathrm{\unnlops{}\,4\,{\smaller TOMTE}\,}}{\Phi_{n+1}, t_+,t_-}\nonumber\\
&&\coloneqq \obs{n+1} \Bigg( 
\ss{n+1}{2}{Q_{n+2}<Q_c \land Q_{n+3}<Q_c}{\Phi_{n+1}}~\f{w}{n+1}{\infty}{\Phi_{n+1}} \Delta_{n} (t_+,t_{n+1})\nonumber\\
&&+~ \ss{n+1}{1}{Q_{n+2}<Q_c}{\Phi_{n+1}} u_1 + \ss{n+1}{1}{Q_{n+2}<Q_c}{\Phi_{n+1}}~u_2 \nonumber\\
&&+~\intOne\ss{n+2}{0}{Q_{n+2}>Q_c}{\Phi_{n+2}} ~u_3
\nonumber\\
&&+~\intOne\ss{n+2}{1}{Q_{n+2}>Q_c \land Q_{n+3}<Q_c }{\Phi_{n+2}} ~ \f{w}{n+1}{\infty}{\Phi_{n+1}} \Delta_{n} (t_+,t_{n+1})~ \nonumber\\
&&\qquad\otimes~\left[\one{n+2}{n+1}- ~\tcboxmath[colback=red!10]{~\f{w}{n+2}{\infty}{\Phi_{n+2}} \Delta_{n+1} (t(\Phi_{n+1},t_{n+2})~}~\right] \nonumber\\
&&+~\intTwo\ss{n+3}{0}{Q_{n+3}>Q_c}{\Phi_{n+3}}  ~\f{w}{n+1}{\infty}{\Phi_{n+1}} \Delta_{n} (t_+,t_{n+1})\nonumber\\
&&\qquad\otimes~ \left[\one{n+3}{n+1}- ~\tcboxmath[colback=orange!10]{~\f{w}{n+2}{\infty}{\Phi_{n+2}} \Delta_{n+1} (t_{n+1},t_{n+2})\cdot \one{n+3}{n+2} ~}~\right]
\Bigg) \nonumber\\
&&+~
\obs{n+2} \Bigg(~\ss{n+2}{0}{Q_{n+2}>Q_c}{\Phi_{n+2}} ~ u_4\nonumber\\
&&+~ \ss{n+2}{1}{Q_{n+2}>Q_c \land Q_{n+3}<Q_c}{\Phi_{n+2}} \nonumber\\
&&\qquad\otimes~\tcboxmath[colback=red!10]{~\f{w}{n+1}{\infty}{\Phi_{n+1}}\f{w}{n+2}{\infty}{\Phi_{n+2}} \Delta_{n} (t_+,t_{n+1})\Delta_{n+1} (t[\Phi_{n+1},t_{n+2})~} \nonumber\\
&&+~ \intOne\ss{n+3}{0}{Q_{n+3}>Q_c}{\Phi_{n+3}} \Big[ ~\tcboxmath[colback=orange!10]{~\f{w}{n+1}{\infty}{\Phi_{n+1}}\f{w}{n+2}{\infty}{\Phi_{n+2}}\Delta_{n} (t_+,t_{n+1}) \Delta_{n+1} (t_{n+1},t_{n+2})\cdot\one{n+3}{n+2}~}
\nonumber\\
&&-~\tcboxmath[colback=blue!10]{~\w{n+1} \w{n+2} \w{n+3} \Delta_{n} (t_+,t_{n+1}) \Delta_{n+1} (t_{n+1},t_{n+2}) \Delta_{n+2} (t_{n+2},t_{n+3})~}~\Big] \Bigg) \nonumber\\
&&+~\ss{n+3}{0}{Q_{n+3}>Q_c}{\Phi_{n+3}} \nonumber\\
&&\quad\otimes~\tcboxmath[colback=blue!10]{~  \w{n+1} \w{n+2}\w{n+3} \Delta_{n} (t_+,t_{n+1}) \Delta_{n+1} (t_{n+1},t_{n+2}) \Delta_{n+2} (t_{n+2},t_{n+3}) ~}\nonumber\\
&&\quad\otimes~\f{\mathcal{F}}{n+3}{\infty}{\Phi_{n+3}, t_{n+3},t_-}~,
\end{eqnarray}
where $t_{n+1}$ was chosen as upper scale in $\Delta_{n+1}$ to ensure non-overlapping resummation regions, as required to maintain the parton-shower accuracy.

\subsection{Removing undesirable $\mathcal{O}(\alpha_s^3)$ terms from the radiation pattern}
\label{sec:fixingweights}

\noindent
To obtain an appropriate radiation pattern, a correct weighting strategy for the terms proportional to
\begin{eqnarray}
\label{eq:lo4reweighting}
&&\ss{n+2}{0}{Q_{n+2}>Q_c}{\Phi_{n+2}}\qquad \textnormal{and}\\
\label{eq:nloexc4reweighting}
&&\ss{n+1}{0+1}{Q_{n+2}<Q_c}{\Phi_{n+1}} =
\s{n+1}{0}{\Phi_{n+1}} + \ss{n+1}{1}{Q_{n+2}<Q_c}{\Phi_{n+1}}
\end{eqnarray}
has be be established. This weighting should not introduce unwanted $\mathcal{O}(\alpha_s^3)$ terms, while ensuring that the resulting terms are regularized as $Q_c\rightarrow 0$. Luckily, the $\obs{n+1}$ contribution to the \unnlops{} matching formula for $n$-parton processes (eq.~\ref{eq:un2lops}) can act as a blueprint, since that calculation can be regarded as an approximation to the $\obs{n+1}$ contribution in \nnnlops{}. In the $\obs{n+1}$ contribution of \unnlops, the tree-level components are multiplied by the  
subtracted parton-shower factors
\begin{eqnarray*}
~\Big[ 1 - \f{w}{n+1}{1}{\Phi_{n+1}} - \f{\Delta}{n}{1}{t_+,t_{n+1}}\Big] \Delta_{n} (t_+,t_{n+1}) \f{w}{n+1}{\infty}{\Phi_{n+1}}~,
\end{eqnarray*}
to avoid over-counting universal virtual corrections~\cite{Hoeche:2014aia}. The same logic can be applied to the parton-shower reweighting for the $\mathcal{O}(\alpha_s^2)$ contributions to eq.\ \ref{eq:nloexc4reweighting}, i.e.
\begin{eqnarray}
&&\ss{n+1}{1}{Q_{n+2}<Q_c}{\Phi_{n+1}}\nonumber\\
&       \xrightarrow{
            \begin{subarray}{l}
               \textnormal{\smaller eq.\ \ref{eq:un2lops}}
            \end{subarray}
        }
&~
\ss{n+1}{1}{Q_{n+2}<Q_c}{\Phi_{n+1}}
~\Delta_{n} (t_+,t_{n+1})
\f{w}{n+1}{\infty}{\Phi_{n+1}}\nonumber\\
&       \xrightarrow{
            \begin{subarray}{l}
                \textnormal{\smaller repair $\mathcal{O}(\alpha_s^1)$ weights}
            \end{subarray}
        }
&~
\ss{n+1}{1}{Q_{n+2}<Q_c}{\Phi_{n+1}}
~\Delta_{n} (t_+,t_{n+1})\f{w}{n+1}{\infty}{\Phi_{n+1}}
~\textnormal{\smaller$\Big[ 1 - \f{w}{n+1}{1}{\Phi_{n+1}} - \f{\Delta}{n}{1}{t_+,t_{n+1}}\Big]$}~.\qquad
\end{eqnarray}
This determines $u_2$. 

The weighting of the $\mathcal{O}(\alpha_s^1)$ pieces of eq.\ \ref{eq:nloexc4reweighting} also need careful consideration. Starting from eq.\ \ref{eq:un2lops}, reweighting with subtracted parton-shower factors is appropriate. However, the subtracted parton-shower terms need to be expanded to second order to avoid spurious $\mathcal{O}(\alpha_s^3)$ terms. This suggests the shifts
\begin{eqnarray}
&&\s{n+1}{0}{\Phi_{n+1}}\nonumber\\
&       \xrightarrow{
            \begin{subarray}{l}
               \textnormal{\smaller eq.\ \ref{eq:un2lops}}
            \end{subarray}
        }
&~
\s{n+1}{0}{\Phi_{n+1}} ~\Delta_{n} (t_+,t_{n+1})~ \f{w}{n+1}{\infty}{\Phi_{n+1}}~\Big[ 1 - \f{w}{n+1}{1}{\Phi_{n+1}} - \f{\Delta}{n}{1}{t_+,t_{n+1}} \Big] \nonumber\\
&       \xrightarrow{
            \begin{subarray}{l}
                \textnormal{\smaller repair $\mathcal{O}(\alpha_s^2)$ weights}
            \end{subarray}
        }
&~
\s{n+1}{0}{\Phi_{n+1}} ~\Delta_{n} (t_+,t_{n+1})\f{w}{n+1}{\infty}{\Phi_{n+1}}\nonumber\\
&&\otimes~\Big[  1 - \f{w}{n+1}{1}{\Phi_{n+1}} - \f{w}{n+1}{2}{\Phi_{n+1}}  - \f{\Delta}{n}{1}{t_+,t_{n+1}} -  \f{\Delta}{n}{2}{t_+,t_{n+1}}\nonumber\\
&&\quad\, + \left(\f{\Delta}{n}{1}{t_+,t_{n+1}}\right)^2 + \left(\f{w}{n+1}{1}{\Phi_{n+1}}\right)^2 + \f{w}{n+1}{1}{\Phi_{n+1}} \f{\Delta}{n}{1}{t_+,t_{n+1}} \Big]
\end{eqnarray}
to ensure an appropriately weighted \nlo{} cross section for exclusive $(n+1)$-parton configurations. The last line removes undesirable $\mathcal{O}(\alpha_s^2)$ terms in the expansion of the all-order-reweighted $\mathcal{O}(\alpha_s^1)$ subtractions, and determines $u_1$.

Finally, the two-jet leading-order contribution (eq.\ \ref{eq:lo4reweighting}) requires care, since it appears with observable dependence $\obs{n+2}$, and, through unitarization and to complement the jet-vetoed cross section, with $\obs{n+1}$ dependence. This leads to several boundary conditions on suitable weights, as discussed in Appendix \ref{app:giggles}.
A suitable strategy is to use
\begin{eqnarray}
&&\obs{n+2}~\ss{n+2}{0}{Q_{n+2}>Q_c}{\Phi_{n+2}} ~u_4 
= 
\obs{n+2}~\ss{n+2}{0}{Q_{n+2}>Q_c}{\Phi_{n+2}} 
\nonumber\\
&&\qquad\otimes~
\Delta_{n} (t_+,t_{n+1}) \Delta_{n+1} (t_{n+1},t_{n+2}) \f{w}{n+1}{\infty}{\Phi_{n+1}} \f{w}{n+2}{\infty}{\Phi_{n+2}}
\nonumber\\
&&\qquad\otimes~\Big( 1 - \f{w}{n+1}{1}{\Phi_{n+1}} - \f{w}{n+2}{1}{\Phi_{n+2}} - \f{\Delta}{n}{1}{t_+,t_{n+1}} - \f{\Delta}{n+1}{1}{t_{n+1},t_{n+2}} \Big)
~,
\label{eq:two-parton-O2}
\end{eqnarray}
as contribution to $\obs{n+2}$ observables, fixing $u_4$. The contribution to $\obs{n+1}$ cross sections contains 
the complement to a jet-vetoed cross section, and the unitarity subtraction of the $\obs{n+2}$ contribution. Thus, 
\begin{eqnarray}
&&\obs{n+1}~\ss{n+2}{0}{Q_{n+2}>Q_c}{\Phi_{n+2}} ~u_3 
= 
\obs{n+1}~\ss{n+2}{0}{Q_{n+2}>Q_c}{\Phi_{n+2}} 
\nonumber\\
&&\qquad\otimes~\Bigg[
~~\Delta_{n} (t_+,t_{n+1})\f{w}{n+1}{\infty}{\Phi_{n+1}}
~
\cdot
\Big( 1 - \f{w}{n+1}{1}{\Phi_{n+1}} - \f{\Delta}{n}{1}{t_+,t_{n+1}}\Big)
\nonumber\\
&&\qquad~~~\,
-
\Delta_{n} (t_+,t_{n+1})\f{w}{n+1}{\infty}{\Phi_{n+1}} \Delta_{n+1} (t_{n+1},t_{n+2}) \f{w}{n+2}{\infty}{\Phi_{n+2}}
\nonumber\\
&&\qquad~~~~~~~
\cdot
\Big( 1 - \f{w}{n+1}{1}{\Phi_{n+1}} - \f{w}{n+2}{1}{\Phi_{n+2}} - \f{\Delta}{n}{1}{t_+,t_{n+1}} - \f{\Delta}{n+1}{1}{t_{n+1},t_{n+2}} \Big)
\Bigg]~.
\label{eq:two-parton-O1}
\end{eqnarray}
This then defines $u_3$, and concludes the discussion of terms that cannot be trivially reweighted.
%\newpage
%\noindent
After these considerations, the complete $(n+1)$-parton spectrum for \nnnlops{} is finally given by
\begin{eqnarray}
&&\ff{\mathcal{F}}{n+1}{\infty}{\mathrm{\unnlops{}\,4\,{\smaller TOMTE}\,}}{\Phi_{n+1}, t_+,t_-}
%\nonumber\\
%&&
\coloneqq \obs{n+1} \Bigg( 
\ss{n+1}{2}{Q_{n+2}<Q_c \land Q_{n+3}<Q_c}{\Phi_{n+1}}~\Delta_{n} (t_+,t_{n+1})\f{w}{n+1}{\infty}{\Phi_{n+1}} \nonumber\\
&&+~
\s{n+1}{0}{\Phi_{n+1}} ~\Delta_{n} (t_+,t_{n+1})\f{w}{n+1}{\infty}{\Phi_{n+1}}
\nonumber\\
&&\qquad\otimes~\Big[                  
1 - \f{w}{n+1}{1}{\Phi_{n+1}} - \f{w}{n+1}{2}{\Phi_{n+1}}  - \f{\Delta}{n}{1}{t_+,t_{n+1}} -  \f{\Delta}{n}{2}{t_+,t_{n+1}}
\nonumber\\
&&\qquad\quad + \left(\f{\Delta}{n}{1}{t_+,t_{n+1}}\right)^2 + \left(\f{w}{n+1}{1}{\Phi_{n+1}}\right)^2 + \f{w}{n+1}{1}{\Phi_{n+1}} \f{\Delta}{n}{1}{t_+,t_{n+1}} \Big]
\nonumber\\
&&+~ \ss{n+1}{1}{Q_{n+2}<Q_c}{\Phi_{n+1}} \Delta_{n} (t_+,t_{n+1})  \f{w}{n+1}{\infty}{\Phi_{n+1}}
%\nonumber\\
%&&\qquad
\otimes~
\Big[ 1 - \f{w}{n+1}{1}{\Phi_{n+1}} - \f{\Delta}{n}{1}{t_+,t_{n+1}}\Big] 
\nonumber\\
&&+~\intOne\ss{n+2}{0}{Q_{n+2}>Q_c}{\Phi_{n+2}} ~\Delta_{n} (t_+,t_{n+1})\f{w}{n+1}{\infty}{\Phi_{n+1}}
\nonumber\\
&&\qquad\otimes~\Bigg[
~
\Big( 1 - \f{w}{n+1}{1}{\Phi_{n+1}} - \f{\Delta}{n}{1}{t_+,t_{n+1}}\Big) \cdot \one{n+2}{n+1}
\nonumber\\
&&\qquad\qquad~~~\,
-
~\tcboxmath[colback=green!10]{\!\!\phantom{\int}\Big( 1 - \f{w}{n+1}{1}{\Phi_{n+1}} - \f{w}{n+2}{1}{\Phi_{n+2}} - \f{\Delta}{n}{1}{t_+,t_{n+1}} - \f{\Delta}{n+1}{1}{t_{n+1},t_{n+2}} \Big)~} 
\nonumber\\
&&\qquad\qquad~~~~~\otimes~
\Delta_{n+1} (t_{n+1},t_{n+2}) \f{w}{n+2}{\infty}{\Phi_{n+2}}
\Bigg]
\nonumber\\
&&+~\intOne\ss{n+2}{1}{Q_{n+2}>Q_c \land Q_{n+3}<Q_c }{\Phi_{n+2}} ~ \Delta_{n} (t_+,t_{n+1}) \f{w}{n+1}{\infty}{\Phi_{n+1}}  ~ 
%\nonumber\\
%&&\qquad
\otimes~\left[ \one{n+2}{n+1} - ~\tcboxmath[colback=red!10]{\!\!\phantom{\int}\f{w}{n+2}{\infty}{\Phi_{n+2}} \Delta_{n+1} (t(\Phi_{n+1},t_{n+2})~}~\right] \nonumber\\
&&+~\intTwo\ss{n+3}{0}{Q_{n+3}>Q_c}{\Phi_{n+3}}  ~ \Delta_{n} (t_+,t_{n+1}) \f{w}{n+1}{\infty}{\Phi_{n+1}} 
%\nonumber\\
%&&\qquad
\otimes~ \left[ \one{n+3}{n+1} - ~\tcboxmath[colback=orange!10]{\!\!\phantom{\int}\f{w}{n+2}{\infty}{\Phi_{n+2}} \Delta_{n+1} (t_{n+1},t_{n+2}) \cdot \one{n+3}{n+2} ~}~\right]
\Bigg) \nonumber\\
&&+~
\obs{n+2} \Bigg(~\ss{n+2}{0}{Q_{n+2}>Q_c}{\Phi_{n+2}} 
\nonumber\\
&&\qquad
\otimes~
\Delta_{n} (t_+,t_{n+1})\Delta_{n+1} (t_{n+1},t_{n+2}) \f{w}{n+1}{\infty}{\Phi_{n+1}}  \f{w}{n+2}{\infty}{\Phi_{n+2}}
\nonumber\\
&&\qquad\otimes~
\tcboxmath[colback=green!10]{\!\!\phantom{\int}\Big[ 1 - \f{w}{n+1}{1}{\Phi_{n+1}} - \f{w}{n+2}{1}{\Phi_{n+2}} - \f{\Delta}{n}{1}{t_+,t_{n+1}} - \f{\Delta}{n+1}{1}{t_{n+1},t_{n+2}} \Big]~}
\nonumber\\
&&+~ \ss{n+2}{1}{Q_{n+2}>Q_c \land Q_{n+3}<Q_c}{\Phi_{n+2}} \nonumber\\
&&\qquad\otimes~\tcboxmath[colback=red!10]{\!\!\phantom{\int}\Delta_{n} (t_+,t_{n+1})\Delta_{n+1} (t_{n+1},t_{n+2}) \f{w}{n+1}{\infty}{\Phi_{n+1}}\f{w}{n+2}{\infty}{\Phi_{n+2}} ~} \nonumber\\
&&+~ \intOne\ss{n+3}{0}{Q_{n+3}>Q_c}{\Phi_{n+3}} \Big[ ~\tcboxmath[colback=orange!10]{\!\!\phantom{\int}\Delta_{n} (t_+,t_{n+1}) \Delta_{n+1} (t_{n+1},t_{n+2})\f{w}{n+1}{\infty}{\Phi_{n+1}}\f{w}{n+2}{\infty}{\Phi_{n+2}}\cdot \one{n+3}{n+2} ~}
\nonumber\\
&&-~\tcboxmath[colback=blue!10]{\!\!\phantom{\int}\Delta_{n} (t_+,t_{n+1}) \Delta_{n+1} (t_{n+1},t_{n+2}) \Delta_{n+2} (t_{n+2},t_{n+3})\w{n+1} \w{n+2} \w{n+3} ~}~\Big] \Bigg) \nonumber\\
&&+~\ss{n+3}{0}{Q_{n+3}>Q_c}{\Phi_{n+3}} 
%\nonumber\\
%&&\quad
\otimes~\tcboxmath[colback=blue!10]{\!\!\phantom{\int}\Delta_{n} (t_+,t_{n+1}) \Delta_{n+1} (t_{n+1},t_{n+2}) \Delta_{n+2} (t_{n+2},t_{n+3}) \w{n+1} \w{n+2}\w{n+3} ~}\nonumber\\
&&\quad\otimes~\f{\mathcal{F}}{n+3}{\infty}{\Phi_{n+3}, t_{n+3},t_-}~.
\label{eq:nnnlops-real}
\end{eqnarray}
Although a somewhat more complicated combination than the \unnlops\ prescription in eq.~\ref{eq:un2lops}, the only two new factors that need to be calculated are {\smaller$\f{w}{n+1}{2}{\Phi_{n+1}}$} and {\smaller$\f{\Delta}{n}{2}{t_+,t_{n+1}}$}. Both factors are simple to generate in the absence of incoming hadrons. Were \pdf{}s to enter the parton-shower evolution, then the generation would become more cumbersome\footnote{The appearance of ratios of \pdf{}s in the Sudakov exponents would lead to contributions from the first-order expansion of the Sudakov factor mixing with the first-order expanded \pdf{} evolution, the product of first-order expanded running-coupling factors with the first-order expansion of \pdf{} evaluations at dynamical factorization scales, and the second-order expansion of \pdf{}s evaluated at dynamical factorization scales. The latter depends on the treatment of running couplings within the \pdf{} fitting/evolution procedure.}. However, it might be feasible to evaluate the pieces for color-singlet production processes at hadron colliders, since the terms are related to the evolution of zero-parton configurations and dynamical scale choices for the one-jet contributions alone. For example, a hybrid approach similar to the \textsc{Minlo} method~\cite{Hamilton:2012rf} could be possible: Employ analytical Sudakov factors and their expansion for the factors related to the evolution of zero-jet configurations ($\Delta_{n}^{(\infty)}$, $\Delta_{n}^{(1)}$, ${\Delta}_{n}^{(2)}$), and use numerical parton-shower results everywhere else. It should be noted that eq.\ \ref{eq:nnnlops-real} is an \nnlopps{} matching formula for processes with an ``additional" parton. The benefits of eq.\ \ref{eq:nnnlops-real} over, e.g., \unnlops\ matching are that the additional parton is allowed to become soft, or collinear to another parton. Furthermore, eq.\ \ref{eq:nnnlops-real} offers a clear, resummation-based, strategy for setting the argument of the running coupling for $\Phi_{n+1}$ states.

\subsection{Unitarizing}

\noindent
After constructing an appropriate radiation pattern, unitarization is necessary to ensure that the target inclusive \nnnlo{} zero-jet is retained after matching. Unitarization further embeds the effects of parton-shower resummation at higher orders into zero-jet exclusive predictions. To this end, all observables in eq.\ \ref{eq:nnnlops-real} are replaced with $\obs{n}$, and the whole contribution is subtracted from the previous result. Using $\obs{n}$ in eq.\ \ref{eq:nnnlops-real} triggers, as desired, the cancellation between the terms highlighted in
\raisebox{1pt}{\tcboxmath[colback=red!10,boxsep=1.5pt]{\cdot}}, \raisebox{1pt}{\tcboxmath[colback=green!10,boxsep=1.5pt]{\cdot}}, \raisebox{1pt}{\tcboxmath[colback=blue!10,boxsep=1.5pt]{\cdot}}
and \raisebox{1pt}{\tcboxmath[colback=orange!10,boxsep=1.5pt]{\cdot}}. After this, the prototype subtraction for unitarization reads\footnote{Using
$\bar\Phi_1=\{t_{i+1},z_{i+1},\phi_{i+1}\},~ \widetilde\Phi_1=\{t_{i+2},z_{i+2},\phi_{i+2}\},~\Phi_1=\{t_{i+3},z_{i+3},\phi_{i+3}\},~
 \Phi_{i+1}=\{ \Phi_{i} \cap \bar\Phi_1\},~ \Phi_{i+2}=\{ \Phi_{i+1} \cap \widetilde\Phi_1\}$ and $\Phi_{i+3}=\{ \Phi_{i+2} \cap \Phi_1\}$, and defining the notation
\begin{eqnarray*}
\intThree \sigma_{i+3}(\Phi_{i+3}) &=&
\int\!\! d\bar  \Phi_1 ~\Theta[t(\Phi_{i+1}) - t^-] ~\Theta[t^+ - t(\Phi_{i+1})]
\int\!\! d\widetilde\Phi_1 ~\Theta[t(\Phi_{i+2}) - t^-] ~\Theta[t^+ - t(\Phi_{i+2})]\\
&&
\int\!\! d      \Phi_1 ~\Theta[t(\Phi_{i+3}) - t^-] ~\Theta[t^+ - t(\Phi_{i+3})]~\sigma_{i+3}(\Phi_{n+3})~.
\end{eqnarray*}
}
\begin{eqnarray}
&&\ss{n}{\infty}{\mathrm{SUBT}}{\Phi_{n}} \coloneqq ~\obs{n}~\Bigg\{
%\nonumber\\
%&&
-\intOne\ss{n+1}{2}{Q_{n+2}<Q_c \land Q_{n+3}<Q_c}{\Phi_{n+1}}~\Delta_{n} (t_+,t_{n+1}) \f{w}{n+1}{\infty}{\Phi_{n+1}} \nonumber\\
&&-~
\intOne\s{n+1}{0}{\Phi_{n+1}} ~\Delta_{n} (t_+,t_{n+1}) \f{w}{n+1}{\infty}{\Phi_{n+1}} \nonumber\\
&&\qquad\otimes~\Big[                  
1 - \f{w}{n+1}{1}{\Phi_{n+1}} - \f{w}{n+1}{2}{\Phi_{n+1}}  - \f{\Delta}{n}{1}{t_+,t_{n+1}} -  \f{\Delta}{n}{2}{t_+,t_{n+1}}
\nonumber\\
&&\qquad\quad + \left(\f{\Delta}{n}{1}{t_+,t_{n+1}}\right)^2 + \left(\f{w}{n+1}{1}{\Phi_{n+1}}\right)^2 + \f{w}{n+1}{1}{\Phi_{n+1}} \f{\Delta}{n}{1}{t_+,t_{n+1}} \Big]
\nonumber\\
&&-~ \intOne\ss{n+1}{1}{Q_{n+2}<Q_c}{\Phi_{n+1}}
~\Big[ 1 - \f{w}{n+1}{1}{\Phi_{n+1}} - \f{\Delta}{n}{1}{t_+,t_{n+1}}\Big] \Delta_{n} (t_+,t_{n+1})  \f{w}{n+1}{\infty}{\Phi_{n+1}}  \nonumber\\
&&-~\intTwo \ss{n+2}{0}{Q_{n+2}>Q_c}{\Phi_{n+2}} ~\Big[  1 - \f{w}{n+1}{1}{\Phi_{n+1}} - \f{\Delta}{n}{1}{t_+,t_{n+1}}\Big] \Delta_{n} (t_+,t_{n+1}) \f{w}{n+1}{\infty}{\Phi_{n+1}} 
\nonumber\\
&&-~\intTwo\ss{n+2}{1}{Q_{n+2}>Q_c \land Q_{n+3}<Q_c }{\Phi_{n+2}} ~ \Delta_{n} (t_+,t_{n+1}) \f{w}{n+1}{\infty}{\Phi_{n+1}} \nonumber\\
&&-~\intThree\ss{n+3}{0}{Q_{n+3}>Q_c}{\Phi_{n+3}}  ~ \Delta_{n} (t_+,t_{n+1}) \f{w}{n+1}{\infty}{\Phi_{n+1}} \Bigg\}
\end{eqnarray}

\newpage

\subsection{Completing the N3LO cross section to obtain the matching formula}

\noindent
The last step in construction the \nnnlops{} matching method is to complement the matching formula with an \nnnlo{} exclusive cross section {\smaller$\ss{n}{0+1+2+3}{\mathrm{EXC}}{\Phi_{n}}$}, and to ensure that the complementary parts (with real-emission configurations above the veto scale $Q_c$) are correctly included. 
This can be achieved by shifting the unitarization subtraction to \emph{not remove} the necessary terms: 
\begin{eqnarray}
&&\ss{n}{\infty}{\mathrm{SUBT}}{\Phi_{n}}\nonumber\\
&&\rightarrow
\ss{n}{\infty}{\mathrm{SUBT+COMPLEMENT}}{\Phi_{n}}~\coloneqq ~\obs{n}~\Bigg\{ \nonumber\\
&&
~~~\intOne\Big[~\textcolor{red}{\one{n+1}{n}}-~ \Delta_{n} (t_+,t_{n+1}\f{w}{n+1}{\infty}{\Phi_{n+1}}\Big]
\ss{n+1}{2}{Q_{n+2}<Q_c \land Q_{n+3}<Q_c}{\Phi_{n+1}}\nonumber\\
&&
+~\intOne\Big[
~\textcolor{red}{\one{n+1}{n}}-
~ \Delta_{n} (t_+,t_{n+1}) \f{w}{n+1}{\infty}{\Phi_{n+1}} ~ \Big(                  
1 - \f{w}{n+1}{1}{\Phi_{n+1}} - \f{w}{n+1}{2}{\Phi_{n+1}}  - \f{\Delta}{n}{1}{t_+,t_{n+1}} -  \f{\Delta}{n}{2}{t_+,t_{n+1}}
\nonumber\\
&&\qquad\quad~ + \left(\f{\Delta}{n}{1}{t_+,t_{n+1}}\right)^2 + \left(\f{w}{n+1}{1}{\Phi_{n+1}}\right)^2 + \f{w}{n+1}{1}{\Phi_{n+1}} \f{\Delta}{n}{1}{t_+,t_{n+1}} \Big)\Big]
~\s{n+1}{0}{\Phi_{n+1}} 
\nonumber\\
&&
+~\intOne\Big[~\textcolor{red}{\one{n+1}{n}} -
~\Delta_{n} (t_+,t_{n+1})  \f{w}{n+1}{\infty}{\Phi_{n+1}} ~\Big( 1 - \f{w}{n+1}{1}{\Phi_{n+1}} - \f{\Delta}{n}{1}{t_+,t_{n+1}}\Big)  \Big]~\ss{n+1}{1}{Q_{n+2}<Q_c}{\Phi_{n+1}} \nonumber\\
&&
+~\intTwo\Big[~\textcolor{red}{\one{n+2}{n}} 
-~\Delta_{n} (t_+,t_{n+1}) \f{w}{n+1}{\infty}{\Phi_{n+1}} ~\Big( 1 - \f{w}{n+1}{1}{\Phi_{n+1}} - \f{\Delta}{n}{1}{t_+,t_{n+1}}\Big) \Big]~\ss{n+2}{0}{Q_{n+2}>Q_c}{\Phi_{n+2}}
\nonumber\\
&&
+~\intTwo\Big[~\textcolor{red}{\one{n+2}{n}} -~ \Delta_{n} (t_+,t_{n+1}) \f{w}{n+1}{\infty}{\Phi_{n+1}}\Big]~\ss{n+2}{1}{Q_{n+2}>Q_c \land Q_{n+3}<Q_c }{\Phi_{n+2}} \nonumber\\
&&
+~\intThree\Big[~\textcolor{red}{\one{n+3}{n}} -~ \Delta_{n} (t_+,t_{n+1}) \f{w}{n+1}{\infty}{\Phi_{n+1}} \Big]~\ss{n+3}{0}{Q_{n+3}>Q_c}{\Phi_{n+3}}~\Bigg\}
\label{eq:subtraction+complement}
\end{eqnarray}
where the necessary factors to complement the exclusive cross sections are highlighted in red. It should be stressed that these factors are introduced because the combination with a jet-vetoed \nnnlo{} cross section is foreseen. The combination with an inclusive \nnnlo{} calculation can be accommodated by simply ignoring the highlighted factors.

\newpage
\noindent
Combining {\smaller$\ss{n}{0+1+2+3}{\mathrm{EXC}}{\Phi_{0}}$} with eq.\ \ref{eq:subtraction+complement} and eq.\ \ref{eq:nnnlops-real} allows to construct the \nnnlops{} matching formula. As before, pairwise canceling terms will indicated with identical (hyperlinked) boxes. This acts as visual help to allow the reader to confirm that the criteria listed in Table \ref{tab:criteria} are indeed fulfilled. The final \nnnlops{} matching formula reads%\footnote{This differs slightly from the preliminary suggestion presented at the workshop ``Taming the accuracy of event generators, CERN, June 2020", which treated the $\ss{n+2}{0}{Q_{n+2}>Q_c}{\Phi_{n+2}}$ terms incorrectly, and used a simpler scheme for running-coupling effects.}
\begin{eqnarray}
&&\ff{\mathcal{F}}{n}{\infty}{\mathrm{{\smaller TOMTE}}\,}{\Phi_n, t_+,t_-} \nonumber\\
&&\coloneqq \obs{n}~\Bigg\{ 
\ss{n}{0+1+2+3}{\mathrm{EXC}}{\Phi_{n}}\nonumber\\
&&
+~
\intOne\ss{n+1}{2}{Q_{n+2}<Q_c \land Q_{n+3}<Q_c}{\Phi_{n+1}}~
\Big[~\one{n+1}{n} -~\tcboxmath[colback=black!00,colframe=red,boxrule=0.5pt,boxsep=1pt,hypertarget=cfredB,hyperlink=cfredA]{\!\!\phantom{\int}
\Delta_{n} (t_+,t_{n+1}) \f{w}{n+1}{\infty}{\Phi_{n+1}}  ~}~\Big]
\nonumber\\
&&
+~
\intOne\s{n+1}{0}{\Phi_{n+1}}~
\Big[~\one{n+1}{n}- 
\tcboxmath[colback=black!00,colframe=myolive,boxrule=0.5pt,boxsep=1pt,hypertarget=cfoliveB,hyperlink=cfoliveA]{
\begin{aligned}
&~\Delta_{n} (t_+,t_{n+1}) \f{w}{n+1}{\infty}{\Phi_{n+1}} ~\nonumber\\
&\quad\cdot\Big(                  
1 - \f{w}{n+1}{1}{\Phi_{n+1}} - \f{w}{n+1}{2}{\Phi_{n+1}}  - \f{\Delta}{n}{1}{t_+,t_{n+1}} -  \f{\Delta}{n}{2}{t_+,t_{n+1}}
\nonumber\\
&\quad~ + \left[\f{\Delta}{n}{1}{t_+,t_{n+1}}\right]^2 + \left[\f{w}{n+1}{1}{\Phi_{n+1}}\right]^2 + \f{w}{n+1}{1}{\Phi_{n+1}} \f{\Delta}{n}{1}{t_+,t_{n+1}} \Big)
\end{aligned}
}
~\Big]
\nonumber\\
&&
+~\intOne\ss{n+1}{1}{Q_{n+2}<Q_c}{\Phi_{n+1}}
~\Big[~\one{n+1}{n} -
~\tcboxmath[colback=black!00,colframe=mypurple,boxrule=0.5pt,boxsep=1pt,hypertarget=cfpurpleB,hyperlink=cfpurpleA]{\!\!\phantom{\int}
\Delta_{n} (t_+,t_{n+1})  \f{w}{n+1}{\infty}{\Phi_{n+1}} ~\Big( 1 - \f{w}{n+1}{1}{\Phi_{n+1}} - \f{\Delta}{n}{1}{t_+,t_{n+1}}\Big) ~}~ \Big] \nonumber\\
&&
+~\intTwo\ss{n+2}{0}{Q_{n+2}>Q_c}{\Phi_{n+2}}~
\Big[~\one{n+2}{n} 
-
~\tcboxmath[colback=mybrown!20,hypertarget=brownB,hyperlink=brownA]{\!\!\phantom{\int}
\Delta_{n} (t_+,t_{n+1}) \f{w}{n+1}{\infty}{\Phi_{n+1}} ~\Big(  1 - \f{w}{n+1}{1}{\Phi_{n+1}} - \f{\Delta}{n}{1}{t_+,t_{n+1}}\Big) \one{n+2}{n+1} ~}~ \Big]
\nonumber\\
&&
+~\intTwo\ss{n+2}{1}{Q_{n+2}>Q_c \land Q_{n+3}<Q_c }{\Phi_{n+2}}
~\Big[~\one{n+2}{n} -
~\tcboxmath[colback=mypurple!20,hypertarget=purpleB,hyperlink=purpleA]{\!\!\phantom{\int}
~ \Delta_{n} (t_+,t_{n+1}) \f{w}{n+1}{\infty}{\Phi_{n+1}} \one{n+2}{n+1} ~}~
\Big] \nonumber\\
&&
+~\intThree\ss{n+3}{0}{Q_{n+3}>Q_c}{\Phi_{n+3}}
~\Big[~\one{n+3}{n} -
~\tcboxmath[colback=mymidgrey!20,hypertarget=greyB,hyperlink=greyA]{~ \Delta_{n} (t_+,t_{n+1}) \f{w}{n+1}{\infty}{\Phi_{n+1}} \one{n+3}{n+1} ~}~ \Big] ~\Bigg\}\nonumber\\
&&+~\obs{n+1}~\Bigg\{ 
\ss{n+1}{2}{Q_{n+2}<Q_c \land Q_{n+3}<Q_c}{\Phi_{n+1}}~\tcboxmath[colback=black!00,colframe=red,boxrule=0.5pt,boxsep=1pt,hypertarget=cfredA,hyperlink=cfredB]{\!\!\phantom{\int}
\Delta_{n} (t_+,t_{n+1}) \f{w}{n+1}{\infty}{\Phi_{n+1}} ~}\nonumber\\
&&+~\s{n+1}{0}{\Phi_{n+1}}~\otimes~
\tcboxmath[colback=black!00,colframe=myolive,boxrule=0.5pt,boxsep=1pt,hypertarget=cfoliveA,hyperlink=cfoliveB]{
\begin{aligned}
&~\Delta_{n} (t_+,t_{n+1}) \f{w}{n+1}{\infty}{\Phi_{n+1}} ~\nonumber\\
&\quad\cdot\Big(                  
1 - \f{w}{n+1}{1}{\Phi_{n+1}} - \f{w}{n+1}{2}{\Phi_{n+1}}  - \f{\Delta}{n}{1}{t_+,t_{n+1}} -  \f{\Delta}{n}{2}{t_+,t_{n+1}}
\nonumber\\
&\quad~ + \left[\f{\Delta}{n}{1}{t_+,t_{n+1}}\right]^2 + \left[\f{w}{n+1}{1}{\Phi_{n+1}}\right]^2 + \f{w}{n+1}{1}{\Phi_{n+1}} \f{\Delta}{n}{1}{t_+,t_{n+1}} \Big)
\end{aligned}
~}
\nonumber\\
&&+~ \ss{n+1}{1}{Q_{n+2}<Q_c}{\Phi_{n+1}} \nonumber\\
&&\qquad\otimes~
\tcboxmath[colback=black!00,colframe=mypurple,boxrule=0.5pt,boxsep=1pt,hypertarget=cfpurpleA,hyperlink=cfpurpleB]{\!\!\phantom{\int}
\Big[ 1 - \f{w}{n+1}{1}{\Phi_{n+1}} - \f{\Delta}{n}{1}{t_+,t_{n+1}}\Big] \Delta_{n} (t_+,t_{n+1})\f{w}{n+1}{\infty}{\Phi_{n+1}} ~}  \nonumber\\
&&+~\intOne\ss{n+2}{0}{Q_{n+2}>Q_c}{\Phi_{n+2}} \nonumber\\
&&\qquad\otimes~ \Delta_{n} (t_+,t_{n+1}) \f{w}{n+1}{\infty}{\Phi_{n+1}}
\nonumber\\
&&\qquad\otimes~\!\Big[~\tcboxmath[colback=mybrown!20,hypertarget=brownA,hyperlink=brownB]{\!\!\phantom{\int} \Big(  1 - \f{w}{n+1}{1}{\Phi_{n+1}} - \f{\Delta}{n}{1}{t_+,t_{n+1}}\Big) \one{n+2}{n+1} ~}
\nonumber\\
&&\qquad~
-
~
\Delta_{n+1} (t_{n+1},t_{n+2}) \f{w}{n+2}{\infty}{\Phi_{n+2}}\nonumber\\
&&\qquad\qquad\otimes
~\tcboxmath[colback=green!10,hypertarget=greenB,hyperlink=greenA]{\!\!\phantom{\int}
\Big( 1 - \f{w}{n+1}{1}{\Phi_{n+1}} - \f{w}{n+2}{1}{\Phi_{n+2}} - \f{\Delta}{n}{1}{t_+,t_{n+1}} - \f{\Delta}{n+1}{1}{t_{n+1},t_{n+2}} \Big)
~}~\Big]
\nonumber\\
&&+~\intOne\ss{n+2}{1}{Q_{n+2}>Q_c \land Q_{n+3}<Q_c }{\Phi_{n+2}} \nonumber\\
&&\qquad\otimes~\Big[
~\tcboxmath[colback=mypurple!20,hypertarget=purpleA,hyperlink=purpleB]{\!\!\phantom{\int}  \Delta_{n} (t_+,t_{n+1}) \f{w}{n+1}{\infty}{\Phi_{n+1}} \one{n+2}{n+1} ~  }
\nonumber\\
&&\qquad~
- ~\tcboxmath[colback=red!10,hypertarget=redB,hyperlink=redA]{\!\!\phantom{\int}  \Delta_{n} (t_+,t_{n+1}) \Delta_{n+1} (t_{n+1},t_{n+2}) \f{w}{n+1}{\infty}{\Phi_{n+1}} \f{w}{n+2}{\infty}{\Phi_{n+2}} ~}~\Big] \nonumber\\
&&+~\intTwo \ss{n+3}{0}{Q_{n+3}>Q_c}{\Phi_{n+3}}\nonumber\\
&&\qquad\otimes~ \Big[
~\tcboxmath[colback=mymidgrey!20,hypertarget=greyA,hyperlink=greyB]{\!\!\phantom{\int} \Delta_{n} (t_+,t_{n+1}) \f{w}{n+1}{\infty}{\Phi_{n+1}} \one{n+3}{n+1}~}
\nonumber\\
&&\qquad~
- ~\tcboxmath[colback=orange!10,hypertarget=orangeB,hyperlink=orangeA]{\!\!\phantom{\int}
\Delta_{n} (t_+,t_{n+1}) \Delta_{n+1} (t_{n+1},t_{n+2}) \f{w}{n+1}{\infty}{\Phi_{n+1}} \f{w}{n+2}{\infty}{\Phi_{n+2}} \one{n+3}{n+2} ~}~\Big]
\Bigg\}
\nonumber\\
&&+~
\obs{n+2} ~\Bigg\{
~\ss{n+2}{0}{Q_{n+2}>Q_c}{\Phi_{n+2}} 
\nonumber\\
&&\qquad\otimes~
\Delta_{n} (t_+,t_{n+1})\Delta_{n+1} (t_{n+1},t_{n+2}) \f{w}{n+1}{\infty}{\Phi_{n+1}}  \f{w}{n+2}{\infty}{\Phi_{n+2}}
\nonumber\\
&&\qquad\otimes~
~\tcboxmath[colback=green!10,hypertarget=greenA,hyperlink=greenB]{\!\!\phantom{\int}\Big[ 1 - \f{w}{n+1}{1}{\Phi_{n+1}} - \f{w}{n+2}{1}{\Phi_{n+2}} - \f{\Delta}{n}{1}{t_+,t_{n+1}} - \f{\Delta}{n+1}{1}{t_{n+1},t_{n+2}} \Big]~}
\nonumber\\
&&+~ \ss{n+2}{1}{Q_{n+2}>Q_c \land Q_{n+3}<Q_c}{\Phi_{n+2}} \nonumber\\
&&\qquad\otimes~\tcboxmath[colback=red!10,hypertarget=redA,hyperlink=redB]{\!\!\phantom{\int}
\Delta_{n} (t_+,t_{n+1})\Delta_{n+1} (t_{n+1},t_{n+2}) \f{w}{n+1}{\infty}{\Phi_{n+1}}\f{w}{n+2}{\infty}{\Phi_{n+2}} ~} \nonumber\\
&&+~ \intOne \ss{n+3}{0}{Q_{n+3}>Q_c}{\Phi_{n+3}} \Big[ ~\tcboxmath[colback=orange!10,hypertarget=orangeA,hyperlink=orangeB]{\!\!\phantom{\int}
\Delta_{n} (t_+,t_{n+1}) \Delta_{n+1} (t_{n+1},t_{n+2}) \f{w}{n+1}{\infty}{\Phi_{n+1}}\f{w}{n+2}{\infty}{\Phi_{n+2}} \one{n+3}{n+2}~}
\nonumber\\
&&-~\tcboxmath[colback=blue!10,hypertarget=blueB,hyperlink=blueA]{\!\!\phantom{\int}
\Delta_{n} (t_+,t_{n+1}) \Delta_{n+1} (t_{n+1},t_{n+2}) \Delta_{n+2} (t_{n+2},t_{n+3})\w{n+1} \w{n+2} \w{n+3}~}~\Big] 
\Bigg\}\nonumber\\
&&+~ \ss{n+3}{0}{Q_{n+3}>Q_c}{\Phi_{n+3}} ~\tcboxmath[colback=blue!10,hypertarget=blueA,hyperlink=blueB]{\!\!\phantom{\int} 
\Delta_{n} (t_+,t_{n+1}) \Delta_{n+1} (t_{n+1},t_{n+2}) \Delta_{n+2} (t_{n+2},t_{n+3})  \w{n+1} \w{n+2}\w{n+3}  ~}\nonumber\\
&&\quad\otimes~\f{\mathcal{F}}{n+3}{\infty}{\Phi_{n+3}, t_{n+3},t_-}~.
\label{eq:nnnlops}
\end{eqnarray}
This is the main result of this note. As is always the case when matching fixed-order calculations to parton showers, this is not necessarily the only possible \nnnlopps{}  matching scheme, provoking the question if the matching scheme uncertainty will outweigh renormalization scale uncertainties at \nnnlo. Scale variations in leading-order parton showers can be large~\cite{Mrenna:2016sih,*Bothmann:2016nao,*Bellm:2016rhh,*Bendavid:2018nar}, so that non-inclusive observables can be burdened with large uncertainties. By design, eq.\ \ref{eq:nnnlops} will immediately apply also when using higher-accuracy showers, since its derivation did not make assumptions on the shower accuracy. Thus, the scale uncertainties will immediately be reduced once higher-order showers become available. Matching scheme variations related to choices of the functional form of renormalization scales have been observed to introduce uncertainties in \nlo{} merging~\cite{Gellersen:2020tdj}. Similarly, fixed-order \nnlo{} calculations have been shown to exhibit a dependence on the functional form of renormalization scales that may be larger than conventional renormalization scale variations by constant factors~\cite{Currie:2018xkj}. The \nnnlops{} method employs very specific functional form of renormalization scales inherited from parton-shower resummation. Other scale setting mechanisms might be possible in \nnnlopps{}  matching, potentially leading to non-negligible scheme dependence. At present, there is no sound way to assess this issue, without first building some intuition about matching at \nnnlo. Thus, to allow for toy studies, the matching formula \ref{eq:nnnlops} has been implemented in the \textsc{Dire} plugin to the \textsc{Pythia} event generator. The fixed-order methods presented in~\cite{Cieri:2018oms,Mondini:2019gid} appear particularly interesting for a future full-fledged \nnnlopps\ implementation, as do the fixed-order components of the resummed calculations~\cite{Camarda:2021ict,Re:2021con}. Since~\cite{Mondini:2019gid} partly relies on an inclusive \nnnlo{} calculation, Appendix \ref{app:inclusive-n3lops} documents a rearrangement of eq.\ \ref{eq:nnnlops} using an inclusive \nnnlo{} calculation for $n$-parton states.

\begin{figure}[ht!]
\centering
  \begin{subfigure}{0.47\textwidth}
  \includegraphics[width=0.85\textwidth]{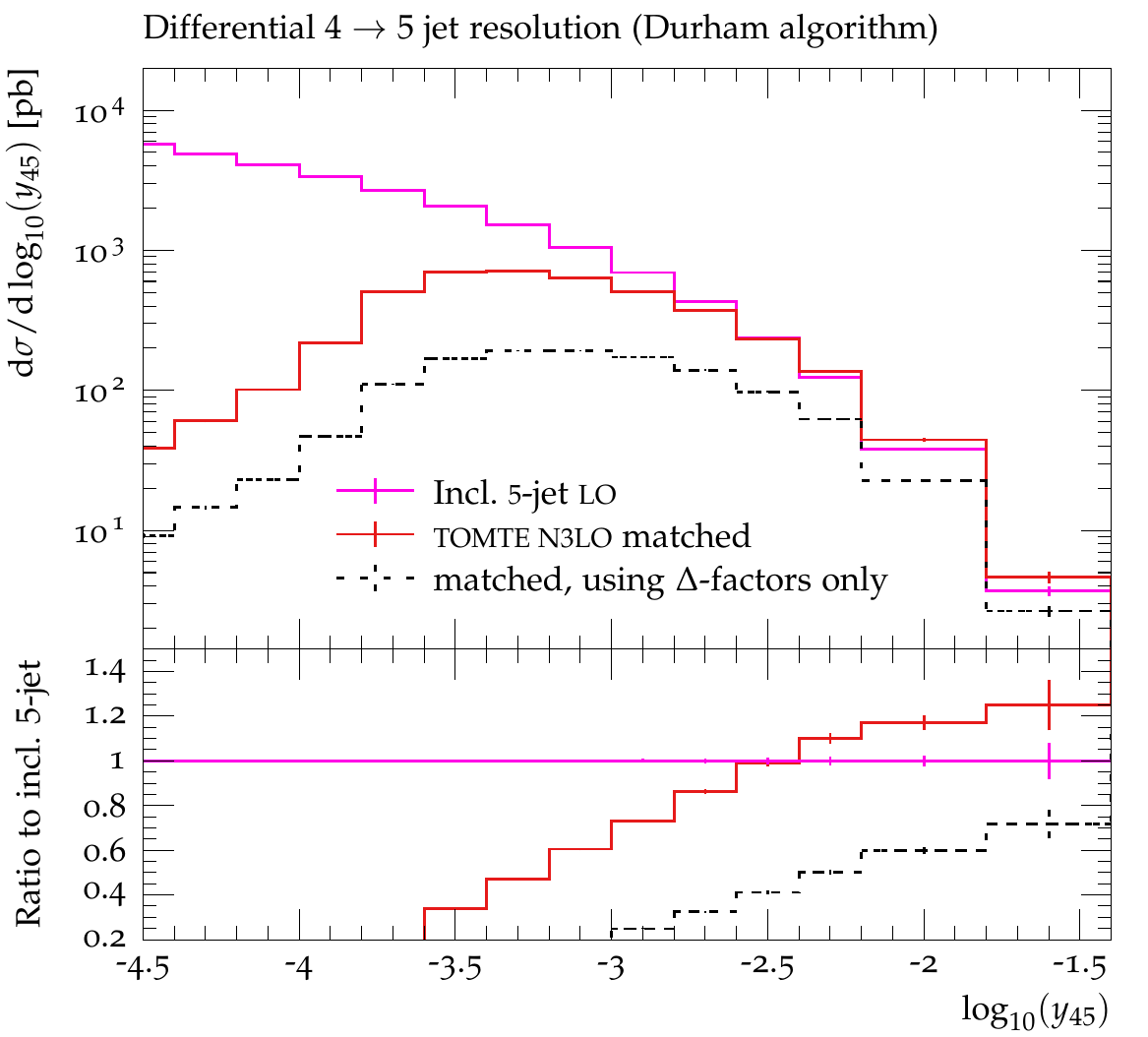}{}
  \caption{\label{fig:3j-closure} Separation of the 4th and 5th hardest jets, as proxy for (inclusive) observables depending on at least five partons.}
  \end{subfigure}
  \hskip 8mm
  \begin{subfigure}{0.47\textwidth}
  \includegraphics[width=0.85\textwidth]{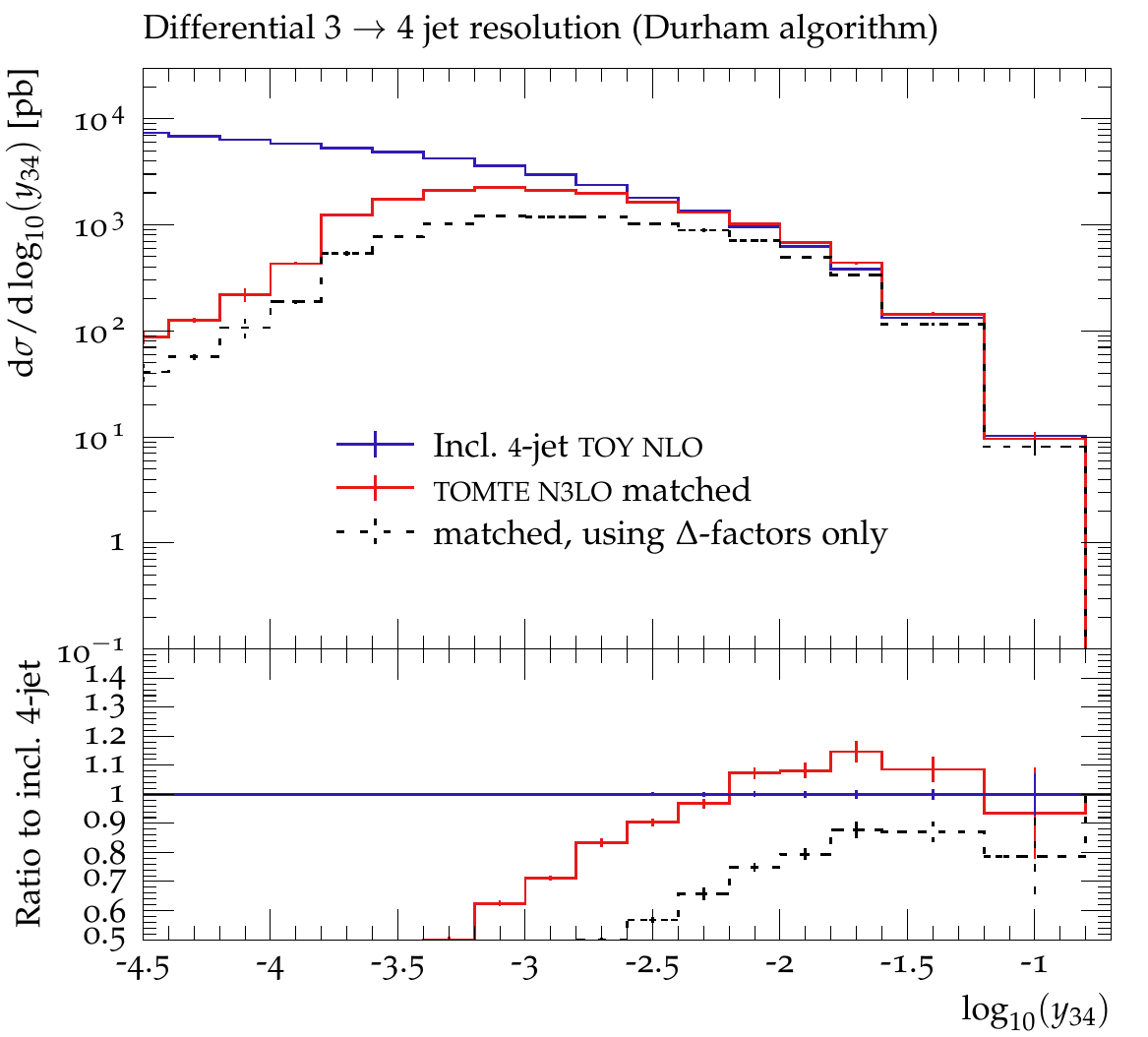}{}
  \caption{\label{fig:2j-closure} Separation of the 3rd and 4th hardest jets, as proxy for (inclusive) observables depending on at least four partons.}
  \end{subfigure}\\
  \begin{subfigure}{0.47\textwidth}
  \includegraphics[width=0.85\textwidth]{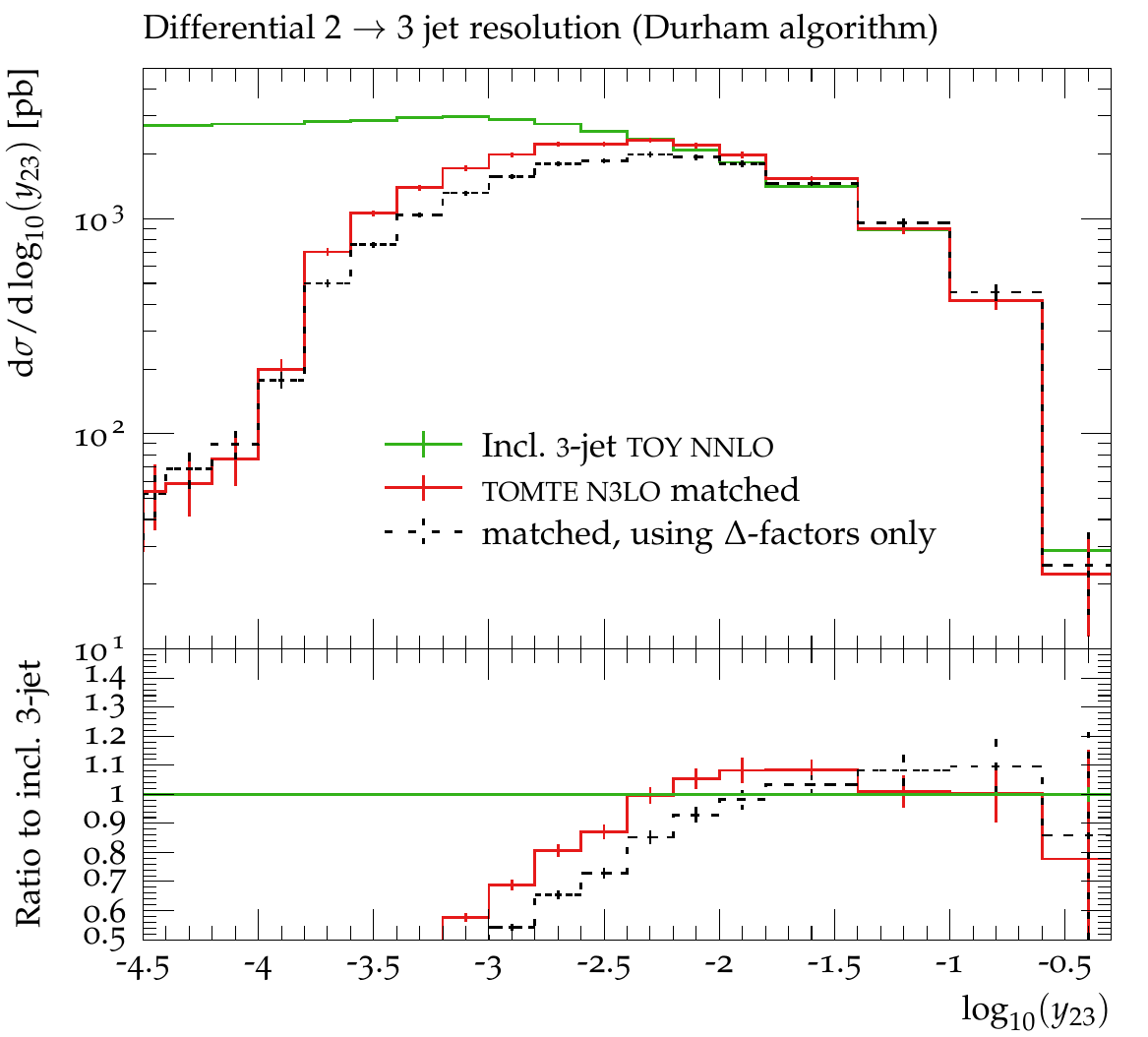}{}
  \caption{\label{fig:1j-closure} Separation of the 2nd and 3rd hardest jets, as proxy for (inclusive) observables depending on at least three partons.}
  \end{subfigure}
  \hskip 8mm
  \begin{subfigure}{0.47\textwidth}
  \includegraphics[width=0.88\textwidth]{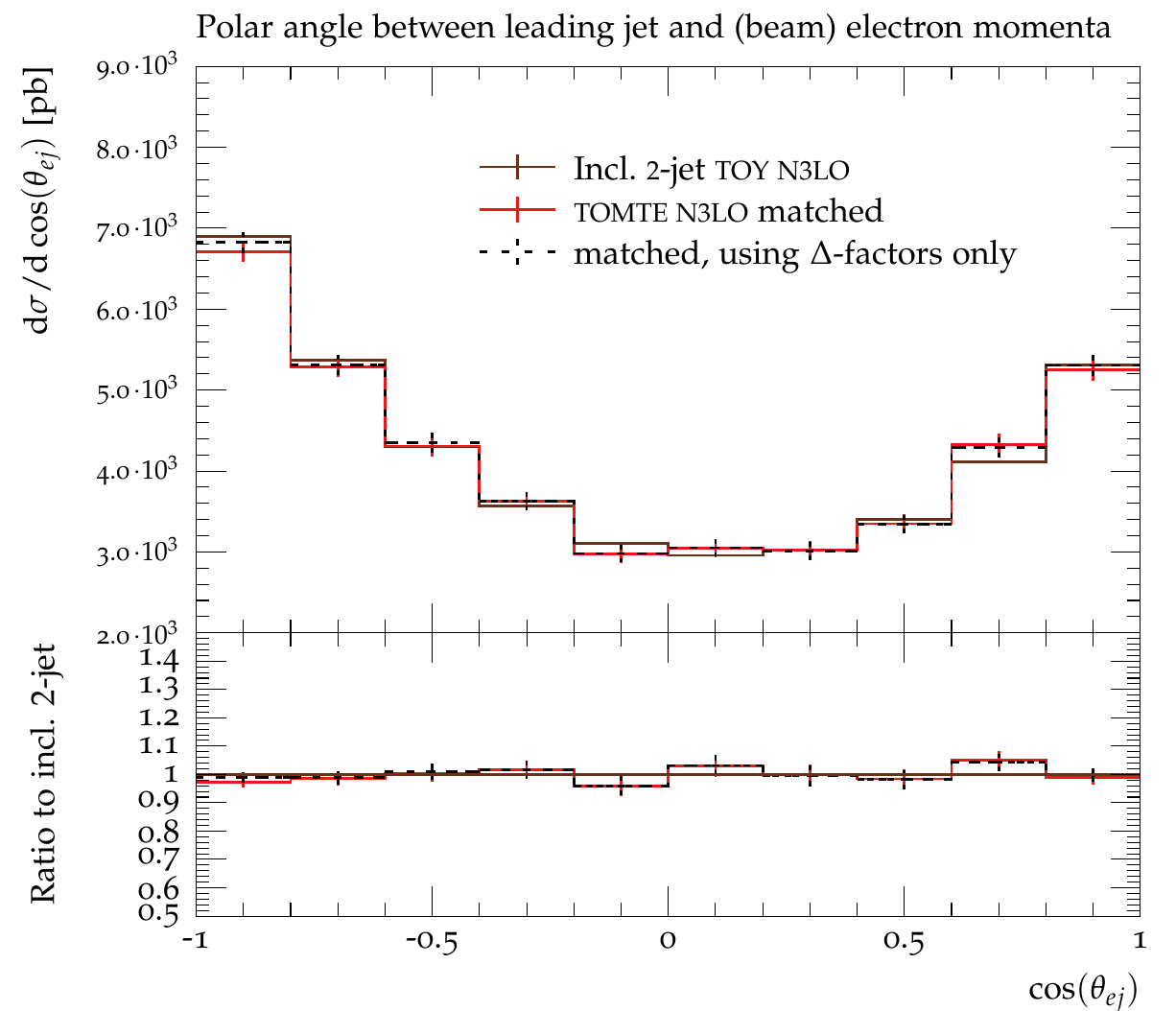}{}
  \caption{\label{fig:0j-closure} Angle of the leading jet and the electron, when always clustering to two jets, as proxy for two-parton observables.}
  \end{subfigure}
\caption{\label{fig:closure} 
Comparison of toy fixed-order curves (with conservative $S(\Phi_{n})$ regularization), toy matching applying only Sudakov factors, and full \nnnlops{} results. Plots were produced with \textsc{Rivet}~\cite{Bierlich:2019rhm}. Bars denote statistical errors.}
\end{figure}
\section{Numerical closure test}
\label{sec:results}
%%%%%%%%%%%%%%%%%%%%%%%%

\noindent
The toy fixed-order calculation constructed in sec.\ \ref{sec:fixed_order} allows for detailed tests of the implementation of the matching formula eq.\ \ref{eq:nnnlops}. All the conditions for \nnnlopps{} accuracy listed in Table \ref{tab:criteria} can be tested in a controlled environment. These tests are presented in Figure \ref{fig:closure}~\footnote{Matched results use inputs with tighter $Q(\Phi_{n})$ cuts. Directly using $S(\Phi_{n})$-regularized toy fixed-order calculations for matching leads to identical \nnnlops{} results, but is less practical for efficiency reasons, due to an excessively high Sudakov rejection rate: any contribution with $Q(\Phi_{n})<Q_c$ (identifying $Q_c$ with the parton-shower cut-off) will be removed by the application of vanishing $\Delta_{n-1} (t,t_{n})$ factors.}. The toy fixed-order calculation makes use of the renormalization scale $\mu=\hat s = M_Z^2$ and $\alpha_s(\mu) = 0.118$, while all parton-shower factors use the running coupling 
\begin{equation*}
\alpha_s(t) = \alpha_s\left( 4\, \frac{ (p_\mathrm{radiator}\cdot p_\mathrm{emission})(p_\mathrm{emission}\cdot p_\mathrm{recoiler}) }{ (p_\mathrm{radiator}\cdot p_\mathrm{emission}) + (p_\mathrm{emission}\cdot p_\mathrm{recoiler}) + (p_\mathrm{radiator}\cdot p_\mathrm{recoiler})  }\right)~.
\end{equation*}
The prefactor in the argument of $\alpha_s$ is chosen so that $\alpha_s(t)\rightarrow \alpha_s(\mu)$ in fixed-order dominated phase-space regions. 

That the tree-level result for observables requiring three or more additional partons is recovered is illustrated by fig.\ \ref{fig:3j-closure}. This shows that for well-separated jets, the fixed-order result is approached adequately. However, the agreement is largely accidental, since the pure fixed-order region is exceedingly small:
the whole distribution receives significant Sudakov suppression (as illustrated by the dashed gray curve in fig.\ \ref{fig:3j-closure}), and the running-coupling effects present in the matched calculation have a significant impact also for well-separated configurations. When approaching phase-space regions with unresolved partons, the all-order factors included in the matched calculation produce a physically meaningful regularize the cross-section. In conclusion, the matched calculation behaves as expected, combining fixed-order accuracy with parton-shower resummation.
The \nlo{} accuracy of observables requiring two additional partons is assessed in fig.\ \ref{fig:2j-closure}. Again, the inclusive fixed-order result is approached as desired, although the pure fixed-order region (measured with the absence of Sudakov effects) is small, and running-coupling effects are large. As before, the matched calculation exhibits the desired Sudakov suppression when approaching phase-space regions containing unresolved partons. The two extremes are again consistently matched. The same conclusions can also be drawn for the \nnlo{} accuracy of inclusive observables requiring at least one additional parton (as shown in fig.\ \ref{fig:1j-closure}), for which the fixed-order region is larger, and the agreement in the hard region is sound. Finally, the \nnnlo{} accuracy of inclusive observables relying only on the particles present at Born level is checked in fig.\ \ref{fig:0j-closure}. For such observables, the matched calculation should recover the toy \nnnlo{} calculation exactly, since no parton-shower factors rescale the inclusive lowest-multiplicity prediction. This is indeed the case, confirming that the implementation of the \nnnlops~\nnnlopps{} matching method is consistent. Extensions to hadron collisions are conceivable, and would benefit from similar closure tests.

%%%%%%%%%%%%%%%%%%%%%%%%%%%%%%%
\section{Summary and Outlook}
\label{sec:outlook}
%%%%%%%%%%%%%%%%%%%%%%%%%%%%%%%%Q

\noindent
This note introduces a straight-forward method to match \nnnlo{} calculations to parton shower resummation, thus allowing for the construction of \nnnlo-precise event generators. The construction of the method is based on the simple idea of unitary matching, and is an extension of the \unnlops\ method. The final formula eq.\ \ref{eq:nnnlops} appears a bit daunting at first glance, but should be easy to understand with the visual help of pairwise cancelling boxes. The \nnnlops\ matching formula had been implemented in the \textsc{Pythia} +  \textsc{Dire} generator, assuming that the fixed-order results can be supplied by external calculations. A toy fixed-order calculation was constructed, and a closure test of the matching scheme was performed.

Although the necessity for \nnnlopps{} accurate event generators is not completely obvious, the development might be an interesting alternative to other avenues of improving event generators. In particular, given the limited application of \nlo-merged predictions combining more than three \nlo{} calculations in experimental analyses, it could be argued that there is little need for merging more than two \nnlo{} calculations. If so, then it would seem that an \nnnlopps{} matching method would be adequate for the foreseeable future. However, if the goal of improved event generators is a decreased uncertainty throughout the spectrum, then improving the fixed-order precision only should not be considered an end on its own, but rather be accompanied by a better understanding of all-order perturbative effects. The derivation of the \nnnlops{} matching formula requires no reference to the actual logarithmic accuracy of the parton shower. Thus, eq.\ \ref{eq:nnnlops} is a valid method to combine \nnnlo{} calculations with high-precision analytic resummation. 

This note offers a first proof of concept for \nnnlopps{} matching. Future research could apply the \nnnlops{} method to processes of interest, using realistic sliced fixed-order calculations. The current proposal could be improved with ease by using an {\smaller MC@NLO}-matched 2-jet component, or tailored to (in spirit similar) Projection-to-Born calculations. Is is also potentially possible to employ a fully differential \nnlopps{} calculation as one-jet component. No fully-differential \nnlopps{} methods for such processes exist to date, since they would require the parton shower to recover the complete singularity structure of \qcd{} at $\mathcal{O}(\alpha_s^{+2})$. Nevertheless, such improvements could materialize within the next years. 

%By its very nature -- combining the ``next-better" fixed-order calculation with event generators -- this note only submits incremental progress towards understanding \qcd{}. Nevertheless, the fact that, for simple processes, \nnnlopps{} matching is feasible with our current understanding of event generators, might be of interest to the \nnnlo{} community.

In any case, for simple processes, \nnnlopps{} matching is feasible with our current understanding of event generators.

\section{Acknowledgments}

This note is supported by funding from the Swedish Research Council, contract numbers 2016-05996 and 2020-04303.

\appendix 

\section{Details of an implementation}
\label{app:details}

\noindent
The main text assumed that all factors in the various matching formulae (eqs. \ref{eq:unlops}, \ref{eq:un2lops} and \ref{eq:nnnlops}) are known and can be generated numerically. The main aim of this appendix is to assess and confirm this assumption for all parton-shower-related terms. The availability of appropriate fixed-order results is still left assumed.

\subsection{Fixed-order cross sections}
\label{app:fixed-order-xsections}

\noindent
The formulae in the main text rely on differential fixed-order predictions, which are then combined with each other and the parton shower to obtained matched predictions. This section gives \emph{symbolic} definitions of such calculations. Unless explicitly stated otherwise, $n$-parton configurations will be assumed to only contain well-separated phase-space regions, i.e.\ 
\begin{eqnarray}
\ss{n}{0+1+\dots}{qualifier}{\Phi_{n}} \equiv \ss{n}{0+1+\dots}{qualifier}{\Phi_{n}}\Theta\left( Q_{n+1} - Q_c\right)~.
\end{eqnarray}
The inclusive \nlo{} cross section is given by
\begin{eqnarray}
&&\ss{n}{0+1}{\mathrm{INC}}{\Phi_{n}} 
=
\s{n}{0}{\Phi_{n}} + \ss{n}{1}{\mathrm{INC}}{\Phi_{n}} \nonumber\\
&&=
 \s{n}{0}{\Phi_{n}} + \left(\s{n}{1}{\Phi_{n}} + \mathcal{I}_{n}(\Phi_n) \right) \nonumber\\
&&\quad+ \upint d\Phi_{1} \left(  w^{\mathrm{FO}}_{n+1}(\Phi_{n+1}, \Phi_n) d\sigma_{n+1}^{(0)}(\Phi_{n+1}) - \mathcal{S}_{n+1}(\widetilde{\Phi}_{n},\Phi_{1})\right)
\end{eqnarray}
where $\mathcal{I}_{n}$ is the analytically integrated analogue of the real-emission subtraction $\mathcal{S}_{n+1}$. The projection rate $w^{\mathrm{FO}}_{n+1}(\Phi_{n+1}, \Phi_n)$ for replacing real-emission kinematics with underlying-Born-kinematics is discussed in Appendix \ref{app:integrals}.
The exclusive (jet-vetoed) next-to-leading order cross section can be obtained from its inclusive counter-part by restricting the phase-space for real-emission contributions:
\begin{eqnarray}
&&\ss{n}{0+1}{Q_{n+1}<Q_c}{\Phi_{n}} 
=
\s{n}{0}{\Phi_{n}} + \ss{n}{1}{Q_{n+1}<Q_c}{\Phi_{n}} \nonumber\\
&&=
 \s{n}{0}{\Phi_{n}} + \left(\s{n}{1}{\Phi_{n}} + \mathcal{I}_{n}(\Phi_n) \right) \nonumber\\
&&\quad+ \upint d\Phi_{1} \left(  w^{\mathrm{FO}}_{n+1}(\Phi_{n+1}, \Phi_n) d\sigma_{n+1}^{(0)}(\Phi_{n+1})\Theta\left( Q_c - Q_{n+1} \right) - \mathcal{S}_{n+1}(\widetilde{\Phi}_{n},\Phi_{1})\right)~.
\end{eqnarray}
Note that the regularizing subtractions should be unaffected by the phase-space constraint.
At \nnlo{}, the inclusive cross section is given, symbolically, by
\begin{eqnarray}
&&\ss{n}{0+1+2}{\mathrm{INC}}{\Phi_{n}} 
=
\s{n}{0}{\Phi_{n}} + \ss{n}{1}{\mathrm{INC}}{\Phi_{n}} + \ss{n}{2}{\mathrm{INC}}{\Phi_{n}}\nonumber\\
&&\ss{n}{2}{\mathrm{INC}}{\Phi_{n}}
=
 \left(\s{n}{2}{\Phi_{n}} + \mathcal{I}^{\mathrm{VV}}_{n}(\Phi_n) \right) \nonumber\\
&&\quad+ \upint d\Phi_{1} \left(  w^{\mathrm{FO}}_{n+1}(\Phi_{n+1}, \Phi_n) d\sigma_{n+1}^{(1)}(\Phi_{n+1}) - \mathcal{S}^{\mathrm{RV}}_{n+1}(\widetilde{\Phi}_{n},\Phi_{1})\right)\nonumber\\
&&\quad+ \upint d\Phi_{2} \left(  w^{\mathrm{FO}}_{n+2}(\Phi_{n+2}, \Phi_n) d\sigma_{n+2}^{(0)}(\Phi_{n+2}) - \mathcal{S}^{\mathrm{RR}}_{n+2}(\widetilde{\Phi}_{n},\Phi_{2})\right)
\end{eqnarray}
where it is assumed that the subtractions $\mathcal{S}^{\mathrm{RV}}_{n+1}$, $\mathcal{S}^{\mathrm{RR}}_{n+2}$ and the integrated subtractions $\mathcal{I}^{\mathrm{VV}}_{n}$ add to zero in infrared-safe observables. The exclusive (jet-vetoed) counterpart can be obtained from the inclusive cross section by restricting the phase-space for real-emission contributions:
\begin{eqnarray}
&&\ss{n}{0+1+2}{Q_{n+1}<Q_c \land Q_{n+2}<Q_c  }{\Phi_{n}} 
=
\s{n}{0}{\Phi_{n}} + \ss{n}{1}{Q_{n+1}<Q_c}{\Phi_{n}} + \ss{n}{2}{Q_{n+1}<Q_c \land Q_{n+2}<Q_c}{\Phi_{n}} \nonumber\\
&&\ss{n}{2}{Q_{n+1}<Q_c \land Q_{n+2}<Q_c}{\Phi_{n}}
=
 \left(\s{n}{2}{\Phi_{n}} + \mathcal{I}^{\mathrm{VV}}_{n}(\Phi_n) \right) \nonumber\\
&&\quad+ \upint d\Phi_{1} \left(  w^{\mathrm{FO}}_{n+1}(\Phi_{n+1}, \Phi_n) d\sigma_{n+1}^{(1)}(\Phi_{n+1})\Theta\left( Q_c - Q_{n+1} \right) - \mathcal{S}^{\mathrm{RV}}_{n+1}(\widetilde{\Phi}_{n},\Phi_{1})\right)\nonumber\\
&&\quad+ \upint d\Phi_{2} \left(  w^{\mathrm{FO}}_{n+2}(\Phi_{n+2}, \Phi_n) d\sigma_{n+2}^{(0)}(\Phi_{n+2})\Theta\left( Q_c - Q_{n+1} \right)\Theta\left( Q_c - Q_{n+2} \right) - \mathcal{S}^{\mathrm{RR}}_{n+2}(\widetilde{\Phi}_{n},\Phi_{2})\right)~.
\end{eqnarray}
The same relation between inclusive and exclusive cross sections persists at \nnnlo{}. In this case, the inclusive cross section is
At NNLO, the inclusive cross section is given, symbolically, by
\begin{eqnarray}
&&\ss{n}{0+1+2+3}{\mathrm{INC}}{\Phi_{n}} 
=
\s{n}{0}{\Phi_{n}} + \ss{n}{1}{\mathrm{INC}}{\Phi_{n}} + \ss{n}{2}{\mathrm{INC}}{\Phi_{n}} + \ss{n}{3}{\mathrm{INC}}{\Phi_{n}}\nonumber\\
&&\ss{n}{3}{\mathrm{INC}}{\Phi_{n}}
=
 \left(\s{n}{3}{\Phi_{n}} + \mathcal{I}^{\mathrm{VVV}}_{n}(\Phi_n) \right) \nonumber\\
&&\quad+ \upint d\Phi_{1} \left(  w^{\mathrm{FO}}_{n+1}(\Phi_{n+1}, \Phi_n) d\sigma_{n+1}^{(2)}(\Phi_{n+1}) - \mathcal{S}^{\mathrm{RVV}}_{n+1}(\widetilde{\Phi}_{n},\Phi_{1})\right)\nonumber\\
&&\quad+ \upint d\Phi_{2} \left(  w^{\mathrm{FO}}_{n+2}(\Phi_{n+2}, \Phi_n) d\sigma_{n+2}^{(1)}(\Phi_{n+2}) - \mathcal{S}^{\mathrm{RRV}}_{n+2}(\widetilde{\Phi}_{n},\Phi_{2})\right)\nonumber\\
&&\quad+ \upint d\Phi_{3} \left(  w^{\mathrm{FO}}_{n+3}(\Phi_{n+3}, \Phi_n) d\sigma_{n+3}^{(0)}(\Phi_{n+3}) - \mathcal{S}^{\mathrm{RRR}}_{n+3}(\widetilde{\Phi}_{n},\Phi_{3})\right)~,
\end{eqnarray}
where the subtractions $\mathcal{S}^{\mathrm{RVV}}_{n+1}$, $\mathcal{S}^{\mathrm{RRV}}_{n+2}$, $\mathcal{S}^{\mathrm{RRR}}_{n+3}$ and the integrated subtractions $\mathcal{I}^{\mathrm{VVV}}_{n}$ add to zero for infrared-safe observables. The exclusive cross-section then reads

\begin{eqnarray}
&&\ss{n}{0+1+2+3}{Q_{n+1}<Q_c \land Q_{n+2}<Q_c \land Q_{n+3}<Q_c}{\Phi_{n}} \nonumber\\
&&\quad=\quad
\s{n}{0}{\Phi_{n}} + \ss{n}{1}{Q_{n+1}<Q_c}{\Phi_{n}} + \ss{n}{2}{Q_{n+1}<Q_c \land Q_{n+2}<Q_c}{\Phi_{n}} + \ss{n}{3}{Q_{n+1}<Q_c \land Q_{n+2}<Q_c \land Q_{n+3}<Q_c}{\Phi_{n}}\nonumber
\end{eqnarray}
where
\begin{eqnarray}
&&\ss{n}{3}{Q_{n+1}<Q_c \land Q_{n+2}<Q_c \land Q_{n+3}<Q_c}{\Phi_{n}}
=
 \left(\s{n}{3}{\Phi_{n}} + \mathcal{I}^{\mathrm{VVV}}_{n}(\Phi_n) \right) \nonumber\\
&&\quad+ \upint d\Phi_{1} \left(  w^{\mathrm{FO}}_{n+1}(\Phi_{n+1}, \Phi_n) d\sigma_{n+1}^{(2)}(\Phi_{n+1})\Theta\left( Q_c - Q_{n+1} \right) - \mathcal{S}^{\mathrm{RVV}}_{n+1}(\widetilde{\Phi}_{n},\Phi_{1})\right)\nonumber\\
&&\quad+ \upint d\Phi_{2} \left(  w^{\mathrm{FO}}_{n+2}(\Phi_{n+2}, \Phi_n) d\sigma_{n+2}^{(1)}(\Phi_{n+2})\Theta\left( Q_c - Q_{n+1} \right)\Theta\left( Q_c - Q_{n+2} \right) - \mathcal{S}^{\mathrm{RRV}}_{n+2}(\widetilde{\Phi}_{n},\Phi_{2})\right)\nonumber\\
&&\quad+ \upint d\Phi_{3} \left(  w^{\mathrm{FO}}_{n+3}(\Phi_{n+3}, \Phi_n) d\sigma_{n+3}^{(0)}(\Phi_{n+3})\Theta\left( Q_c - Q_{n+1} \right)\Theta\left( Q_c - Q_{n+2} \right)\Theta\left( Q_c - Q_{n+3} \right) - \mathcal{S}^{\mathrm{RRR}}_{n+3}(\widetilde{\Phi}_{n},\Phi_{3})\right)~.\quad
\end{eqnarray}

%Q_{n}>Q_c \land Q_{n+1}<Q_c 

\subsection{Histories}
\label{app:histories}

\noindent
Parton-shower histories are crucial for calculating the matching terms. This note also employs parton-shower histories for the construction of cuts and integrated contributions in the toy fixed-order calculation described in sec.\ \ref{sec:fixed_order}. Some background on histories may thus be helpful.

Showers generate multi-parton configurations through a sequence of parton branchings. A key realization of shower-centered matching -- and, particularly, merging -- methods is to invert this picture when treating fixed-order calculations that contain multiple partons as input to the event generator. This leads to the concept of parton-shower histories, which are then utilized for many of the necessary tasks.

\begin{figure}[ht!]
\begin{tikzpicture}[remember picture]
\node (phi1) [box,inner sep=2pt, inner ysep=2pt,scale=0.7] {
\begin{tikzpicture}
  \begin{feynman} [every blob={/tikz/fill=gray!10,/tikz/inner sep=2pt}]
    \vertex (a) {\smaller $q$};
    \vertex [right=of a, blob] (b) {$\Phi_1$};
    \vertex [right=of b] (c) {\smaller $\bar q$};
    \vertex [above=of b,yshift=-0.4cm] (d);
    \vertex [below=of b,yshift=0.2cm] (e) {\smaller $g$};
    \diagram* {
      (a) -- (b),
      (b) -- (c),
      (d) -- [boson, momentum=\(q\)] (b),
      (b) -- [gluon] (e)
    };
  \end{feynman}
\end{tikzpicture}
};
\node (phi01) [box, below left=2mm of phi1.south,xshift=-5mm,scale=0.6,fill=mymidgrey!20] {
\begin{tikzpicture}
  \begin{feynman} [every blob={/tikz/fill=gray!10,/tikz/inner sep=2pt}]
    \vertex (a) {\smaller $q$};
    \vertex [right=of a, blob] (b) {$\Phi_0^1$};
    \vertex [left=of b, xshift=0.6cm] (a1);
    \vertex [right=of b] (c) {\smaller $\bar q$};
    \vertex [above=of b,yshift=-0.4cm] (d);
    \vertex [below=of a1,yshift=0.7cm] (e) {\smaller $g$};
    \diagram* {
      (a) -- (b),
      (b) -- (c),
      (d) -- [boson, momentum=\(q\)] (b),
      (a1) -- [gluon] (e)
    };
  \end{feynman}
\end{tikzpicture}
};
\node (phi02) [box, below right=2mm of phi1.south,xshift=5mm,scale=0.6,fill=mybrown!20] {
\begin{tikzpicture}
  \begin{feynman} [every blob={/tikz/fill=gray!10,/tikz/inner sep=2pt}]
    \vertex (a) {\smaller $q$};
    \vertex [right=of a, blob] (b) {$\Phi_0^2$};
    \vertex [right=of b, xshift=-0.6cm] (c1);
    \vertex [right=of b] (c) {\smaller $\bar q$};
    \vertex [above=of b,yshift=-0.4cm] (d);
    \vertex [below=of c1,yshift=0.7cm] (e) {\smaller $g$};
    \diagram* {
      (a) -- (b),
      (b) -- (c),
      (d) -- [boson, momentum=\(q\)] (b),
      (c1) -- [gluon] (e)
    };
  \end{feynman}
\end{tikzpicture}
};
\node (h1) [circ,inner sep=3pt, inner ysep=3pt,draw=black, fill=green!05, left=40mm of phi1.south] {
$\mathbb{H}(\Phi_1)$
};
\node (eq) [nobox, right=1mm of h1] {
=
};
\draw[style={font=\sffamily\small}]
    ($(phi1.west)+(-0.1,0.0)$) edge ($(phi01.north)+(0.0,0.1)$);
\draw[style={font=\sffamily\small}]
    ($(phi1.east)+(0.1,0.0)$) edge ($(phi02.north)+(0.0,0.1)$);
\end{tikzpicture}
\caption{\label{fig:hist1} Histories for one additional parton: The gluon can be recombined with the other partons in two distinct ways.}
\end{figure}
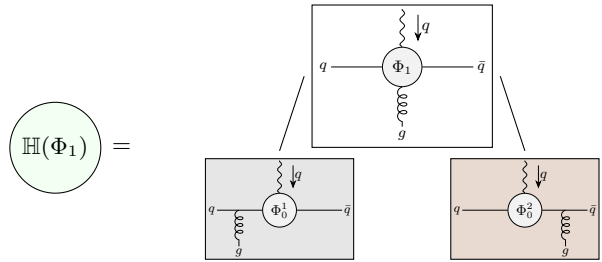

A parton shower history is the set of all possible sequences of branchings that may have lead to the multi-parton state. Figure \ref{fig:hist1} shows the shower history for the production of a $qg\bar q$ final state: Two distinct sequences -- starting from different states $\Phi_0^{[i]}$, and transitioning at different evolution variable $t^{[i]}$ -- can lead to the same final state $\Phi_1$. Thus, the parton-shower prediction for an observable $\obs{1}$ is an admixture of the rate of two paths. Assuming that $\Phi_1$ can be parametrized by the dimensionless variables $(\tau,\zeta,\varphi)$, and that the phase-space points fulfill $\tau<\tau^+$, then the prediction reads,
\begin{eqnarray}
\obs{1}^{\mathrm{PS}}(\Phi_1) &=& 
\int^{\tau^+}{\rm d}\bar{\tau}\,
\int{\rm d} \zeta
\int\frac{{\rm d} \varphi}{2\pi}\\
&&\otimes
~\Bigg[
\mathcal{J}_0^{[1]}\left|\mathcal{M}_0(\Phi_0^{[1]})\right|^2\mathcal{J}_1^{[1]}
\Delta_0(t(\tau^+),t^{[1]}(\bar{\tau}))
\frac{\alpha_s(t^{[1]}(\bar{\tau}))}{\alpha_s(\mu)}
\frac{\alpha_s(\mu)}{2\pi}
\,\frac{P_1^{[1]}(z^{[1]}(\zeta))}{q^{2[1]}(\bar{\tau})}\nonumber\\
&&~+~
\mathcal{J}_0^{[2]}\left|\mathcal{M}_0(\Phi_0^{[2]})\right|^2\mathcal{J}_1^{[2]}\Delta_0(t(\tau^+),t^{[2]}(\bar{\tau}))
\frac{\alpha_s(t^{[2]}(\bar{\tau}))}{\alpha_s(\mu)}
\frac{\alpha_s(\mu)}{2\pi}
\,\frac{P_1^{[2]}(z^{[2]}(\zeta))}{q^{2[2]}(\bar{\tau})}
\Bigg]\nonumber
\label{eq:ps-one-rate}
\end{eqnarray}
with the symbols defined in Table~\ref{tab:ps-symbols}. Note that a spurious $\alpha_s(\mu)$-factor was inserted for later convenience.
Equation \ref{eq:ps-one-rate} highlights that parton-shower resummation produces a particular admixture of functional forms of the argument of $\alpha_s$, and of Sudakov factors with differing integration regions ($[t(\tau^+),t^{[1]}(\bar{\tau})]$ vs.\ $[t(\tau^+),t^{[2]}(\bar{\tau})]$). The same admixture of all-order factors should be employed when calculating parton-shower factors for fixed-order matching. The parton-shower accuracy of the prediction may otherwise be in danger. 

\begin{table}[t!]
\renewcommand{\arraystretch}{1.5}
\begin{tabular}{ |l l p{0.8\textwidth}|}
\hline
superscript ``$[1]$'' \quad & : & 
factors pertaining to transitions where parton ``$1$" emits (\raisebox{1pt}{\tcboxmath[colback=mymidgrey!20,boxsep=1.5pt]{\cdot}} in fig.\ \ref{fig:hist1})
\\
superscript ``$[2]$'' \quad & : & 
factors pertaining to transitions where parton ``$2$" emits (\raisebox{1pt}{\tcboxmath[colback=mybrown!20,boxsep=1.5pt]{\cdot}} in fig.\ \ref{fig:hist1})
\\
$\left|\mathcal{M}_0(\Phi_0^{[i]})\right|^2$ \quad & : & the lowest-multiplicity transition probability, evaluated at the phase space point $\Phi_0^{[i]}$\\
$\mathcal{J}_0^{[i]}$ \quad & : & other factors (of $\hat s,\pi\dots$) such that the combination with $|\mathcal{M}_0(\Phi_0^{[i]})|^2$ results in $\sigma_{0}^{(0)}$   \\
$\mathcal{J}_1^{[i]}$ \quad & : & Jacobian factors to the phase-space mapping $\Phi_0^{[i]} \cap \{\tau,\zeta,\varphi\} \rightarrow  \Phi_0^{[i]} \cap \{t^{[i]},z^{[i]},\varphi\} \rightarrow \Phi_1$, and other factors to ensure that the splitting kernel $P^{[i]}(z^{[i]}(\zeta))$ will be divided by the \emph{virtuality} of the intermediate propagator $q^{2[i]}(\bar{\tau})$. This note employs the mapping $\Phi_0^{[i]} \cap \{t^{[i]},z^{[i]},\varphi\} \rightarrow \Phi_1$ of~\cite{Hoche:2015sya}\\
$t^{[i]}(\bar{\tau})$ \quad &: & the evolution scale assigned to parton $i$ emitting at integration point $\bar\tau$. This note uses the definition of $t^{[i]}$ found in~\cite{Hoche:2015sya}\\
$z^{[i]}(\zeta)$ \quad &: &  the auxiliary variable (e.g.\ the light-cone momentum fraction) assigned to parton $i$ emitting at integration point $\zeta$. The definition of $z^{[i]}$ in terms of two-particle invariants is taken from~\cite{Hoche:2015sya}\\
$P_1^{[i]}(z^{[i]}(\zeta))$\quad &: & the (dimensionless) splitting kernel determining the rate of emissions off parton $i$.\\
\hline
\end{tabular}
\caption{\label{tab:ps-symbols} Definition of the symbols used in eq.~\ref{eq:ps-one-rate}.}
\end{table}

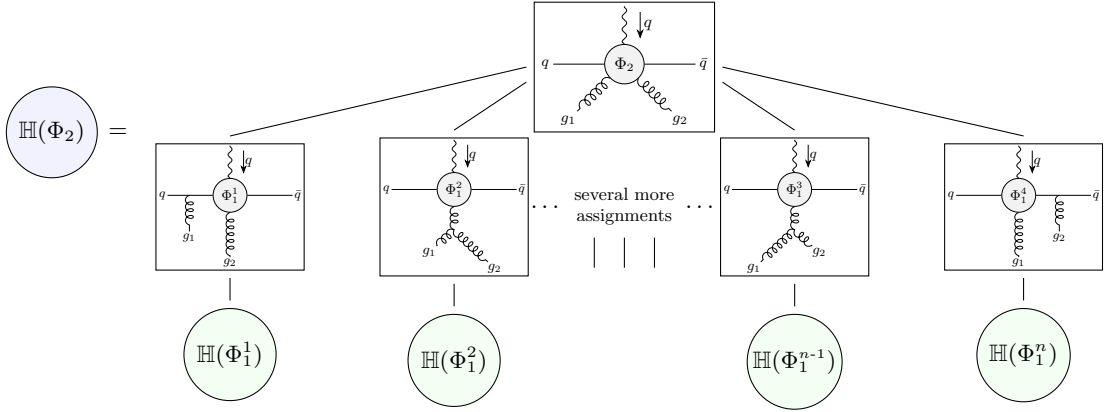
\begin{figure}[ht!]
\begin{tikzpicture}[remember picture]
\node (phi2) [box,inner sep=2pt, inner ysep=2pt,scale=0.7] {
\begin{tikzpicture}
  \begin{feynman} [every blob={/tikz/fill=gray!10,/tikz/inner sep=2pt}]
    \vertex (a) {\smaller $q$};
    \vertex [right=of a, blob] (b) {$\Phi_2$};
    \vertex [right=of b] (c) {\smaller $\bar q$};
    \vertex [above=of b,yshift=-0.4cm] (d);
    \vertex [below left=of b] (e) {\smaller $g_1$};
    \vertex [below right=of b] (f) {\smaller $g_2$};
    \diagram* {
      (a) -- (b),
      (b) -- (c),
      (d) -- [boson, momentum=\(q\)] (b),
      (b) -- [gluon] (e),
      (b) -- [gluon] (f)
    };
  \end{feynman}
\end{tikzpicture}
};
\node (phi12) [box, below left=1mm of phi2.south west,scale=0.6] {
\begin{tikzpicture}
  \begin{feynman} [every blob={/tikz/fill=gray!10,/tikz/inner sep=2pt}]
    \vertex (a) {\smaller $q$};
    \vertex [right=of a, blob] (b) {$\Phi_1^2$};
    \vertex [right=of b] (c) {\smaller $\bar q$};
    \vertex [above=of b,yshift=-0.4cm] (d);
    \vertex [below=of b,yshift=0.6cm] (g);
    \vertex [below left=5mm of g] (e) {$g_1$};
    \vertex [below right=10mm of g] (f) {$g_2$};
    \diagram* {
      (a) -- (b),
      (b) -- (c),
      (d) -- [boson, momentum=\(q\)] (b),
      (b) -- [gluon] (g),
      (g) -- [gluon] (e),
      (g) -- [gluon] (f)
    };
  \end{feynman}
\end{tikzpicture}
};
\node (phi13) [box, below right=1mm of phi2.south east,scale=0.6] {
\begin{tikzpicture}
  \begin{feynman} [every blob={/tikz/fill=gray!10,/tikz/inner sep=2pt}]
    \vertex (a) {\smaller $q$};
    \vertex [right=of a, blob] (b) {$\Phi_1^3$};
    \vertex [right=of b] (c) {\smaller $\bar q$};
    \vertex [above=of b,yshift=-0.4cm] (d);
    \vertex [below=of b,yshift=0.6cm] (g);
    \vertex [below left=10mm of g] (e) {$g_1$};
    \vertex [below right=5mm of g] (f) {$g_2$};
    \diagram* {
      (a) -- (b),
      (b) -- (c),
      (d) -- [boson, momentum=\(q\)] (b),
      (b) -- [gluon] (g),
      (g) -- [gluon] (e),
      (g) -- [gluon] (f)
    };
  \end{feynman}
\end{tikzpicture}
};

\node (dots1) [nobox, below=7mm of phi2.south,scale=0.75,text width=2cm,align=center] {
several more\\
assignments
};
\path (phi12) -- node{\ldots} (dots1);
\path (dots1) -- node{\ldots} (phi13);
\draw[style={font=\sffamily\small}]
    ($(dots1.south)+(-0.4,-0.15)$) edge ($(dots1.south)+(-0.4,-0.55)$);
\draw[style={font=\sffamily\small}]
    ($(dots1.south)+(0.0,-0.15)$) edge ($(dots1.south)+(0.0,-0.55)$);
\draw[style={font=\sffamily\small}]
    ($(dots1.south)+(0.4,-0.15)$) edge ($(dots1.south)+(0.4,-0.55)$);

\node (phi11) [box, left=of phi12.west,scale=0.6] {
\begin{tikzpicture}
  \begin{feynman} [every blob={/tikz/fill=gray!10,/tikz/inner sep=2pt}]
    \vertex (a) {\smaller $q$};
    \vertex [right=of a, blob] (b) {$\Phi_1^1$};
    \vertex [left=of b, xshift=0.6cm] (a1);
    \vertex [right=of b] (c) {\smaller $\bar q$};
    \vertex [above=of b,yshift=-0.4cm] (d);
    \vertex [below=of a1,yshift=0.7cm] (e) {\smaller $g_1$};
    \vertex [below=of b] (f) {\smaller $g_2$};
    \diagram* {
      (a) -- (b),
      (b) -- (c),
      (d) -- [boson, momentum=\(q\)] (b),
      (a1) -- [gluon] (e),
      (b) -- [gluon] (f)
    };
  \end{feynman}
\end{tikzpicture}
};
\node (phi14) [box, right=of phi13.east,scale=0.6] {
\begin{tikzpicture}
  \begin{feynman} [every blob={/tikz/fill=gray!10,/tikz/inner sep=2pt}]
    \vertex (a) {\smaller $q$};
    \vertex [right=of a, blob] (b) {$\Phi_1^4$};
    \vertex [right=of b, xshift=-0.6cm] (a1);
    \vertex [right=of a1, xshift=-0.8cm] (c) {\smaller $\bar q$};
    \vertex [above=of b,yshift=-0.4cm] (d);
    \vertex [below=of a1,yshift=0.7cm] (e) {\smaller $g_2$};
    \vertex [below=of b] (f) {\smaller $g_1$};
    \diagram* {
      (a) -- (b),
      (b) -- (c),
      (d) -- [boson, momentum=\(q\)] (b),
      (a1) -- [gluon] (e),
      (b) -- [gluon] (f)
    };
  \end{feynman}
\end{tikzpicture}
};
\node (h2) [circ,inner sep=3pt, inner ysep=3pt,draw=black, fill=blue!05, left=70mm of phi2.south] {
$\mathbb{H}(\Phi_2)$
};
\node (h11) [circ,inner sep=3pt, inner ysep=3pt,draw=black, fill=green!05, below=5mm of phi11] {
$\mathbb{H}(\Phi_1^1)$
};
\node (h12) [circ,inner sep=3pt, inner ysep=3pt,draw=black, fill=green!05, below=5mm of phi12] {
$\mathbb{H}(\Phi_1^2)$
};
\node (h13) [circ,inner sep=1.2pt, inner ysep=1.2pt,draw=black, fill=green!05, below=5mm of phi13] {
$\mathbb{H}(\Phi_1^\textnormal{\smaller$n$-1})$
};
\node (h14) [circ,inner sep=3pt, inner ysep=3pt,draw=black, fill=green!05, below=5mm of phi14] {
$\mathbb{H}(\Phi_1^n)$
};
\node (eq) [nobox, right=1mm of h2] {
=
};
\draw[style={font=\sffamily\small}]
    ($(phi2.west)+(-0.1,0.0)$) edge ($(phi11.north)+(0.0,0.1)$);
\draw[style={font=\sffamily\small}]
    ($(phi2.east)+(0.1,0.0)$) edge ($(phi14.north)+(0.0,0.1)$);
\draw[style={font=\sffamily\small}]
    ($(phi2.west)+(-0.1,-0.2)$) edge ($(phi12.north)+(0.0,0.1)$);
\draw[style={font=\sffamily\small}]
    ($(phi2.east)+(0.1,-0.2)$) edge ($(phi13.north)+(0.0,0.1)$);
\draw[style={font=\sffamily\small}]
    ($(phi11.south)+(0.0,-0.1)$) edge ($(h11.north)+(0.0,0.1)$);
\draw[style={font=\sffamily\small}]
    ($(phi12.south)+(0.0,-0.1)$) edge ($(h12.north)+(0.0,0.1)$);
\draw[style={font=\sffamily\small}]
    ($(phi13.south)+(0.0,-0.1)$) edge ($(h13.north)+(0.0,0.1)$);
\draw[style={font=\sffamily\small}]
    ($(phi14.south)+(0.0,-0.1)$) edge ($(h14.north)+(0.0,0.1)$);
\end{tikzpicture}
\caption{\label{fig:hist2} Histories for two additional gluons: An individual gluon can be recombined with the other partons in several ways, leading to several underlying states with one additional gluons. Each of these underlying states leads to a parton shower history illustrated by fig.\ \ref{fig:hist1}, i.e.\ the history exhibits a recursive structure.}
\end{figure}

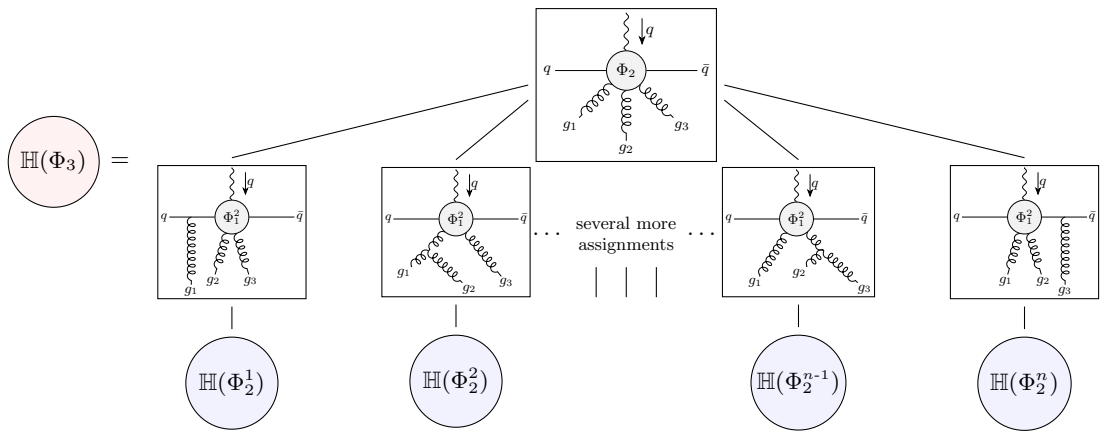
\begin{figure}[ht!]
\begin{tikzpicture}[remember picture]
\node (phi3) [box,inner sep=2pt, inner ysep=2pt,scale=0.7] {
\begin{tikzpicture}
  \begin{feynman} [every blob={/tikz/fill=gray!10,/tikz/inner sep=2pt}]
    \vertex (a) {\smaller $q$};
    \vertex [right=of a, blob] (b) {$\Phi_2$};
    \vertex [right=of b] (c) {\smaller $\bar q$};
    \vertex [above=of b,yshift=-0.4cm] (d);
    \vertex [below left=of b] (e) {\smaller $g_1$};
    \vertex [below=of b] (f) {\smaller $g_2$};
    \vertex [below right=of b] (g) {\smaller $g_3$};
    \diagram* {
      (a) -- (b),
      (b) -- (c),
      (d) -- [boson, momentum=\(q\)] (b),
      (b) -- [gluon] (e),
      (b) -- [gluon] (f),
      (b) -- [gluon] (g)
    };
  \end{feynman}
\end{tikzpicture}
};
\node (phi12) [box, below left=1mm of phi3.south west,scale=0.6] {
\begin{tikzpicture}
  \begin{feynman} [every blob={/tikz/fill=gray!10,/tikz/inner sep=2pt}]
    \vertex (a) {\smaller $q$};
    \vertex [right=of a, blob] (b) {$\Phi_1^2$};
    \vertex [right=of b] (c) {\smaller $\bar q$};
    \vertex [above=of b,yshift=-0.4cm] (d);
    \vertex [below left=10mm of b,xshift=0.1cm] (e);
    \vertex [below right=20mm of b,xshift=-0.3cm] (f) {$g_3$};
    \vertex [below left=5mm of e] (h) {$g_1$};
    \vertex [below right=10mm of e] (i) {$g_2$};
    \diagram* {
      (a) -- (b),
      (b) -- (c),
      (d) -- [boson, momentum=\(q\)] (b),
      (b) -- [gluon] (e),
      (b) -- [gluon] (f),
      (e) -- [gluon] (h),
      (e) -- [gluon] (i)
    };
  \end{feynman}
\end{tikzpicture}
};

\node (phi13) [box, below right=1mm of phi3.south east,scale=0.6] {
\begin{tikzpicture}
  \begin{feynman} [every blob={/tikz/fill=gray!10,/tikz/inner sep=2pt}]
    \vertex (a) {\smaller $q$};
    \vertex [right=of a, blob] (b) {$\Phi_1^2$};
    \vertex [right=of b] (c) {\smaller $\bar q$};
    \vertex [above=of b,yshift=-0.4cm] (d);
    \vertex [below left=20mm of b,xshift=0.5cm] (f) {$g_1$};
    \vertex [below right=10mm of b,xshift=-0.1cm] (e);
    \vertex [below left=5mm of e] (h) {$g_2$};
    \vertex [below right=10mm of e] (i) {$g_3$};
    \diagram* {
      (a) -- (b),
      (b) -- (c),
      (d) -- [boson, momentum=\(q\)] (b),
      (b) -- [gluon] (e),
      (b) -- [gluon] (f),
      (e) -- [gluon] (h),
      (e) -- [gluon] (i)
    };
  \end{feynman}
\end{tikzpicture}
};
\node (dots1) [nobox, below=7mm of phi3.south,scale=0.75,text width=2cm,align=center] {
several more\\
assignments
};
\path (phi12) -- node{\ldots} (dots1);
\path (dots1) -- node{\ldots} (phi13);
\draw[style={font=\sffamily\small}]
    ($(dots1.south)+(-0.4,-0.15)$) edge ($(dots1.south)+(-0.4,-0.55)$);
\draw[style={font=\sffamily\small}]
    ($(dots1.south)+(0.0,-0.15)$) edge ($(dots1.south)+(0.0,-0.55)$);
\draw[style={font=\sffamily\small}]
    ($(dots1.south)+(0.4,-0.15)$) edge ($(dots1.south)+(0.4,-0.55)$);
\node (phi11) [box, left=of phi12.west,scale=0.6] {
\begin{tikzpicture}
  \begin{feynman} [every blob={/tikz/fill=gray!10,/tikz/inner sep=2pt}]
    \vertex (a) {\smaller $q$};
    \vertex [right=of a, blob] (b) {$\Phi_1^2$};
    \vertex [right=of b] (c) {\smaller $\bar q$};
    \vertex [above=of b,yshift=-0.4cm] (d);
    \vertex [below left=20mm of b,xshift=1cm] (e) {$g_2$};
    \vertex [below right=20mm of b,xshift=-1cm] (f) {$g_3$};
    \vertex [left=9mm of b] (a1);
    \vertex [below=of a1] (h) {$g_1$};
    \diagram* {
      (a) -- (b),
      (b) -- (c),
      (d) -- [boson, momentum=\(q\)] (b),
      (b) -- [gluon] (e),
      (b) -- [gluon] (f),
      (a1) -- [gluon] (h)
    };
  \end{feynman}
\end{tikzpicture}
};
\node (phi14) [box, right=of phi13.east,scale=0.6] {
\begin{tikzpicture}
  \begin{feynman} [every blob={/tikz/fill=gray!10,/tikz/inner sep=2pt}]
    \vertex (a) {\smaller $q$};
    \vertex [right=of a, blob] (b) {$\Phi_1^2$};
    \vertex [right=of b] (c) {\smaller $\bar q$};
    \vertex [above=of b,yshift=-0.4cm] (d);
    \vertex [below left=20mm of b,xshift=1cm] (e) {$g_1$};
    \vertex [below right=20mm of b,xshift=-1cm] (f) {$g_2$};
    \vertex [right=9mm of b] (a1);
    \vertex [below=of a1] (h) {$g_3$};
    \diagram* {
      (a) -- (b),
      (b) -- (c),
      (d) -- [boson, momentum=\(q\)] (b),
      (b) -- [gluon] (e),
      (b) -- [gluon] (f),
      (a1) -- [gluon] (h)
    };
  \end{feynman}
\end{tikzpicture}
};
\node (h3) [circ,inner sep=3pt, inner ysep=3pt,draw=black, fill=red!05, left=70mm of phi3.south] {
$\mathbb{H}(\Phi_3)$
};
\node (h21) [circ,inner sep=3pt, inner ysep=3pt,draw=black, fill=blue!05, below=5mm of phi11] {
$\mathbb{H}(\Phi_2^1)$
};
\node (h22) [circ,inner sep=3pt, inner ysep=3pt,draw=black, fill=blue!05, below=5mm of phi12] {
$\mathbb{H}(\Phi_2^2)$
};
\node (h23) [circ,inner sep=1.2pt, inner ysep=1.2pt,draw=black, fill=blue!05, below=5mm of phi13] {
$\mathbb{H}(\Phi_2^\textnormal{\smaller$n$-1})$
};
\node (h24) [circ,inner sep=3pt, inner ysep=3pt,draw=black, fill=blue!05, below=5mm of phi14] {
$\mathbb{H}(\Phi_2^n)$
};
\node (eq) [nobox, right=1mm of h3] {
=
};
\draw[style={font=\sffamily\small}]
    ($(phi3.west)+(-0.1,0.0)$) edge ($(phi11.north)+(0.0,0.1)$);
\draw[style={font=\sffamily\small}]
    ($(phi3.east)+(0.1,0.0)$) edge ($(phi14.north)+(0.0,0.1)$);
\draw[style={font=\sffamily\small}]
    ($(phi3.west)+(-0.1,-0.2)$) edge ($(phi12.north)+(0.0,0.1)$);
\draw[style={font=\sffamily\small}]
    ($(phi3.east)+(0.1,-0.2)$) edge ($(phi13.north)+(0.0,0.1)$);
\draw[style={font=\sffamily\small}]
    ($(phi11.south)+(0.0,-0.1)$) edge ($(h21.north)+(0.0,0.1)$);
\draw[style={font=\sffamily\small}]
    ($(phi12.south)+(0.0,-0.1)$) edge ($(h22.north)+(0.0,0.1)$);
\draw[style={font=\sffamily\small}]
    ($(phi13.south)+(0.0,-0.1)$) edge ($(h23.north)+(0.0,0.1)$);
\draw[style={font=\sffamily\small}]
    ($(phi14.south)+(0.0,-0.1)$) edge ($(h24.north)+(0.0,0.1)$);
\end{tikzpicture}
\caption{\label{fig:hist3} Histories for three additional gluons: An individual gluon can be recombined with the other partons in several ways, leading to several underlying states with two  additional gluons. Each of these underlying states leads to a parton shower history illustrated by fig.\ \ref{fig:hist2}.}
\end{figure}

For a clear picture of how to generate parton-shower factors for matching, it is useful to perform a ``matrix-element correction" shift on the splitting kernels
\begin{eqnarray}
&&\frac{P^{[i]}(z^{[i]}(\zeta))}{q^{2[i]}(\bar{\tau})}
\longrightarrow
\frac{P_1^{[i]\mathrm{MEC}}(z^{[i]}(\zeta))}{q^{2[i]}(\bar{\tau})}
=
\frac{P_1^{[i]}(z^{[i]}(\zeta))}{q^{2[i]}(\bar{\tau})}
\otimes
\textnormal{\smaller$
\frac{
\left|\mathcal{M}_1(\Phi_1)\right|^2
}{
\left|\mathcal{M}_0(\Phi_0^{[1]})\right|^2
\frac{\alpha_s(\mu)}{2\pi}
\frac{P_1^{[1]}(z^{[1]}(\zeta))}{q^{2[1]}(\bar{\tau})}
+
\left|\mathcal{M}_0(\Phi_0^{[2]})\right|^2
\frac{\alpha_s(\mu)}{2\pi}
\frac{P_1^{[2]}(z^{[2]}(\zeta))}{q^{2[2]}(\bar{\tau})}
}$}~.
\label{eq:mec}
\end{eqnarray}
After this shift, the parton shower will recover the complete tree-level transition probability $|\mathcal{M}_1(\Phi_1)|^2$ when starting from the prior distribution $|\mathcal{M}_0(\Phi_0^{[i]})|^2$,
\begin{eqnarray*}
\sum\displaylimits_{i=1}^2 \left|\mathcal{M}_0(\Phi_0^{[i]})\right|^2 \frac{\alpha_s(\mu)}{2\pi} \frac{P^{[i]\mathrm{MEC}}(z^{[i]}(\zeta))}{q^{2[i]}(\bar{\tau})} = \left|\mathcal{M}_1(\Phi_1)\right|^2~.
\end{eqnarray*}
Introducing the shift \ref{eq:mec} in \ref{eq:ps-one-rate} leads to
\begin{eqnarray}
&&\obs{1}^{\mathrm{PS}}(\Phi_1) =
\int^{\tau^+}{\rm d}\bar{\tau}\,
\int{\rm d} \zeta
\int\frac{{\rm d} \varphi}{2\pi}
\otimes \left|\mathcal{M}_1(\Phi_1)\right|^2\\
&&\otimes
~\Bigg[
\mathcal{J}_0^{[1]}\mathcal{J}_1^{[1]}
\Delta_0(t(\tau^+),t^{[1]}(\bar{\tau}))
\frac{\alpha_s(t^{[1]}(\bar{\tau}))}{\alpha_s(\mu)}
\tcboxmath[colback=mymidgrey!20]{
\textnormal{\smaller$
\frac{
\left|\mathcal{M}_0(\Phi_0^{[1]})\right|^2
\frac{\alpha_s(\mu)}{2\pi}
\frac{P_1^{[1]}(z^{[1]}(\zeta))}{q^{2[1]}(\bar{\tau})}
}{
\left|\mathcal{M}_0(\Phi_0^{[1]})\right|^2
\frac{\alpha_s(\mu)}{2\pi}
\frac{P_1^{[1]}(z^{[1]}(\zeta))}{q^{2[1]}(\bar{\tau})}
+
\left|\mathcal{M}_0(\Phi_0^{[2]})\right|^2
\frac{\alpha_s(\mu)}{2\pi}
\frac{P_1^{[2]}(z^{[2]}(\zeta))}{q^{2[2]}(\bar{\tau})}
}
$}
}
\nonumber\\
&&\,\,+~
\mathcal{J}_0^{[2]}
\mathcal{J}_1^{[2]}\Delta_0(t(\tau^+),t^{[2]}(\bar{\tau}))
\frac{\alpha_s(t^{[2]}(\bar{\tau}))}{\alpha_s(\mu)}
\,
\tcboxmath[colback=mybrown!20]{
\textnormal{\smaller$
\frac{
\left|\mathcal{M}_0(\Phi_0^{[2]})\right|^2
\frac{\alpha_s(\mu)}{2\pi}
\frac{P_1^{[2]}(z^{[2]}(\zeta))}{q^{2[2]}(\bar{\tau})}
}{
\left|\mathcal{M}_0(\Phi_0^{[1]})\right|^2
\frac{\alpha_s(\mu)}{2\pi}
\frac{P_1^{[1]}(z^{[1]}(\zeta))}{q^{2[1]}(\bar{\tau})}
+
\left|\mathcal{M}_0(\Phi_0^{[2]})\right|^2
\frac{\alpha_s(\mu)}{2\pi}
\frac{P_1^{[2]}(z^{[2]}(\zeta))}{q^{2[2]}(\bar{\tau})}
}
$}
}
~\Bigg]\nonumber
\label{eq:ps-one-rate-mec}
\end{eqnarray}
It is now permissible to assume a that the phase-space points $\Phi_1$ were originally distributed according to the tree-level transition probability $|\mathcal{M}_1(\Phi_1)|^2$. Then, eq.\ \ref{eq:ps-one-rate-mec} gives a unique prescription for obtaining the correct (admixture of) parton-shower factors for $\Phi_1$:
\begin{itemize}
\item Construct the parton-shower history in fig.\ \ref{fig:hist1}.
\item Calculate the factors highlighted by \raisebox{1pt}{\tcboxmath[colback=mymidgrey!20,boxsep=1.5pt]{\cdot}} and \raisebox{1pt}{\tcboxmath[colback=mybrown!20,boxsep=1.5pt]{\cdot}}, and include the factors for both paths according to these proportions. For example, this can be achieved by probabilistically picking one path according to these factors\footnote{Explicit summation is also possible, though computationally more intensive, since more higher-order factors need to be evaluated per phase-space point.}.
\item Calculate the $\Delta_0(t(\tau^+),t^{[i]}(\bar{\tau}))
\frac{\alpha_s(t^{[i]}(\bar{\tau}))}{\alpha_s(\mu)}$ for the chosen path ($i$), and include this as rescaling of the cross section for the phase-space point $\Phi_1$.
\end{itemize}
This reasoning extends beyond this one-emission example~\cite{Lonnblad:2001iq,Lonnblad:2011xx}. Relevant histories for two-gluon and three-gluon states are shown in fig.\ \ref{fig:hist2} and fig.\ \ref{fig:hist3}, respectively. The weight of a path in a history for more than one additional parton is given by the product of the weights for all the individual transitions in the path, potentially including shifted rates due to matching or merging fixed-order matrix elements, and including matrix-element correction factors and ordering constraints~\cite{Fischer:2017yja}. If the history $\mathbb{H}(\Phi_n)$ is the collection of all paths $p_0,p_1,\dots,p_m$
from any $\Phi_0$ to $\Phi_n$, then the mixing weight for path $p_a$ becomes
\begin{eqnarray}
\label{eq:pickPathWeight}
w^{p_a}_n\left(\Phi_n\right)=
\frac{
\left|\,\mathcal{M}_0\left(\Phi_0^{\left[p_a\right]}\right)\right|^2
\prod\displaylimits_{i=1}^{n}
\frac{P_i^{[p_a]}\left(z_i^{\left[p_a\right]}\right) }{ q_i^{2\,\left[p_a\right]}}
}{
\sum\limits_{p_b\in\mathbb{H}(\Phi_n)}
\left|\,\mathcal{M}_0\left(\Phi_0^{p_b}\right)\right|^2
\prod\displaylimits_{i=1}^{n} 
\frac{P_i^{[p_b]}\left(z_i^{\left[p_b\right]}\right) }{ q_i^{2\,\left[p_b\right]}}
}~.
\end{eqnarray}
where the variables $z_i^{[r]}$ and $q_i^{2\,[r]}$ are calculated from $\Phi_{i}$ and knowledge of its production from $\Phi_{i-1}^{[r]}$. All parton-shower factors necessary for matching $n$-parton configurations are calculated by constructing the full parton-shower history of $\Phi_n$ and employing the method above. For example, the calculation of the \raisebox{1pt}{\tcboxmath[colback=red!10,boxsep=1.5pt]{\cdot}} terms in eq.~\ref{eq:nnnlops} require sequences $(\Phi_0,t_+)\rightarrow(\Phi_1,t_1)\rightarrow(\Phi_2,t_2)$, while the calculation of the \raisebox{1pt}{\tcboxmath[colback=blue!10,boxsep=1.5pt]{\cdot}} terms assumes knowledge of $(\Phi_0,t_+)\rightarrow(\Phi_1,t_1)\rightarrow(\Phi_2,t_2)\rightarrow(\Phi_3,t_3)$. 

The sequences for products of all-order factors are thus constructed in analogy to the CKKW-L merging scheme~\cite{Lonnblad:2001iq}, as admixture of contributions from all paths in the history. Subtracted factors (e.g.\ \raisebox{1pt}{\tcboxmath[colback=green!10,boxsep=1.5pt]{\cdot}} in eq.~\ref{eq:nnnlops}) also rely on the construction of the parton-shower history, since the all-order factors being subtracted are determined from the history. More details on parton-shower factors can be found in Appendix \ref{app:weights}.

\subsection{Definition of parton-shower accuracy}

\label{app:shower-accuracy}

\noindent
The logarithmic accuracy of parton showers is notoriously difficult to define, since parton showers are tools to produce \emph{event samples}, i.e.\ are not observable-specific. Thus, defining the accuracy by discussing individual observables or small sets of observables can be misleading. Furthermore, the parton shower is, ultimately, a numerical algorithm. This algorithm should remain self-consistent even in matching procedures.

This note thus employs an operational definition of the term ``parton-shower accuracy" that is of algorithmic nature, and stricter than typical log-counting arguments: The all-order factors of the parton-shower should be reproduced exactly, such that no measurement could distinguish the parton-shower and the matched prediction if all fixed-order cross section were calculated using the same approximations employed to derive parton-shower splitting kernels. When using exact fixed-order calculations, three levels of defining a criterion for ``maintaining the parton-shower accuracy" may be considered:
\begin{enumerate}
\item \emph{Strict parton-shower accuracy criterion}\\
All parton-shower all-order factors are reproduced identically, particularly including their admixture (cf.\ eqs.\ \ref{eq:ps-one-rate-mec} and \ref{eq:pickPathWeight}), and their (approximated fixed-order) prefactors. No other sources of higher-order contributions exist. This would also entail that sub-leading or finite fixed-order contributions should not multiply all-order factors, and could only enter as corrections at a fixed coupling power. This strict condition may be considered undesirable, since it e.g.\ ignores arguments about the treatment of hard virtual corrections~\cite{Parisi:1979se}. Consequently, no matching or merging method in the literature fulfills this overly strict condition. This strict criterion is not used in this note.
\item \emph{Balanced parton-shower accuracy criterion}\\
All parton-shower all-order factors are reproduced identically, particularly including their admixture (cf.\ eqs.\ \ref{eq:ps-one-rate-mec} and \ref{eq:pickPathWeight}). The fixed-order prefactors of all-order
factors may differ from strict parton-shower result by sub-leading contributions. Thus, complete fixed-order calculations (including finite and power-suppressed contributions) may multiply parton-shower all-order factors. This is the norm for matching methods. This note will apply this criterion when assessing changes to the all-order behavior due to matching.
\item \emph{Weak parton-shower accuracy criterion}\\
Parton-shower all-order factors are reproduced in the strongly ordered limit, where a single parton-shower path saturates the rate to produce the configuration. This condition allows subtleties about the admixture of all-order factors, and is thus closely related an accuracy definition by enumerating logarithms of the shower evolution variable. In this approximation, it may be possible to replace numerical parton-shower factors by analytic expressions, as e.g.\ advocated in~\cite{Catani:2001cc}. This weak criterion is not invoked in this note.
\end{enumerate}
This note will employ the \emph{balanced parton-shower accuracy criterion}. It follows that the ``symbolic parton-shower factor notation" of the main text is defined by
\begin{eqnarray}
\nonumber
&&\prod\limits_{i=n}^{n+m} \Delta_{i} (t_i,t_{i+1}) \f{w}{i+1}{\infty}{\Phi_{i+1}}\\
&&
\nonumber
=
\sum\limits_{p_a\in\mathbb{H}(\Phi_{n+m})}
\frac{
\left|\,\mathcal{M}_n\left(\Phi_n^{\left[p_a\right]}\right)\right|^2
\prod\displaylimits_{i=n+1}^{n+m}
\frac{P_i^{[p_a]}\left(z_i^{\left[p_a\right]}\right) }{ q_i^{2\,\left[p_a\right]}}
}{
\sum\limits_{p_b\in\mathbb{H}(\Phi_{n+m})}
\left|\,\mathcal{M}_n\left(\Phi_n^{\left[p_b\right]}\right)\right|^2
\prod\displaylimits_{i=n+1}^{n+m} 
\frac{P_i^{[p_b]}\left(z_i^{\left[p_b\right]}\right) }{ q_i^{2\,\left[p_b\right]}}
}\\
\nonumber
&&\qquad\qquad\otimes~
\Delta_{n} (t_+,t(\Phi_{n+1}^{\left[p_a\right]}))
\Delta_{n+1} (t(\Phi_{n+1}^{\left[p_a\right]}),t(\Phi_{n+2}^{\left[p_a\right]}))
\cdots
\Delta_{n+m} (t(\Phi_{n+m-1}^{\left[p_a\right]}),t(\Phi_{n+m}^{\left[p_a\right]}))
\\
&&\qquad\qquad\otimes~
\f{w}{n+1}{\infty}{\Phi_{n+1}^{\left[p_a\right]}}
\f{w}{n+2}{\infty}{\Phi_{n+2}^{\left[p_a\right]}}
\cdots
\f{w}{n+m+1}{\infty}{\Phi_{n+m+1}^{\left[p_a\right]}}
\end{eqnarray}
Terms multiplied by an expansion of shower factors should be calculated by multiplying each term in the sum by its own expansion.

In order to claim compliance with parton-shower accuracy, the value of such ``parton-shower factors" for a fixed $\Phi_{i+1}$ should be \emph{numerically identical}, irrespective of $\Phi_{i+1}$ having been sampled by the shower, or if  $\Phi_{i+1}$ had been pretabulated (by a fixed-order calculation) prior to showering, and the ``shower factor" had been calculated post facto.

\subsection{Generation of real-emission integrals and bias correction factors}
\label{app:integrals}

\noindent
Understanding the mixed weighting of the ``radiation pattern" $|\mathcal{M}_1(\Phi_1)|^2$ in eq.\ \ref{eq:ps-one-rate-mec} also leads to a definite method how to generate the (all-order) subtractions necessary for unitary matching. Since the shower produces an admixture of factors for fixed emission states $\Phi_1$, and the mixture is applied when matching an externally generated radiation pattern, the mixing should also be employed when projecting $\Phi_1$ onto lower-multiplicity phase space points $\Phi_0^{[i]}$. This ensures eq.\ \ref{eq:unlops} (as an extension of eq.\ \ref{eq:ps2}) are accurately reproduced. The integration over the spectrum that is necessary to produce the $\upoint$ (and $\upoiint$, $\upoiiint$) integrals in subtractions for unitarization, and to produce the $\upint\! \mathrm{d}\Phi_{\mathrm{R}}$ integrations for the toy fixed-order calculation are thus produced numerically. 
To give a concrete example, the $(n+1)$-parton dependence of the third-to-last line of eq.\ \ref{eq:un2lops} (highlighted by
\raisebox{1pt}{\tcboxmath[colback=blue!10,boxsep=1.5pt]{\cdot}}) can be obtained by
\begin{itemize}
\item[a)] tabulating two-gluon phase space points and constructing the history in fig.\ \ref{fig:hist2},
\item[b)] calculating the probability $w^q_{n+2}$ of each path $q$, and picking a path according to its probability,
\item[c)] multiplying the all-order factors for the chosen path $p$, and performing the replacement $\Phi_{n+2}\rightarrow \Phi_{n+1}^{[p]}$, and subtracting the result.
\end{itemize}
This method will lead to the desired result
\begin{eqnarray}
&&\intOne \ss{n+2}{0}{Q_{n+2}>Q_c}{\Phi_{n+2}}
~\f{w}{n+1}{\infty}{\Phi_{n+1}}\f{w}{n+2}{\infty}{\Phi_{n+2}} \Delta_{n} (t_+,t_{n+1}) \Delta_{n+1} (t_{n+1},t_{n+2})~\obs{n+1}\\
&&=
\upint \s{i+j}{k}{\Phi_{i+j}} \!\!
\sum\limits_{p\in\mathbb{H}(\Phi_{n+2})} \!\!
w^{p}_{n+2}\left(\Phi_{n+2}\right)
\f{w}{n+1}{\infty}{\Phi_{n+1}^{[p]}}\f{w}{n+2}{\infty}{\Phi_{n+2}} \Delta_{n} (t_+,t(\Phi_{n+1}^{[p]})) \Delta_{n+1} (t(\Phi_{n+1}^{[p]}),t(\Phi_{n+2}))~\obs{n+1} (\Phi_{n+1}^{[p]})~. \nonumber
\end{eqnarray}
Note that the requirement to recover the correct admixture relies on the weights $w^p_{n+2}$, which depend on the details of both the $(n+1)$- and the $(n+2)$-parton configurations. That the probability to assign an $(n+1)$-parton states depends on the $(n+1)$-parton configuration is necessary from the shower perspective. However, this is not appropriate for generating the complements to exclusive cross sections, since it can introduce artificial deformations of inclusive spectra. Take  \raisebox{1pt}{\tcboxmath[colback=orange!10,boxsep=1.5pt]{\cdot}} in eq.\ \ref{eq:un2lops} as an example. Since the sum of all $w^p_{n+2}$ is unity, any method to assign $(n+1)$-parton states is equivalent if the degrees of freedom of the$(n+1)$th parton are integrated over. This is however not the case for differential distributions. In fact, using eq.\ \ref{eq:pickPathWeight} to assign underlying states will favor $(n+1)$-parton states with small inter-parton separation, due to collinear enhancements. Thus, the transverse momentum spectrum of the $(n+1)$th parton will be deformed such that low transverse momenta become more likely than high values. Similar deformations were observed in~\cite{Lonnblad:2012ix}, and erroneously attributed to mismatched phase-space mappings, casting doubt on that implementation. Furthermore, in the integration of real corrections to form {\smaller$\overline{\mathrm{B}}$}-cross sections in the \textsc{Powheg} method, similar artifacts may naively arise when using Catani-Seymour subtraction, and if the Born matrix element values vary significantly across phase space, and would have to be regularized~\cite{Frixione:2007vw}.

The introduction of correction factors $\one{i+j}{i}$ allows to overcome such issues. These factors should ensure that $a)$ inclusive fixed-order $i$-parton cross sections do not accumulate undue biases, and that $b)$ a potential admixture of all-order factors for transitions from $0$ to $i$ additional partons is identical for all contributions to the inclusive cross section. 
For example, the factor $\one{n+3}{n+2}$ in the \raisebox{1pt}{\tcboxmath[colback=orange!10,boxsep=1.5pt]{\cdot}} terms in eq.\ \ref{eq:nnnlops} arranges that the integration matches the real-emission integral in the fixed-order prefactors of \raisebox{1pt}{\tcboxmath[colback=red!10,boxsep=1.5pt]{\cdot}} terms. At the same time, it establishes that the method to produce all-order factors in \raisebox{1pt}{\tcboxmath[colback=orange!10,boxsep=1.5pt]{\cdot}} and \raisebox{1pt}{\tcboxmath[colback=red!10,boxsep=1.5pt]{\cdot}} is equivalent.

It will be assumed that all parton-shower factors multiplying a specific fixed-order contribution will be calculated simultaneously. This calculation will proceed by constructing the history of phase-space points entering the fixed-order term. To obtain the correct admixture of shower factors, a path is chosen probabilistically according to eq.\ \ref{eq:pickPathWeight}. If the path is also used to replace phase-space points with underlying configurations (e.g.\ for the sake of creating a subtraction), then an undesirable bias (relative to the fixed-order inclusive calculation) has been introduced. For \nnnlops{} matching, the bias correction factors $\one{n+1}{n},\,\one{n+2}{n},\,\one{n+3}{n}\,\one{n+2}{n+1}\,\one{n+3}{n+1}$ and $\one{n+3}{n+2}$ have to be defined.

The most straight-forward bias-correction factors is $\one{n+1}{n}$, since its coefficients in eq.\ \ref{eq:nnnlops} does not include parton-shower factors.
Introducing the symbol $w^{\mathrm{FO}}_{n+m}$ for the rate at which a real-emission phase-space point $\Phi_{n+m}$ would have been replaced by an underlying phase-space point $\Phi_n^{[p]}$ in an inclusive fixed-order calculation, this bias correction is
\begin{eqnarray}
\one{n+1}{n} &=& 
\frac{w^{\mathrm{FO}}_{n+1}\left(\Phi_{n+1},\Phi_n^{[p]}\right)}{\sum\limits_{q\in\mathbb{H}(\Phi_{n+1})} w^q_{n+1}\left(\Phi_{n+1}\right) \delta \left(\Phi_{n}^{[p]} - \Phi_{n}^{[q]}\right) }~,
\end{eqnarray}
This is obtained by reading off from the expected $w^{\mathrm{FO}}$ weighting of the complementary low-separation real-emission contributions in $\ss{n}{1}{Q_{n}>Q_c \land Q_{n+1}<Q_c }{\Phi_{n}^{[p]}}$, and dividing by the rate at which the shower method would have suggested the replacement $\Phi_{n+m}\rightarrow\Phi_n^{[p]}$. The sum in the denominator is necessary if multiple shower paths might lead to the underlying state $\Phi_n^{[p]}$\footnote{For example, both paths Fig.\ \ref{fig:hist1} may lead to the leftmost underlying state in Fig.\ \ref{fig:hist2}.}, so that the sum of the paths needs to be replaced.

The coefficients of the factors $\one{n+2}{n}$ and $\one{n+3}{n}$ do not contain parton-shower factors either. They depend on the details of how the inclusive \nnnlo{} prediction in defined, and in particular on whether new functions ``$w^{\mathrm{FO}}_{n+m}$" are used. In the toy calculation described in sec.\ \ref{sec:fixed_order}, no new fixed-order biases beyond $w^{\mathrm{FO}}_{n+m}$ are introduced, since the generation proceeds in successive multiplicity steps. This means that $\one{n+2}{n}$ and $\one{n+3}{n}$ take the forms
\begin{eqnarray}
\one{n+2}{n} &=& 
\frac{w^{\mathrm{FO}}_{n+2}\left(\Phi_{n+2},\Phi_{n+1}^{[p]}\right) w^{\mathrm{FO}}_{n+1}\left(\Phi_{n+1}^{[p]},\Phi_{n}^{[p]}\right) }{\sum\limits_{q\in\mathbb{H}(\Phi_{n+2})} w^q_{n+2}\left(\Phi_{n+2}\right) \delta \left(\Phi_{n+1}^{[p]} - \Phi_{n+1}^{[q]}\right) \delta \left(\Phi_{n}^{[p]} - \Phi_{n}^{[q]}\right) }
\\
\one{n+3}{n} &=& 
\frac{w^{\mathrm{FO}}_{n+3}\left(\Phi_{n+3},\Phi_{n+2}^{[p]}\right) 
      w^{\mathrm{FO}}_{n+2}\left(\Phi_{n+2}^{[p]},\Phi_{n+1}^{[p]}\right)
      w^{\mathrm{FO}}_{n+1}\left(\Phi_{n+1}^{[p]},\Phi_{n}^{[p]}\right)
 }{\sum\limits_{q\in\mathbb{H}(\Phi_{n+3})} w^q_{n+3}\left(\Phi_{n+3}\right) \delta \left(\Phi_{n+2}^{[p]} - \Phi_{n+2}^{[q]}\right) \delta \left(\Phi_{n+1}^{[p]} - \Phi_{n+1}^{[q]}\right) \delta \left(\Phi_{n}^{[p]} - \Phi_{n}^{[q]}\right) }
 ~,\qquad
\end{eqnarray}
which are again obtained by dividing the expected $w^{\mathrm{FO}}$ weighting by the rate at which the shower method would have replaced the state.

All other bias-correction factors ($\one{n+2}{n+1},\one{n+3}{n+2},\one{n+3}{n+1}$) should also guarantee the correctly weighted mixture of parton-shower factors for the underlying states. This is luckily already guaranteed by analogous definitions
\begin{eqnarray}
\one{n+2}{n+1} &=& \frac{w^{\mathrm{FO}}_{n+2}\left(\Phi_{n+2},\Phi_{n+1}^{[p]}\right)}{\sum\limits_{q\in\mathbb{H}(\Phi_{n+2})} w^q_{n+2}\left(\Phi_{n+2}\right) \delta \left(\Phi_{n+1}^{[p]} - \Phi_{n+1}^{[q]}\right) }\\ 
\one{n+3}{n+2} &=& \frac{w^{\mathrm{FO}}_{n+3}\left(\Phi_{n+3},\Phi_{n+2}^{[p]}\right)}{\sum\limits_{q\in\mathbb{H}(\Phi_{n+3})} w^q_{n+3}\left(\Phi_{n+3}\right) \delta \left(\Phi_{n+2}^{[p]} - \Phi_{n+2}^{[q]}\right) } \\
\one{n+3}{n+1} &=& 
\frac{w^{\mathrm{FO}}_{n+3}\left(\Phi_{n+3},\Phi_{n+2}^{[p]}\right) w^{\mathrm{FO}}_{n+2}\left(\Phi_{n+2}^{[p]},\Phi_{n+1}^{[p]}\right) }{\sum\limits_{q\in\mathbb{H}(\Phi_{n+3})} w^q_{n+3}\left(\Phi_{n+3}\right) \delta \left(\Phi_{n+2}^{[p]} - \Phi_{n+2}^{[q]}\right) \delta \left(\Phi_{n+1}^{[p]} - \Phi_{n+1}^{[q]}\right)  }
~,\qquad\quad
\end{eqnarray}
This can be seen by combining the rate with which the parton-shower would have ``mixed in" the chosen path and bias-correction factor, for example leading to
\begin{eqnarray*}
&&w^p_{n+2}\left(\Phi_{n+2}\right) \one{n+2}{n+1}\\
&& =\quad
w^{\mathrm{FO}}_{n+2}\left(\Phi_{n+2},\Phi_{n+1}^{[p]}\right)
\frac{
w^p_{n+2}\left(\Phi_{n+2}\right)
%\sum\limits_{r\in\mathbb{H}(\Phi_{n+2})} w^r_{n+2}\left(\Phi_{n+2}\right) 
%\delta \left(\Phi_{n+1}^{[p]} - \Phi_{n+1}^{[r]}\right)
%\delta \left(\Phi_{n}^{[p]} - \Phi_{n}^{[r]}\right)
}{\sum\limits_{q\in\mathbb{H}(\Phi_{n+2})} w^q_{n+2}\left(\Phi_{n+2}\right) \delta \left(\Phi_{n+1}^{[p]} - \Phi_{n+1}^{[q]}\right)}\nonumber\\
&&=\quad
w^{\mathrm{FO}}_{n+2}\left(\Phi_{n+2},\Phi_{n+1}^{[p]}\right)
\frac{C\left(\Phi_{n+2},\Phi_{n+1}^{[p]}\right)\, w^{p\in\mathbb{H}\left(  \Phi_{n+1}^{[p]} \right)}_{n+1}  }
{C\left(\Phi_{n+2},\Phi_{n+1}^{[p]}\right) \sum\limits_{t\in\mathbb{H}\left(\Phi_{n+1}^{[p]}\right)} w^t_{n+1} \left(  \Phi_{n+1}^{[p]}  \right) }
\quad=\quad
w^{\mathrm{FO}}_{n+2}
w^{p\in\mathbb{H}\left(  \Phi_{n+1}^{[p]} \right)}_{n+1}\left( \Phi_{n+1}^{[p]} \right) ~,
\end{eqnarray*}
by virtue of eq.\ \ref{eq:pickPathWeight}, thus leading to the desired result. The same conclusion applies to $\one{n+3}{n+2}$ and $\one{n+3}{n+1}$. 
While the factors $\one{i+j}{i}$ guarantee the correct fixed-order behavior, they unfortunately complicate the assessment of changes to the parton-shower accuracy for exclusive $(i>n)$-parton observables. To determine their impact, it is useful to note that terms related to $\one{i+j}{i}$ typically enter exclusive predictions in the form
\begin{eqnarray}
&&
\intOne \s{i+j}{k}{\Phi_{i+j}} \left[\one{i+j}{i} \f{f}{i}{\infty}{\Phi_i} - \f{f}{i}{\infty}{\Phi_i} \f{f}{i+j}{\infty}{\Phi_{i+j}}\right]\\
&&\nonumber
=
~
\tcboxmath[colback=black!00,colframe=red!50,boxrule=0.5pt,boxsep=1pt,hypertarget=cfoliveA,hyperlink=cfoliveB]{~
\intOne \s{i+j}{k}{\Phi_{i+j}} (\one{i+j}{i} - 1 ) \otimes \f{f}{i}{\infty}{\Phi_i} ~}
+ ~
\tcboxmath[colback=black!00,colframe=black!50,boxrule=0.5pt,boxsep=1pt,hypertarget=cfoliveA,hyperlink=cfoliveB]{~
\intOne \s{i+j}{k}{\Phi_{i+j}} \f{f}{i}{\infty}{\Phi_i} \left[ 1 -  \f{f}{i+j}{\infty}{\Phi_{i+j}}\right]~} \quad
\end{eqnarray}
with $\f{f}{i}{\infty}{\Phi_i}$ a placeholder for parton-shower factors relating to transitions from $0$-additional-parton to (at most) $i$-additional-parton configurations, and
$\f{f}{i+j}{\infty}{\Phi_{i+j}}$ for factors for transitions from $i$- to $(i+j)$-additional-parton states.

The \raisebox{1pt}{\tcboxmath[colback=black!00,colframe=black!50,boxrule=0.5pt,boxsep=1.5pt]{\cdot}} term is the typical parton-shower contribution, for which the first term in the expansion of $\f{f}{i+j}{\infty}{\Phi_{i+j}}$ removes hard real-emission terms, while the remaining higher orders resum the effect of the jet veto. Thus, this term does not impair the accuracy. The $(\one{i+j}{i}- 1)$ factor does not contain any large logarithms. It measures the difference between the fixed-order bias and the admixture of shower paths, and would vanish for an ideal fixed-order calculation that employs the method of Appendix \ref{app:histories} to produce the real-emission integrals in the inclusive cross sections, or if there is a single dominant path from a phase-space point $\Phi_{i}^{[p]}$ to $\Phi_{i+j}$.  
Thus, log-counting suggests that the term can be omitted when discussing the all-order accuracy of exclusive observables. The term however changes the ``accuracy of the parton shower"  in the strictest sense, i.e.\ when requiring that the admixture of all-order factors is completely equivalent to the shower. The \raisebox{1pt}{\tcboxmath[colback=black!00,colframe=red!50,boxrule=0.5pt,boxsep=1.5pt]{\cdot}}  term basically suggests that in a matched calculation, the definition of a fixed-order exclusive cross section should ideally be determined by the parton-shower method. Thus, the change in all-order accuracy is at the same level as attaching parton-shower resummation to finite parts of fixed-order cross sections -- and may thus be deemed acceptable. A detailed study of the impact of different $\Phi_{i}$-selection procedures could be an interesting addition to a future publication.

The functional form of the $w^{\mathrm{FO}}_{i+j}$ factor may differ between fixed-order methods that produce inclusive cross sections. The toy calculation described in Appendix \ref{app:toycalc} will employ a non-trivial $w^{\mathrm{FO}}_{i+j}$ to allow the validation of the correctness of the $\one{i+j}{i}$ implementations.

\subsection{Details on generating the toy fixed-order calculation}

\label{app:toycalc}

\noindent
The toy third-order calculation used in this note in constructed from tree-level event samples generated with \textsc{Madgraph5}\_\textsc{aMc@nlo}. It should again be stressed that this is a toy calculation that only serves to assess the implementation of the \nnnlops\ formula eq.\ \ref{eq:nnnlops}. Since tree-level events are used, the generation of samples containing additional gluons requires regularization cuts $S(\Phi_{n})>S_c$. Thus, minimal cuts are applied on the projection of the (sum of) gluon four-momenta onto the four-momenta of the other partons. The notation $S(\Phi_{n})>S_c$ implies very inclusive cuts,
\begin{eqnarray*}
& p_u p_{g_i}, p_{\bar{u}} p_{g_i}, p_{g_i} p_{g_{j\neq i}} > S_c~,~1\leq i,j \leq n\\
& p_u (p_{g_i} + p_{g_{j\neq i}}), p_{\bar{u}} (p_{g_i} + p_{g_{j\neq i}}), (p_{g_k}  + p_{g_{i\neq j, k}} + p_{g_{j\neq i,k}})/2  > S_c~,~1\leq i,j,k \leq n\\
& p_u (p_{g_{i\neq j, k}} + p_{g_{j\neq i,k}} + p_{g_{k\neq i,j}} ), p_{\bar{u}} (p_{g_{i\neq j, k}} + p_{g_{j\neq i,k}} + p_{g_{k\neq i,j}} ) > S_c~,~1\leq i,j,k \leq n~,
\end{eqnarray*}
with $S_c=0.1~\gev^2$. These cuts should be minimal enough to ensure that the majority of the radiation spectra is retained, and allow to assemble reasonable approximations for jet-vetoed fixed-order calculations.

These tree-level samples are then used to construct the toy jet-vetoed cross sections defined in eqs.\ \ref{eq:toynlo2}, \ref{eq:toynnlo2} and \ref{eq:toynnnlo2}. Parton-shower histories are used to perform the $\mathrm{d}\Phi_{\mathrm{R}}$ integration, and to apply the jet veto constraints $Q(\Phi_{n})$\,{\smaller$>$}\,$Q_c$\footnote{Note that the final form of the toy calculations only depend on constraints $Q(\Phi_{n})$\,{\smaller$>$}\,$Q_c$ that remove configurations too close to the phase-space boundaries}:
\begin{enumerate}
\item A small value of $Q_c$ is chosen. The results in sec.\ \ref{sec:results} use $Q_c=1~\gev^2$. 
\item The parton-shower history for a pre-tabulated phase-space point $\Phi_{n}$ is constructed (cf.\ Appendix \ref{app:histories} and
figs.\ \ref{fig:hist1}, \ref{fig:hist2} and \ref{fig:hist3})
\item Only paths that satisfy $t_{n}$\,{\smaller$>$}\,$Q_c$ are retained, where $t_{n}$ is the parton-shower evolution variable assigned to the transition $\Phi_{n-1}^{[i]} \rightarrow \Phi_{n}$\footnote{Another strategy would be to ignore this constraint and retain the paths, resulting in all phase-space points passing the slicing cut. The state would then be employed in matching, and reweighted with parton-shower factors. At this stage, the path selection weight in eq.~\ref{eq:pickPathWeight} would favor paths with $t_{n}$\,{\smaller$<$}\,$Q_c$, so that no-emission factors $\Delta_{n} (t_{start},t(\Phi_{n}) \textnormal{\smaller$<$}\,Q_c)$ would multiply $\Phi_{n}$ states. Such contributions would be absent in the parton shower if $Q_c$ is identified with the parton-shower cut-off (as is assumed in this note), since the evolution would have terminated before being able to produce an emission at $t(\Phi_{n})$, i.e.\ $\Delta_{n} (t_{start},t(\Phi_{n}) \textnormal{\smaller$<$}\,Q_c)\equiv 0$ when applied to $\Phi_{n}$ states. Thus, retaining the paths will have no impact on the final, matched result for $\Phi_{n}$ states, nor on their unitarity subtraction. The path may still contribute to the complement for exclusive $(n-m)$-parton calculations (where $0< m\leq n$), if $t_{n-m}$\,{\smaller$>$}\,$Q_c$ holds. Inspecting the
definitions of exclusive cross sections in section \ref{sec:fixed_order} reveals that the effect of including these contributions to the complements is to cancel contributions to the exclusive cross section derived from paths with
$t(\Phi_{n}) \textnormal{\smaller$<$}\,Q_c)$. Thus, the final result is identical to systematically discarding such paths. Only the selection of paths with $t_{n}$\,{\smaller$>$}\,$Q_c$ will lead to a non-vanishing shower weight, i.e.\ contribute to the final result. The validity of this argument has been verified with the toy implementation to sub-percent level, for several tens of observables.}. Note that this criterion depends on the path.
\item If no paths exist that fulfill the constraint, the event is discarded.
\item If valid paths exist, and no further constraints should be applied, and the event should not be processed further, then include the phase-space point in the toy calculation.
\item If valid paths exist, and the further cut $Q(\Phi_{n-1})$\,{\smaller$>$}\,$Q_c$ should be applied (to produce the
\raisebox{1pt}{\tcboxmath[colback=black!00,colframe=mypurple,boxrule=0.5pt,boxsep=1.5pt]{\cdot}} integrals in eqs.\ \ref{eq:toynlo2}, \ref{eq:toynnlo2} and \ref{eq:toynnnlo2}), then retain only the paths that satisfy the constraint. If no paths exist that fulfill the constraint, the event is discarded. Otherwise, the $\mathrm{d}\Phi_{\mathrm{R}}$ integral is performed and an appropriate phase-space point $\Phi_{n-1}$ included in the toy calculation.
\end{enumerate}
The $\mathrm{d}\Phi_{\mathrm{R}}$ integrals are obtained with the method described in Appendix \ref{app:integrals}, but omitting bias correction factors. To avoid undue dependence on the parton shower splitting kernels, the requested $\Phi_{n-1}^{[p]}$ points are chosen amongst $\{\Phi_{n-1}^{[1]},\Phi_{n-1}^{[2]},\dots\}$ according to the simpler weights
\begin{eqnarray}
\label{eq:intRweights}
w^{\mathrm{FO}}_n\left(\Phi_n,\Phi_{n-1}^{[p]}\right) =
\frac{
P_n^{[p]}(z^{[p]}) / t^{[p]}
}{
\sum_q P_n^{[q]}(z^{[q]}) / t^{[q]}
}~,
\end{eqnarray}
where the variables $z^{[r]}$ and $t^{[r]}$ are calculated from $\Phi_{n}$ and knowledge of its production from $\Phi_{n-1}^{[r]}$, and $\sum_q$ sums over all possible replacements. This differs slightly from the prescription for producing unitarity subtractions, since the latter mixes underlying states according to the weight in eq.\ \ref{eq:pickPathWeight}. The prescription \ref{eq:intRweights} is chosen to be less sensitive to the details of lower-multiplicity phase-space points, to allow non-trivial validation, and to increase the similarity to other methods for generating toy calculations such as~\cite{Rubin:2010xp}.

The \raisebox{1pt}{\tcboxmath[colback=black!00,colframe=mypurple,boxrule=0.5pt,boxsep=1.5pt]{\cdot}} contribution to eq.\ \ref{eq:toynlo2} offers a tangible example for the $\mathrm{d}\Phi_{\mathrm{R}}$ integration. This contribution can be obtained by \emph{a)} tabulating three-gluon phase space points, \emph{b)} constructing the history in fig.\ \ref{fig:hist3}, \emph{c)} retaining only histories for which the $Q$-constraints are fulfilled (see discussion above), \emph{d)} calculating the weight \ref{eq:intRweights} of each valid $\Phi_2^{[q]}$, \emph{e)} picking one $\Phi_2^{[p]}$ probabilistically, \emph{f)} performing the replacement $\Phi_3\rightarrow \Phi_2^{[p]}$, and subtracting the result.

\begin{table}[t!]
\renewcommand{\arraystretch}{1.5}
\begin{tabular}{ |c c c c c c|}
\hline
$a_2^{q\bar q}$~=~  -50, \qquad\qquad & $a_2^{q g_1}$~=~ 10, \qquad\qquad & $a_2^{q g_2}$~=~ 50, \qquad\qquad & $a_2^{\bar q g_1}$~=~ 10, \qquad\qquad & $a_2^{\bar q g_2}$~=~ 50, \qquad\qquad & $a_2^{gg}$~=~ 100 \\
$a_1^{q\bar q}$~=~ 50, \qquad\quad &$a_1^{q g}$~=~ 100, \qquad\quad &$a_1^{\bar q g}$~=~ 100, \qquad\quad &$b_1^{q\bar q}$~=~ 0, \qquad\quad &$b_1^{q g}$~=~ -7000, \qquad\quad &$b_1^{\bar q g}$~=~ -7000 \\
$a_0^{qe}$~=~ 2, \qquad\quad &$a_0^{\bar qe}$~=~ 10, \qquad\quad &$b_0^{qe}$~=~ -100, \qquad\quad &$b_0^{\bar q e}$~=~ -100, \qquad\quad &$c_0^{qe}$~=~ 10000, \qquad\quad &$c_0^{\bar q e}$~=~ 10000 \\
\hline
\end{tabular}
\caption{\label{tab:toycalc-coefficients} Values of the coefficients $a$, $b$ and $c$ used in eqs.\ \ref{eq:toynlo2}, \ref{eq:toynnlo2} and \ref{eq:toynnnlo2}.}
\end{table}

For the matching implementation, it is further useful to retain the clustering scale $t^{[p]}$ for each $\Phi_{n-1}^{[p]}$. Otherwise, it is challenging to match the Sudakov reweighting between reweighted (toy) fixed-order inputs and \nnnlops{}-produced complements. The generation of the latter proceeds simultaneously to the generation of other $\Phi_n$-dependent factors, of which some require selecting parton shower paths of ordered emission sequences to $\Phi_n$. Thus, to produce matching Sudakov factors for complements and unitarity subtractions, the former should sum over the same paths. To obtain an exact match of the all-order factors generated when reweighting (toy) fixed-order inputs and their complements, the reweighting of the former should consider the value of the clustering scale $t^{[p]}$ when selecting valid shower paths. For the closure test calculation used in this note, not invoking the constraint can lead to $\leq 5$\% differences when approaching phase-space regions with multiple collinear partons. Retaining an additional clustering scale information is uncomfortable, since it can blurs the line between the fixed-order and the shower programs. However, the documentation of this scale in an inclusive fixed-order calculation should be technically feasible, and allowed within the accuracy of the calculation. Note that auxiliary (veto or shower starting) scale information is routinely deemed acceptable in {\smaller POWHEG} and {\smaller MC@NLO} matching. 

\subsection{Assessing the impact of bias correction factors}

\begin{figure}[ht!]
\centering
  \begin{subfigure}{0.45\textwidth}
  \includegraphics[width=\textwidth]{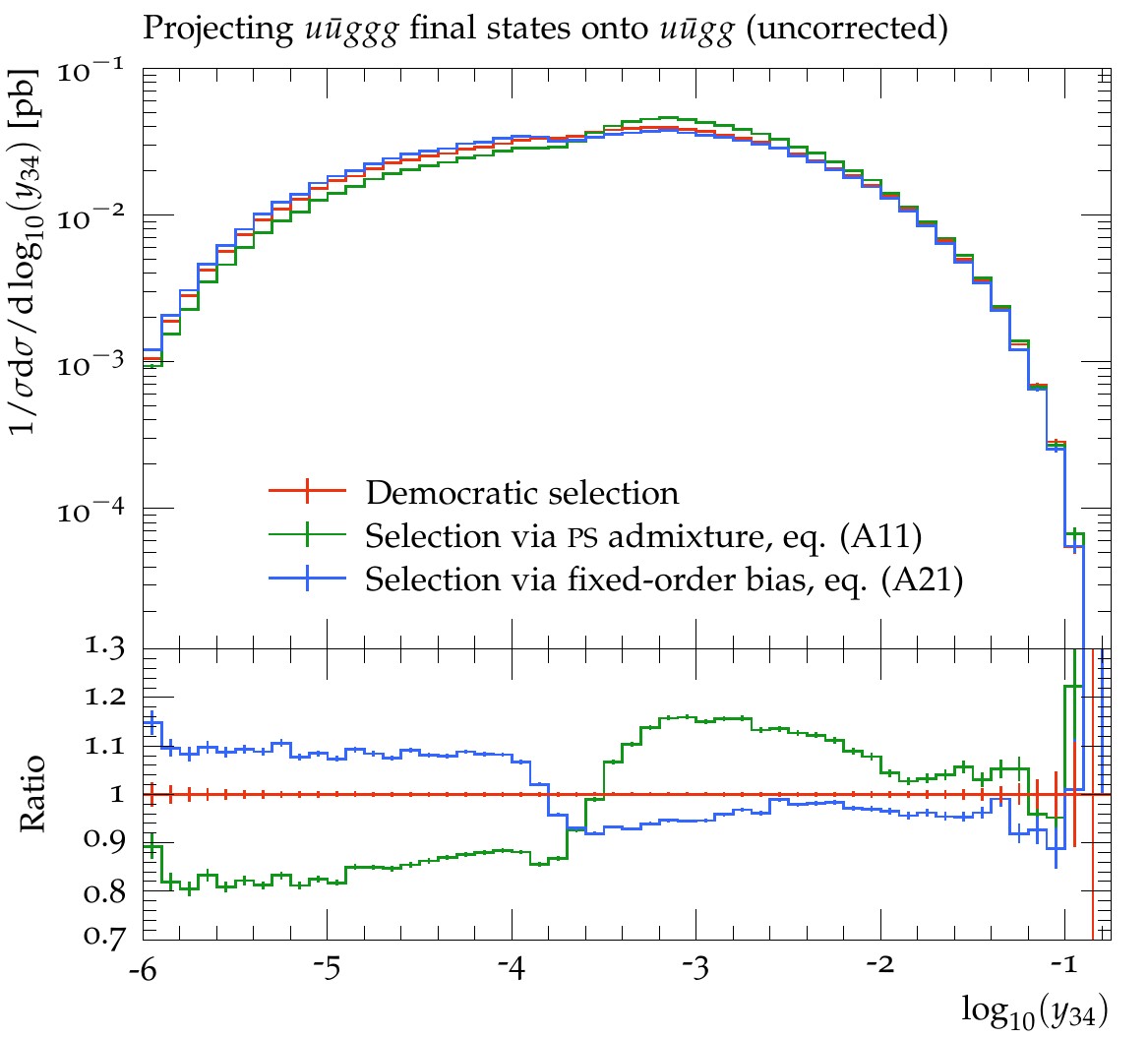}{}
  \caption{\label{fig:biased_clustering}
  Impact of different clustering strategies on four-parton observables.}
  \end{subfigure}\hfill
  \begin{subfigure}{0.45\textwidth}
  \includegraphics[width=\textwidth]{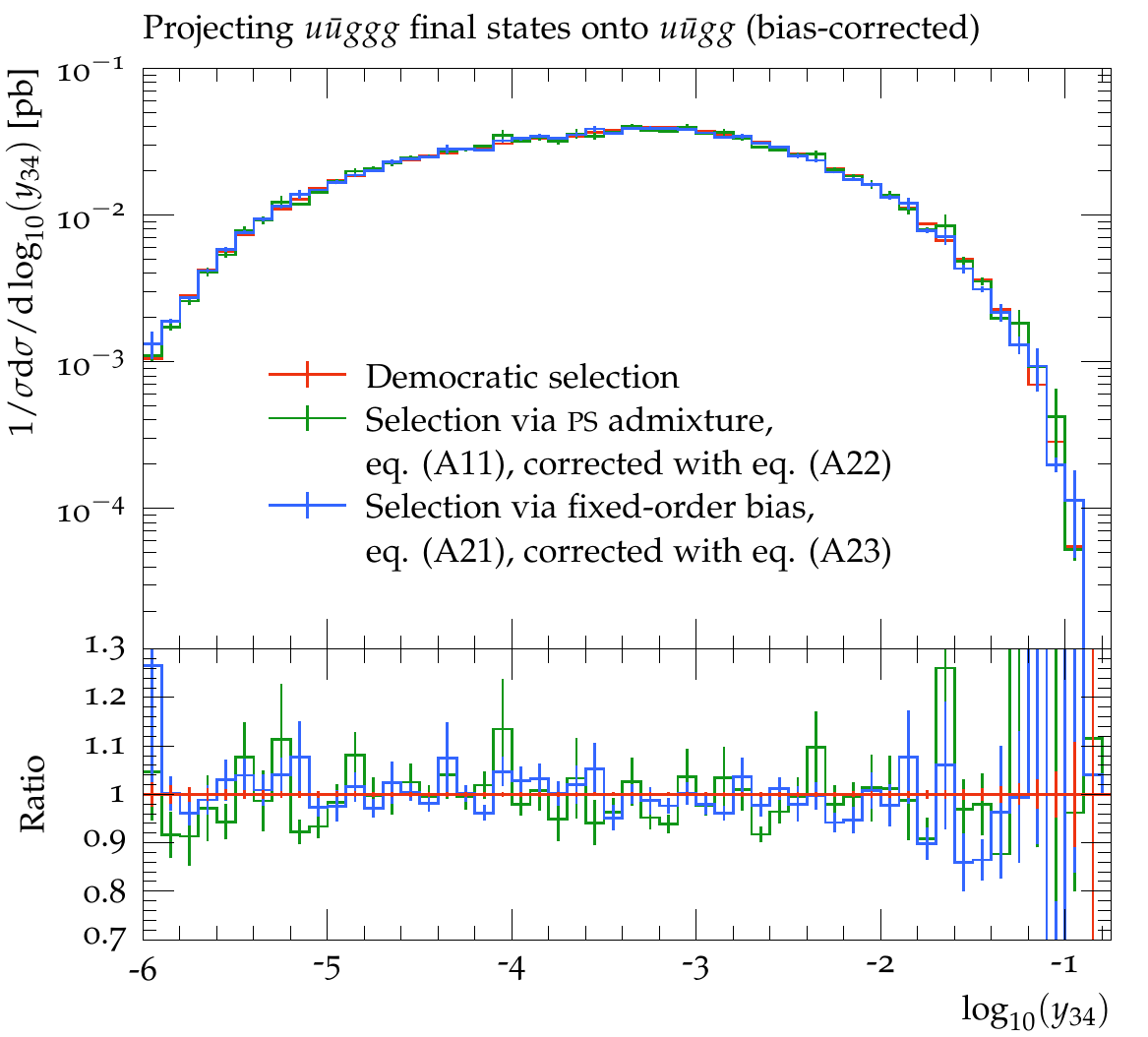}{}
  \caption{\label{fig:unbiased_clustering}
  Four-parton observable results, using $\one{3}{2}$ factors to map onto a
  democratic clustering strategy.}
  \end{subfigure}
\end{figure}

\noindent
The sample generation discussed at the end of the previous section also allows non-trivial studies of the spectrum deformation effects (and the correction via $\one{3}{2}$ factors) discussed at the end of Appendix \ref{app:histories}. The current section will focus on such deformations for one instructive example. For other lepton (hadron) collider processes, deformations of similar (larger) size are expected. Figure \ref{fig:biased_clustering} highlights that, given $q\bar q g g g$ states, the method to assign underlying $q\bar q g g$ configurations does indeed lead to a non-negligible bias. The differences are similar in size to the naive expectation for \nlo{} corrections. The deformation of the spectra is somewhat unexpected. Using \ref{eq:intRweights} results smaller $y_{34}$ jet separation than the democratic clustering approach, yet the naive expectation is that the probability-based clustering removes the least-separated partons first, potentially yielding harder four-parton states. The bias introduced by eq.\ \ref{eq:pickPathWeight} is not completely obvious either, since it produces larger $y_{34}$ separation, while the naive expectation is that the dependence of eq.\ \ref{eq:pickPathWeight} on (an approximation of) the four-parton matrix element would tilt the spectrum in favor of less separated partons.
% The deformation of the $y_{23}$ spectrum is less pronounced -- the bias is reduced for observables that do not sensitive to the detailed distribution of four-parton states. Nevertheless, some deformations remain. 
The clustering bias is non-trivial. 

These undesirable biases can be removed by including appropriate corrective weights,
\begin{eqnarray}
\label{eq:one_complement_weights_plot_ps}
&&\one{3}{2}\mathrm{(PS)} = \frac{1/\textnormal{number of possible replacements of a $5$-parton state with a $4$-parton state}}{
\sum\limits_{q\in\mathbb{H}(\Phi_{n+3})} w^q_{n+3}\left(\Phi_{n+3}\right) \delta \left(\Phi_{n+2}^{[p]} - \Phi_{n+2}^{[q]}\right) 
}\\
\label{eq:one_complement_weights_plot_fo}
&&\one{3}{2}\mathrm{(FO)} = \frac{1/\textnormal{number of possible replacements of a $5$-parton state with a $4$-parton state}}{
w^{\mathrm{FO}}_{n+3}\left(\Phi_{n+3},\Phi_{n+2}^{[p]}\right)
}~,
\end{eqnarray}
as shown in Figure \ref{fig:unbiased_clustering}, were different strategies are mapped to a democratic selection baseline. Any corrective weight, to convert mapping strategies into each other, may be obtained once the parton-shower history has been constructed. The price to pay for this flexibility is, as the comparison between Figures \ref{fig:biased_clustering} and \ref{fig:unbiased_clustering} indicates, a slower statistical convergence (given an identical number of events) in the latter due to an additional source of event weights. This is acceptable, since the weights are necessary to reinstate formal correctness. In the development of the \pythia{}+\dire{} implementation of \nnnlops{}, the removal of biases can explicitly be verified, due to having full control over the toy fixed-order calculations. Overall, it would seem prudent to assess the bias (potentially) introduced by the unitarization strategy also for other unitarized matching and merging approaches.

\subsection{Parton-shower factors and trial showers}
\label{app:weights}

\noindent
Parton-shower histories are paramount in defining the boundary conditions under which parton-shower and matching factors can be calculated. This appendix will be concerned with the technical implementation of the matching factors appearing in eq.\ \ref{eq:nnnlops}. For the sake of brevity, the main text uses a symbolic notation for parton-shower factors. This symbolic notation for products of parton-shower factors $\f{f}{n}{\infty}{\Phi_{n}} \f{f}{n+1}{\infty}{\Phi_{n+1}} \cdots \f{f}{n+m}{\infty}{\Phi_{n+m}}$ is, as required by the \emph{balanced parton-shower accuracy criterion}, defined by the mixture
\begin{eqnarray}
\label{eq:symbolic-shower-factors-notation}
&&\f{f}{n}{\infty}{\Phi_{n}}
\f{f}{n+1}{\infty}{\Phi_{n+1}}
\cdots
\f{f}{n+m}{\infty}{\Phi_{n+m}}\\
&&\nonumber
=
\sum\limits_{p_a\in\mathbb{H}(\Phi_{n+m})}
\frac{
\left|\,\mathcal{M}_n\left(\Phi_n^{\left[p_a\right]}\right)\right|^2
\prod\displaylimits_{i=n+1}^{n+m}
\frac{P_i^{[p_a]}\left(z_i^{\left[p_a\right]}\right) }{ q_i^{2\,\left[p_a\right]}}
}{
\sum\limits_{p_b\in\mathbb{H}(\Phi_{n+m})}
\left|\,\mathcal{M}_n\left(\Phi_n^{\left[p_b\right]}\right)\right|^2
\prod\displaylimits_{i=n+1}^{n+m} 
\frac{P_i^{[p_b]}\left(z_i^{\left[p_b\right]}\right) }{ q_i^{2\,\left[p_b\right]}}
}
\f{f}{n}{\infty}{\Phi_n^{\left[p_a\right]}}
\f{f}{n+1}{\infty}{\Phi_{n+1}^{\left[p_a\right]}}
\cdots
\f{f}{n+m}{\infty}{\Phi_{n+m}^{\left[p_a\right]}}~,
\end{eqnarray}
where the sums extend over all possible parton shower paths. A detailed example for one contribution in eq.\ \ref{eq:nnnlops} is 
\begin{eqnarray}
\nonumber
&&\Delta_{n} (t_+,t_{n+1})\Delta_{n+1} (t_{n+1},t_{n+2}) \f{w}{n+1}{\infty}{\Phi_{n+1}}  \f{w}{n+2}{\infty}{\Phi_{n+2}}
\\
\nonumber
&&\otimes~
\Big[ 1 - \f{w}{n+1}{1}{\Phi_{n+1}} - \f{w}{n+2}{1}{\Phi_{n+2}} - \f{\Delta}{n}{1}{t_+,t_{n+1}} - \f{\Delta}{n+1}{1}{t_{n+1},t_{n+2}} \Big]\\
&&
\nonumber
=
\sum\limits_{p_a\in\mathbb{H}(\Phi_{n+2})}
\frac{
\left|\,\mathcal{M}_n\left(\Phi_n^{\left[p_a\right]}\right)\right|^2
\prod\displaylimits_{i=n+1}^{n+2}
\frac{P_i^{[p_a]}\left(z_i^{\left[p_a\right]}\right) }{ q_i^{2\,\left[p_a\right]}}
}{
\sum\limits_{p_b\in\mathbb{H}(\Phi_{n+2})}
\left|\,\mathcal{M}_n\left(\Phi_n^{\left[p_b\right]}\right)\right|^2
\prod\displaylimits_{i=n+1}^{n+2} 
\frac{P_i^{[p_b]}\left(z_i^{\left[p_b\right]}\right) }{ q_i^{2\,\left[p_b\right]}}
}\\
\nonumber
&&\qquad\qquad\otimes~
\Delta_{n} (t_+,t(\Phi_{n+1}^{\left[p_a\right]}))\Delta_{n+1} (t(\Phi_{n+1}^{\left[p_a\right]}),t(\Phi_{n+2}^{\left[p_a\right]})) \f{w}{n+1}{\infty}{(\Phi_{n+1}^{\left[p_a\right]})}  \f{w}{n+2}{\infty}{(\Phi_{n+2}^{\left[p_a\right]})}
\\
\nonumber
&&\qquad\qquad\otimes~
\Big[ 1 - \f{w}{n+1}{1}{\Phi_{n+1}^{\left[p_a\right]}} - \f{w}{n+2}{1}{\Phi_{n+2}^{\left[p_a\right]}} - \f{\Delta}{n}{1}{t_+,t(\Phi_{n+1}^{\left[p_a\right]})} - \f{\Delta}{n+1}{1}{t(\Phi_{n+1}^{\left[p_a\right]}),t(\Phi_{n+2}^{\left[p_a\right]})} \Big]
\end{eqnarray}
This mixture can be constructed with the algorithm given in Appendix \ref{app:histories}. From here on, the discussion will assume that a particular path $p_a$ of transitions has been chosen, and the path index $p_a$ will be omitted. The sequence of reconstructed states will be $\Phi_0\rightarrow\Phi_1\rightarrow\cdots\rightarrow\Phi_n$, and the evolution scales at which transitions occur are $t_+,t_1,\dots,t_n$ (where $t_+$ is a parton-shower starting scale). 
The factors $\f{w}{i}{\infty}{\Phi_{i}}$ implement the effect of the running coupling, and are thus straight-forward to calculate, 
\begin{eqnarray}
\f{w}{i}{\infty}{\Phi_{i}}=\alpha_s(t_i) / \alpha_s(\mu)~,
\end{eqnarray}
where $\mu=M_Z^2$ is used. For second-order running, the expansion of this term reads
\begin{eqnarray}
\f{w}{i}{0}{\Phi_{i}} &=& 1~,\quad
\f{w}{i}{1}{\Phi_{i}} ~=~ \alpha_s(\mu) \frac{\beta_0}{4\pi}\ln\left(\frac{\mu}{t_i}\right)~,\quad
\f{w}{i}{2}{\Phi_{i}} ~=~ \left(\f{w}{i}{1}{\Phi_{i}}\right)^2 + \left(\frac{\alpha_s(\mu)}{4\pi}\right)^2\beta_1\ln\left(\frac{\mu}{t_i}\right) ~.
\end{eqnarray}
The second-order expansion always appears in the combination
\begin{eqnarray}
&&-\f{w}{i}{2}{\Phi_{i}} + \left(\f{w}{i}{1}{\Phi_{i}}\right)^2 =
- \left(\frac{\alpha_s(\mu)}{4\pi}\right)^2\beta_1\ln\left(\frac{\mu}{t_i}\right)~.
\end{eqnarray}
For first-order running, the $\beta_1$-dependent term can be omitted. All $w_i$-related factors are simple analytical expressions.

The calculation of parton-shower Sudakov factors using analytic results is typically tedious, due to complicated $z$-integration limits in eq.\ \ref{eq:pssudakov}. Luckily, since the sequence of states and evolution scales is known, it is possible to use \emph{trial showering} to generate Sudakov factors and their expansions. The factors $\Delta_{m} (t_+,t_-)$ may be generated by:

\vskip 1ex
\fbox{\parbox{0.9\textwidth}{\begin{itemize}
\item[1)] Initialize $\Delta_{\mathrm{trial}}=0$, and define the number of trial shower sampling points $N$.
\item[2)] Initialize the parton-shower on the state $\Phi_{m}$, and with a maximal evolution scale $t_+$. 
\item[3)] Generate a parton-shower sequence. If the first parton-shower emission occurs at an evolution scale $ t < t_-$, shift $\Delta_{\mathrm{trial}}\rightarrow \Delta_{\mathrm{trial}}+1$.
\item[4)] Discard the sequence, and repeat steps $2)$ -- $4)$ a total of $N$ times. Finally, set $\Delta_{m} (t_+,t_-)=\Delta_{\mathrm{trial}}/N$.
\end{itemize}}}

\vskip 2ex
\noindent
Trial showering allows access to the details of individual parton-shower emissions that produce $\Delta_{m} (t_+,t_-)$, and thus also allows to generate terms in the expansion of Sudakov factors. The most prevalent term required for \nnnlops\ matching is 
\begin{eqnarray}
\label{eq:delta1}
\f{\Delta}{m}{1}{t_+,t_-} = - \int\limits^{t_+}_{t_-} \frac{dt}{t} dz d\phi \frac{\alpha_s(\mu)}{2\pi} P^{(0)}(t,z,\phi)
\end{eqnarray}
where $P^{(0)}(t,z,\phi)$ is the sum of the first-order expansion of all parton-shower splitting kernels that may lead to a transition from $\Phi_m$\footnote{For gluons emissions from a quark line, the leading-order \textsc{Dire} parton shower employs the splitting kernel $P_{qq}= \left(1+\frac{\alpha_s}{2\pi}K\right)\,2\,(1-z) / \left((1-z)^2 + t/m^2_{\mathrm{dipole}}\right) - (1+z)$, where $K$ is chosen to recover the second-order corrections to soft-gluon emission. Consequently, $P^{(0)}(t,z,\phi)= 2\,(1-z) / \left((1-z)^2 + t/m^2_{\mathrm{dipole}}\right) - (1+z)$, and 
$P^{(1)}(t,z,\phi)= 2\,K\,(1-z) / \left((1-z)^2 + t/m^2_{\mathrm{dipole}}\right)$.}. Note the evaluation of $\alpha_s$ at the reference scale $\mu$.
This term can be generated by extracting the average number of emissions from the trial shower~\cite{Lonnblad:2012ix}:

\vskip 1ex
\fbox{\parbox{0.9\textwidth}{
\begin{itemize}
\item[1)] Initialize $\Delta_{\mathrm{trial}}^{(1)}=0$, and define the number of trial shower sampling points $N$.
\item[2)] Initialize the parton-shower on the state $\Phi_{m}$, and with a maximal evolution scale $t_+$. 
\item[3)] Generate a parton-shower emission. If the emission occurs at an evolution scale $t_{\mathrm{em}} > t_-$, shift $\Delta_{\mathrm{trial}}^{(1)}\rightarrow \Delta_{\mathrm{trial}}^{(1)}+\frac{1}{w}$ where $\frac{1}{w}=\frac{\alpha_s(\mu)}{\alpha_s(t_{\mathrm{em}})}$. The ratio of couplings is necessary if the trial shower is performed with dynamic $\alpha_s$ argument, and guarantees that the desired fixed-$\alpha_s(\mu)$-result is extracted. 
\item[4)] Discard the emission, reset the maximal evolution scale to $t_{\mathrm{em}}$ and repeat step $3)$ until $t_+ < t_-$.
\item[5)] Repeat steps $2)$ -- $4)$ for a total of $N$ times. Finally, set $\Delta_{m}^{(1)} (t_+,t_-)=-\Delta_{\mathrm{trial}}^{(1)}\,/\,N$.
\end{itemize}
}}

\vskip 2ex
\noindent
This method is employed heavily in \nlo{} merging methods, which also feature first-order shower expansions. The \nnnlops\ matching method further requires the generation of the second-order expansion
\begin{eqnarray}
\label{eq:delta2}
\f{\Delta}{m}{2}{t_+,t_-}
&=&
- \int\limits^{t_+}_{t_-} \frac{dt}{t} dz d\phi \left(\frac{\alpha_s(\mu)}{2\pi}\right)^2 P^{(1)}(t,z,\phi)
+
 \frac{1}{2}\left[ -\int\limits^{t_+}_{t_-} \frac{dt}{t} dz d\phi \frac{\alpha_s(\mu)}{2\pi} P^{(0)}(t,z,\phi) \right]^2\nonumber\\
&&-~
\int\limits^{t_+}_{t_-} \frac{dt}{t} 
dz d\phi \left(\frac{\alpha_s(\mu)}{2\pi}\right)^2 \frac{\beta_0}{2}\ln\left(\frac{\mu}{t}\right) P^{(0)}(t,z,\phi)~.
\end{eqnarray}
This term always appears in combination with two other subtractions, 
\begin{eqnarray}
&&-\f{\Delta}{n}{2}{t_+,t_{n+1}} + \left[\f{\Delta}{n}{1}{t_+,t_{n+1}}\right]^2  + \f{w}{n+1}{1}{\Phi_{n+1}} \f{\Delta}{n}{1}{t_+,t_{n+1}}\nonumber\\
&&=
 \int\limits^{t_+}_{t_{n+1}} \frac{dt}{t} dz d\phi \left(\frac{\alpha_s(\mu)}{2\pi}\right)^2 P^{(1)}(t,z,\phi)
+
~ \frac{1}{2}\left[~\int\limits^{t_+}_{t_{n+1}}\! \frac{dt}{t} dz d\phi \frac{\alpha_s(\mu)}{2\pi} P(t,z,\phi) \right]^2\nonumber\\
&&\qquad~+
\int\limits^{t_+}_{t_{n+1}} \!\frac{dt}{t} 
dz d\phi \left(\frac{\alpha_s(\mu)}{2\pi}\right)^2 \frac{\beta_0}{2}\ln\left(\frac{\mu}{t}\right) P(t,z,\phi)
-
\int\limits^{t_+}_{t_{n+1}} \!\frac{dt}{t} 
dz d\phi \left(\frac{\alpha_s(\mu)}{2\pi}\right)^2 \frac{\beta_0}{2}\ln\left(\frac{\mu}{t_{n+1}}\right) P(t,z,\phi)\nonumber\\
&&=
\int\limits^{t_+}_{t_{n+1}} \frac{dt}{t} dz d\phi \left(\frac{\alpha_s(\mu)}{2\pi}\right)^2 P^{(1)}(t,z,\phi)
\\
&&~
+
\tcboxmath[colback=red!10]{~\int\limits^{t_+}_{t_{n+1}}\! \frac{dt}{t} dz d\phi \frac{\alpha_s(\mu)}{2\pi} P(t,z,\phi)
\!\!
\int\limits^{t}_{t_{n+1}}\! \frac{d\bar t}{\bar t} d\bar z d\bar\phi \frac{\alpha_s(\mu)}{2\pi} P(\bar t,\bar z,\bar\phi) ~}
+ 
\tcboxmath[colback=green!10]{~
\int\limits^{t_+}_{t_{n+1}} \!\frac{dt}{t} 
dz d\phi \left(\frac{\alpha_s(\mu)}{2\pi}\right)^2 \frac{\beta_0}{2} \ln\left(\frac{t_{n+1}}{t}\right) P(t,z,\phi)~}\nonumber
\end{eqnarray}
The first term is most easily generated jointly with \ref{eq:delta1}, since this avoids separating the splitting kernel into two components.
The calculation of the term highlighted by \raisebox{1pt}{\tcboxmath[colback=green!10,boxsep=1.5pt]{\cdot}} is a straight-forward extension of generating the average number of emissions -- only the replacement $\frac{1}{w}\rightarrow \frac{\alpha_s(\mu)}{\alpha_s(t_{\mathrm{em}})}\frac{\alpha_s(\mu)}{2\pi} \frac{\beta_0}{2}\ln\left( t_{n+1} / t_{\mathrm{em}}\right)$ is necessary to produce the contribution. The term in \raisebox{1pt}{\tcboxmath[colback=red!10,boxsep=1.5pt]{\cdot}} is the rate of an emission with $\bar t \in [t, t_{n+1}]$ given a previous emission at evolution scale $t$ with $t \in [{t_+}, {t_{n+1}}]$. This may also be extracted from generating trial emissions:

\vskip 1ex
\fbox{\parbox{0.9\textwidth}{
\begin{itemize}
\item[1)] Define the number sampling points $N$, initialize $\mathrm{DE}_{\mathrm{trial}}=0$, and initialize an empty set $\mathrm{e}_{\mathrm{trial}}$ 
\item[2)] Initialize the parton-shower on the state $\Phi_{m}$, and with a maximal evolution scale $t_+$. 
\item[3)] Generate a parton-shower emission. If the emission occurs at an evolution scale $t_{\mathrm{em}} > t_{n+1}$, insert the weight $\frac{1}{w}=\frac{\alpha_s(\mu)}{\alpha_s(t_{\mathrm{em}})}$ at the end of the set, $\mathrm{e}_{\mathrm{trial}}\rightarrow \mathrm{e}_{\mathrm{trial}} \cap \frac{1}{w}$.
\item[4)] Discard the emission, reset the maximal evolution scale to $t_{\mathrm{em}}$ and repeat step $3)$ until $t_+ < t_{n+1}$.\\
\vskip -7ex
\begin{align*}
\hskip -4.5ex
\mathrm{5)~Shift}\qquad 
\mathrm{DE}_{\mathrm{trial}}=\sum\displaylimits_{i=1, i \in \mathrm{e}_{\mathrm{trial}}}^{ n(\mathrm{e}_{\mathrm{trial}})} \mathrm{e}_{\mathrm{trial}}^{(i)} \otimes \sum_{j=i+1}^{n(\mathrm{e}_{\mathrm{trial}})} \mathrm{e}_{\mathrm{trial}}^{(j)}\qquad \qquad \qquad \qquad \qquad \qquad \qquad \qquad \qquad \qquad \qquad
\end{align*}
\item[6)]
\vskip -1ex
 Repeat steps $2)$ -- $5)$ for a total of $N$ times. Finally, set \raisebox{1pt}{\tcboxmath[colback=red!10,boxsep=1.5pt]{\cdot}} $= \mathrm{DE}_{\mathrm{trial}}\,/\,N$.
\end{itemize}
}}

\vskip 2ex
\noindent
For later convenience, it is useful to collect the complete form of the $\mathcal{O}(\alpha_s^2)$ subtraction that is required for all-order weighted tree-level $n+1$-parton cross section $\s{n+1}{0}{\Phi_{n+1}}$, which reads
\begin{eqnarray}
&& - \f{w}{n+1}{2}{\Phi_{n+1}} + \left[\f{w}{n+1}{1}{\Phi_{n+1}}\right]^2
  + \f{w}{n+1}{1}{\Phi_{n+1}} \f{\Delta}{n}{1}{t_+,t_{n+1}}
  -  \f{\Delta}{n}{2}{t_+,t_{n+1}}
  + \left[\f{\Delta}{n}{1}{t_+,t_{n+1}}\right]^2 
 \nonumber\\
&=&
- \left(\frac{\alpha_s(\mu)}{4\pi}\right)^2\beta_1\ln\left(\frac{\mu}{t_{n+1}}\right)
+ 
\int\limits^{t_+}_{t_{n+1}} \!\frac{dt}{t} 
dz d\phi \left(\frac{\alpha_s(\mu)}{2\pi}\right)^2 \frac{\beta_0}{2} \ln\left(\frac{t_{n+1}}{t}\right) P(t,z,\phi)\nonumber\\
&&+
\int\limits^{t_+}_{t_{n+1}}\! \frac{dt}{t} dz d\phi \frac{\alpha_s(\mu)}{2\pi} P(t,z,\phi)
\!\!
\int\limits^{t}_{t_{n+1}}\! \frac{d\bar t}{\bar t} d\bar z d\bar\phi \frac{\alpha_s(\mu)}{2\pi} P(\bar t,\bar z,\bar\phi)
+ \int\limits^{t_+}_{t_{n+1}} \frac{dt}{t} dz d\phi \left(\frac{\alpha_s(\mu)}{2\pi}\right)^2 P^{(1)}(t,z,\phi)
\label{eq:2nd-order-exp}
\end{eqnarray}

\section{Auxiliary discussions on matching}
\label{app:giggles}

Matching fixed-order results and parton showering always leads to a certain matching scheme dependence, in particular due to choices in the treatment of terms beyond the all-order parton-shower accuracy. This appendix offers supplementary arguments for the choices in the main text, and provides additional details on the accuracy of the \nnnlops{} matching formula.

\subsection{Configurations without ordered parton-shower interpretation}
\label{app:unordered}

\noindent
Although the phase-space generation of modern parton showers is based on exact phase-space factorization formulae, and thus should allow full coverage, there are various restrictions that prevent parton showers from reaching all phase space regions. Possible restrictions are due to the parton-shower starting scale $t_+$, and due to the requirement that parton-shower emissions are ordered by decreasing values of the evolution variable. The former restriction is absent for $e^+e^-\rightarrow$ jets, since the starting scale is typically chosen to be $\hat s = (p_{e^+}+p_{e^-})^2$. The ordering constraint is ameliorated if the shower includes the emission of multi-parton clusters, since the partons within one such cluster need not adhere to specific ordering constraints. This is e.g.\ the case for the two-parton emission clusters of the \nlo{} parton shower presented in~\cite{Hoche:2017iem,Dulat:2018vuy}. Configurations that cannot be reached by the parton shower will be called \emph{non-shower configurations}.

\begin{table}[ht!]
\setlength{\tabcolsep}{5pt}
\renewcommand{\arraystretch}{1.5}
\begin{tabular}{|ll c p{0.25\textwidth} p{0.35\textwidth}| }
\hline
& Ordering & & Interpretation & Treatment\\
$a_1$) &\textcolor{red}{$t_+ > t(\Phi_{n+1})$} & : & Potentially unresolved emission  & Sudakov between $t_+$ and $t(\Phi_{n+1})$\\
$b_1$) & $t(\Phi_{n+1}) > t_+$ & : & Hard jet & Fixed-order contribution for $\Phi_{n+1}$\\
\hline
\end{tabular}
\caption{\label{tab:one-parton-unordered} Possible evolution variable orderings for one-parton states. For $e^+e^-\rightarrow$ jets, only the ordering in red applies, while any ordering is possible for generic hadron-collider processes.}
\end{table}

Configurations that cannot be reached by showering require special considerations when matching to the parton shower. For \nnnlopps{} matching, configurations with one, two or three additional partons need to be considered. Possible one-parton states are listed in Table \ref{tab:one-parton-unordered}. In states with evolution variable above the shower starting scale ($t(\Phi_{n+1}) > t_+$), the additional parton cannot become unresolved, so that no Sudakov resummation is necessary. Thus, the $b_1$ contributions enter as fixed-order corrections. However, a sensible scale choice for the argument of $\alpha_s$ is mandatory. The factorization scale value also needs to be considered carefully, since it doubles as the starting scale of a subsequent shower. A detailed scale setting mechanism for generic multi-parton states that is independent of the shower accuracy and is suitable for matched calculations can be found in~\cite{Fischer:2017yja}.

Beyond these consideration -- which apply irrespective of the desired accuracy -- it is important to determine how non-shower states contribute the fixed-order prediction. If e.g.\ non-shower configurations are included in inclusive fixed-order cross sections, then a double-counting of the contributions when matching several multiplicities can be avoided by simply including non-shower configurations in the unitarization procedure. The $b_1$ contributions in Table \ref{tab:one-parton-unordered} would e.g.\ by projected onto a $\Phi_{n}$ phase-space point, and included as fixed-order subtraction. Alternatively, if non-shower states are not (or should not) be included in inclusive fixed-order predictions, then the contributions can be included without explicit unitarization. 

\begin{table}[ht!]
\setlength{\tabcolsep}{5pt}
\renewcommand{\arraystretch}{1.5}
\begin{tabular}{|llcp{0.25\textwidth}p{0.35\textwidth}|}
\hline
&Ordering & & Interpretation & Treatment\\
$a_2$) &\textcolor{red}{$t_+>t(\Phi_{n+1})>t(\Phi_{n+2})$}  & : & Two ordered potentially unresolved emissions & Sudakov between $t_+$ and $t(\Phi_{n+1})$, and $t(\Phi_{n+1})$ and $t(\Phi_{n+2})$\\
$b_2$) &$t(\Phi_{n+1})>t_+>t(\Phi_{n+2})$  & : & Hard jet and single potentially unresolved emission & Fixed-order contribution for $\Phi_{n+1}$, Sudakov between $t(\Phi_{n+1})$ and $t(\Phi_{n+2})$\\
$d_2$) &$t(\Phi_{n+2})>t_+>t(\Phi_{n+1})$  & : & Two hard jets & Fixed-order contribution for $\Phi_{n+2}$\\
$e_2$) &\textcolor{red}{$t_+>t(\Phi_{n+2})>t(\Phi_{n+1})$}  & : & Potentially unresolved double emission & Sudakov for double-emission between $t_+$ and $t_{\mathrm{eff}}(t(\Phi_{n+1}), t(\Phi_{n+2}))$\\
$f_2$) &$t(\Phi_{n+1})>t(\Phi_{n+2})>t_+$  & : & Two hard jets & Fixed-order contribution for $\Phi_{n+2}$\\
$g_2$) &$t(\Phi_{n+2})>t(\Phi_{n+1})>t_+$  & : & Two hard jets & Fixed-order contribution for $\Phi_{n+2}$\\
\hline
\end{tabular}
\caption{\label{tab:two-parton-unordered} Possible evolution variable orderings for two-parton states. For $e^+e^-\rightarrow$ jets, only the ordering sequences in red apply, while any ordering is possible for generic hadron-collider processes.}
\end{table}

States with two additional partons allow for many more evolution scale orderings, see Table \ref{tab:two-parton-unordered}. Non-shower states with one and two partons need to be considered, as well as the emission of a ``soft" two-parton cluster. The former again demand suitable choices of renormalization and factorization scales. The effective scales introduced in~\cite{Fischer:2017yja} can serve this purpose. The evolution variable of double-emission contributions in an \nlo{} parton shower offers a theoretically appealing scale definition for configurations with $t(\Phi_{n+2})>t(\Phi_{n+1})$, since it may be applied irrespective of the value of $t_+$. In Table \ref{tab:two-parton-unordered}, it was assumed that showering off 
hard $\Phi_{n+1}$ corrections would be initiated at $t(\Phi_{n+1})$, and that any scale hierarchies above $t_+$ are not large enough to warrant resummation. The latter assumption could easily be relaxed.  

The inclusion of non-shower configurations with two hard jets ($d_2$, $f_2$, $g_2$) in inclusive predictions follows similar arguments as outlined for one-parton states above. If the contributions are included, in integrated form, in inclusive cross sections, then the contributions should be included in the unitarization procedure. For the configurations $f_2$ and $g_2$, a fixed-order subtraction would have to be generated by projected the configurations onto $\Phi_{n+1}$ phase-space points, and then subtracting. Similarly, an all-order subtraction of $b_2$ should be generated with $\Phi_{n+1}$ dependence. Note that as an all-order reweighted contribution, $b_2$ should enter on equal footing to the other two-parton tree-level contributions in eq.\ \ref{eq:nnnlops}, i.e.\ the Sudakov factor $t(\Phi_{n+1})$ and $t(\Phi_{n+2})$ should be subtracted appropriately. The two-jet configurations $d_2$ include $\Phi_{n+2}$ states are fully resolved, while $t_+\gg t(\Phi_{n+1})$ may produce unresolved $(n+1)$-parton states. Guidance on their treatment can be obtained from the double-emission shower of~\cite{Hoche:2017iem,Dulat:2018vuy}, which allows $t(\Phi_{n+2})>t(\Phi_{n+1})$. There, the impact of double-emission clusters on inclusive observables cancels between $n$- and $(n+2)$-parton states, since both emissions become unresolved simultaneously. Similarly, configuration $d_2$ may be regarded as emission of a hard two-parton cluster. Consequently, the necessary fixed-order subtraction for $d_2$ should be generated with $\Phi_{n}$ dependence. Other treatments of $d_2$ are conceivable, and might become relevant when \nnlopps{} and \nnnlopps{} matching mature. If inclusive fixed-order corrections do not contain non-shower states, then non-shower two-jet contributions can be included without explicit unitarization. Configurations $a_2$ and $e_2$ are reachable by an \nlo{} parton shower and thus do not require special attention.

\begin{table}[t!]
\setlength{\tabcolsep}{5pt}
\renewcommand{\arraystretch}{1.5}
\begin{tabular}{|llcp{0.23\textwidth}p{0.33\textwidth}|}
\hline
&Ordering & & Interpretation & Treatment\\
$a_3$)&\textcolor{red}{$t_+>t(\Phi_{n+1})>t(\Phi_{n+2})>t(\Phi_{n+3})$}  & :& Three ordered potentially unresolved emissions & Sudakov between $t_+$ and $t(\Phi_{n+1})$, $t(\Phi_{n+1})$ and  $t(\Phi_{n+2})$, and $t(\Phi_{n+2})$ and $t(\Phi_{n+3})$ \\
$b_3$)&$t(\Phi_{n+1})>t_+>t(\Phi_{n+2})>t(\Phi_{n+3})$  & :& Hard jet + two ordered potentially unresolved emissions & Fixed-order contribution for $\Phi_{n+1}$, Sudakov between $t(\Phi_{n+1})$ and $t(\Phi_{n+2})$ and $t(\Phi_{n+2})$ and $t(\Phi_{n+3})$ \\
$c_3$)&$t(\Phi_{n+2})>t_+>t(\Phi_{n+1})>t(\Phi_{n+3})$  & :& Two hard jets + potentially unresolved emission & Fixed-order contribution for $\Phi_{n+2}$, Sudakov between $t_{\mathrm{eff}}(t(\Phi_{n+1}), t(\Phi_{n+2}))$ and $t(\Phi_{n+3})$ \\
$d_3$)&\textcolor{red}{$t_+>t(\Phi_{n+2})>t(\Phi_{n+1})>t(\Phi_{n+3})$}  & :& Potentially unresolved double emission, followed by potentially unresolved emission & Sudakov from $t_+$ to  $t_{\mathrm{eff}}(t(\Phi_{n+1}), t(\Phi_{n+2}))$, and $t_{\mathrm{eff}}(t(\Phi_{n+1}), t(\Phi_{n+2}))$ and $t(\Phi_{n+3})$ \\
$e_3$)&$t(\Phi_{n+1})>t(\Phi_{n+2})>t_+>t(\Phi_{n+3})$  & :& Two hard jets + potentially unresolved emission & Fixed-order contribution for $\Phi_{n+2}$, Sudakov between $t(\Phi_{n+2})$ and $t(\Phi_{n+3})$ \\
$f_3$)&$t(\Phi_{n+2})>t(\Phi_{n+1})>t_+>t(\Phi_{n+3})$  & :& Two hard jets + potentially unresolved emission & Fixed-order contribution for $\Phi_{n+2}$, Sudakov between $t_{\mathrm{eff}}(t(\Phi_{n+1}), t(\Phi_{n+2}))$ and $t(\Phi_{n+3})$ \\
$g_3$)&$t(\Phi_{n+2})>t(\Phi_{n+1})>t(\Phi_{n+3})>t_+$  & :& Three hard jets & Fixed-order contribution for $\Phi_{n+3}$ \\
$h_3$)&$t(\Phi_{n+1})>t(\Phi_{n+2})>t(\Phi_{n+3})>t_+$  & :& Three hard jets & Fixed-order contribution for $\Phi_{n+3}$ \\
$i_3$)&$t(\Phi_{n+3})>t(\Phi_{n+2})>t(\Phi_{n+1})>t_+$  & :& Three hard jets & Fixed-order contribution for $\Phi_{n+3}$ \\
$j_3$)&$t(\Phi_{n+2})>t(\Phi_{n+3})>t(\Phi_{n+1})>t_+$  & :& Three hard jets & Fixed-order contribution for $\Phi_{n+3}$ \\
$k_3$)&$t(\Phi_{n+1})>t(\Phi_{n+3})>t(\Phi_{n+2})>t_+$  & :& Three hard jets & Fixed-order contribution for $\Phi_{n+3}$ \\
$l_3$)&$t(\Phi_{n+3})>t(\Phi_{n+1})>t(\Phi_{n+2})>t_+$  & :& Three hard jets & Fixed-order contribution for $\Phi_{n+3}$\\
$m_3$)&$t(\Phi_{n+3})>t_+>t(\Phi_{n+2})>t(\Phi_{n+1})$  & :& Three hard jets & Fixed-order contribution for $\Phi_{n+3}$\\
$n_3$)&\textcolor{red}{$t_+>t(\Phi_{n+3})>t(\Phi_{n+2})>t(\Phi_{n+1})$}  & :& Potentially unresolved triple emission & Sudakov for triple-emission between $t_+$ and $t_{\mathrm{eff}}(t(\Phi_{n+1}), t(\Phi_{n+2}),t(\Phi_{n+3}))$ \\
$o_3$)&$t(\Phi_{n+2})>t(\Phi_{n+3})>t_+>t(\Phi_{n+1})$  & :& Three hard jets & Fixed-order contribution for $\Phi_{n+3}$\\
$p_3$)&$t(\Phi_{n+3})>t(\Phi_{n+2})>t_+>t(\Phi_{n+1})$  & :& Three hard jets & Fixed-order contribution for $\Phi_{n+3}$\\
$q_3$)&\textcolor{red}{$t_+>t(\Phi_{n+2})>t(\Phi_{n+3})>t(\Phi_{n+1})$}  & :& Potentially unresolved triple emission & Sudakov for triple-emission between $t_+$ and $t_{\mathrm{eff}}(t(\Phi_{n+1}), t(\Phi_{n+2}),t(\Phi_{n+3}))$\\
$r_3$)&$t(\Phi_{n+2})>t_+>t(\Phi_{n+3})>t(\Phi_{n+1})$  & :& Three hard jets & Fixed-order contribution for $\Phi_{n+3}$\\
$s_3$)&$t(\Phi_{n+1})>t_+>t(\Phi_{n+3})>t(\Phi_{n+2})$  & :& Hard jet + potentially unresolved double emission & Fixed-order contribution for $\Phi_{n+1}$, Sudakov for double-emission between $t_{n+1}$ and $t_{\mathrm{eff}}(t(\Phi_{n+2}), t(\Phi_{n+3}))$ \\
$t_3$)&\textcolor{red}{$t_+>t(\Phi_{n+1})>t(\Phi_{n+3})>t(\Phi_{n+2})$}  & :& Potentially unresolved emission, followed by potentially unresolved double emission & Sudakov between $t_+$ and $t_{n+1}$, and Sudakov for double-emission between $t(\Phi_{n+1})$ and $t_{\mathrm{eff}}(t(\Phi_{n+2}), t(\Phi_{n+3}))$ \\
$u_3$)&$t(\Phi_{n+3})>t(\Phi_{n+1})>t_+>t(\Phi_{n+2})$  & :& Three hard jets & Fixed-order contribution for $\Phi_{n+3}$\\
$v_3$)&$t(\Phi_{n+1})>t(\Phi_{n+3})>t_+>t(\Phi_{n+2})$  & :& Three hard jets & Fixed-order contribution for $\Phi_{n+3}$ \\
$w_3$)&\textcolor{red}{$t_+>t(\Phi_{n+3})>t(\Phi_{n+1})>t(\Phi_{n+2})$}  & :& Potentially unresolved triple emission & Sudakov for triple-emission between $t_+$ and $t_{\mathrm{eff}}(t(\Phi_{n+1}), t(\Phi_{n+2}),t(\Phi_{n+3}))$ \\
$x_3$)&$t(\Phi_{n+3})>t_+>t(\Phi_{n+1})>t(\Phi_{n+2})$  & :& Three hard jets & Fixed-order contribution for $\Phi_{n+3}$\\
\hline
\end{tabular}
\caption{\label{tab:three-parton-unordered} Possible evolution variable orderings for three-parton states. For $e^+e^-\rightarrow$ jets, only the ordering sequences in red apply, while any ordering is possible for generic hadron-collider processes.}
\end{table}

States with three additional partons allow for yet more evolution scale orderings (Table \ref{tab:three-parton-unordered}). Note that
table \ref{tab:three-parton-unordered} assumes that showers off hard $\Phi_{n+1}$ corrections would commence at $t(\Phi_{n+1})$, and showering from hard two-jet configurations was started at $t(\Phi_{n+2})$ or $t_{\mathrm{eff}}(t(\Phi_{n+1}), t(\Phi_{n+2}))$. Furthermore, any scale hierarchies above $t_+$ are not considered large enough to warrant resummation. 

Multiple ordering combinations reachable by showering ($a_3$, $d_3$, $n_3$, $q_3$, $t_3$, $w_3$) are possible. Several of these would only be accessible in an \nnlo{} parton shower that includes the emission of correlated triple-parton clusters. Although this is currently beyond reach, the extension of~\cite{Hoche:2017iem,Dulat:2018vuy} suggests that the evolution variable of correlated triple-emission may be $\propto p_a(p_1+p_2+p_3) p_b(p_1+p_2+p_3) / (p_ap_b)$, where $p_a$ and $p_b$ are the momenta of (high-energy) radiator and recoiler after the branching, and $p_{1,2,3}$ are the emission momenta. Such a definition would constitute an appealing renormalization, factorization, and shower starting scale for non-shower configurations with three hard jets. Alternatively, the method of~\cite{Fischer:2017yja} might be used irrespectively of the shower accuracy.

The treatment of non-shower states with one or two partons was discussed above. Similar configurations appear also at three-parton level ($b_3$, $c_3$, $e_3$, $f_3$, $s_3$), and will be weighted by all-order Sudakov factors. The latter will depend on the (effective) scales assigned to the fixed-order configurations. If inclusive fixed-order cross sections include these non-shower configurations in integrated form, then unitarization through all-order subtraction is necessary. This proceeds by subtracting the contributions after projection onto $(n+2)$-parton states, with the notable exception of $s_3$. That contribution contains potentially unresolved $(n+2)$- and $(n+3)$-parton configurations with $t(\Phi_{n+3})>t(\Phi_{n+2})$. Again taking guidance from \nlo{} parton showering, the impact of these configurations on inclusive observables should cancel between $(n+3)$- and $(n+1)$-parton configurations, so that the necessary all-order subtraction should be generated with $\Phi_{n+1}$-dependence.

Finally, configurations containing three hard jets can be treated as fixed-order corrections. If such configurations were included (again in integrated form) in inclusive cross sections, then fixed-order unitarity subtractions are required to retain the correct inclusive results. Depending on the ordering, these subtractions enter with $\Phi_{n+2}$-dependence ($g_3$,$h_3$,$j_3$,$o_3$,$r_3$), $\Phi_{n+1}$-dependence ($k_3$,$v_3$), or $\Phi_{n}$-dependence ($i_3$, $l_3$,$m_3$,$p_3$,$u_3$,$x_3$). The last treatment is inspired by a hypothetical \nnlo{} shower that would include the emission of internally disordered triple-parton clusters. 

\subsection{Differences in matching at NNLO}

%\subsubsection{Comments on eq.\ \ref{eq:un2lops}}
\begin{center}
{\textit{Comments on eq.\ \ref{eq:un2lops}}}
\end{center}

This note used the \nnlopps{} matching formula eq.\ \ref{eq:un2lops} as a starting point to derive \nnnlopps{}  matching. This equation differs slightly from the \unnlops\ prescriptions in the literature. The first change is the more abundant inclusion of running-coupling factors, especially the running-coupling rescaling of virtual and real corrections \emph{if} additional partons were already present at tree-level. The reason for this choice is encapsulated in eqs.\ \ref{eq:ps2} and \ref{eq:unlops}. These define the relation between the emission pattern and the no-emission (Sudakov) factor: the latter is determined by the difference of the integral of the former and unity. Conversely, the integral of the emission pattern is determined by the difference of the Sudakov factor from unity. Theoretical arguments strongly favor a dynamic coupling evaluation in the Sudakov exponent~\cite{Amati:1978by}. The same is true if the Sudakov factor is produced from an emission spectrum via explicitly unitarization. Hence, the integrated emission spectrum should employ a dynamic coupling evaluation. This also applies to {\smaller(N)NLO} fixed-order results. Thus, it is arguably more consistent with eq.\ \ref{eq:ps2} to include running-coupling factors in the rescaling of virtual and real corrections -- though the inclusion of running-coupling effects is beyond the formal accuracy of the method.

The second change relative to~\cite{Hoeche:2014aia} is that eq.\ \ref{eq:un2lops} demands and exclusive $n+1$-parton \nlo{} cross section, which is complemented by tree-level $n+2$-parton distributions, whereas~\cite{Hoeche:2014aia} employed an MC@NLO-matched $n+1$-parton calculation. The main reason for this differences is technical: The fixed-order cross sections in eq.\ \ref{eq:un2lops} are not dependent on the parton shower, so that it is straight-forward to produce toy fixed-order calculations for the inputs to eq.\ \ref{eq:un2lops}. The sliced approach of eq.\ \ref{eq:un2lops} admits a more dynamical scale-setting method for $n+2$-parton contributions than would be possible in an MC@NLO calculation, potentially resulting in a more ``physical" result especially in the presence of large hardness hierarchies in $n+2$-parton states. On the other hand, using an MC@NLO-matched calculation avoids the slicing of $n+1$-parton configurations into two disjoint ``bins".

\begin{center}
{\textit{Comments on fixed-multiplicity ``bins" in the matching formula}}
\end{center}

The matching formulae used in this note (eqs.\ \ref{eq:unlops}, \ref{eq:un2lops}, \ref{eq:nnnlops}) are combinations of weighted $n$-parton, $n+1$-parton and possibly $n+2$-parton and $n+3$-parton states. This is indicated by the ``observable dependence" $\mathcal{O}_{m}$
($m=n\dots n+3$). The highest parton multiplicity sample will be distributed over even higher multiplicities by the action of the parton shower. In all other cases, there is no ``smearing" of the description of $m$-parton states to higher multiplicities $m+1$. This ``binned" approach is common for matching~\cite{Lonnblad:2012ix,Alioli:2012fc,Hoeche:2014aia}. However, it has the disadvantage that virtual corrections for $m$-parton states naively do not migrate to real-emission observables, at odds with the arguments of~\cite{Parisi:1979se}. The implementation in this note accepts this feature. A method to ameliorate the ``binned" behavior was discussed in~\cite{Hoche:2014dla}.

\subsection{Accuracy of the T{\smaller{OMTE}} matching formula}

\noindent
The formula eq.\ \ref{eq:nnnlops} allows the combination of \nnnlo{} calculations with parton showering, and hence with event generation. It fulfills all the criteria set out in Table \ref{tab:criteria}. This appendix provides a more fine-grained discussion of the choices leading to the \nnnlops{} method, and of its accuracy. Any confirmation of the desired accuracy 
relies on accurately reproducing fixed-order results for inclusive cross sections, and producing a combination of fixed-order and resummed results for exclusive cross sections.

%\subsubsection{More details on matching the tree-level two-jet contribution}
\begin{center}
{\textit{More details on matching the tree-level two-jet contribution}}
\end{center}

The matching of this contribution relies on its interplay with several other contributions, and thus requires careful consideration of any terms beyond the formal accuracy that are introduced by parton-shower factors. Firstly, the term forms the Born contribution to the two-parton \nlo{} cross section. This suggests that it should contain all-order prefactors matching the prefactors of the real- and virtual corrections, i.e.\ dynamic arguments of $\alpha_s$ for both partons, and Sudakov factors to ensure a suitable description in the presence of scale hierarchies between the partons. On the other hand, the term should complement the the exclusive one-parton \nlo{} cross section {\smaller$\ss{n+1}{1}{Q_{n+2}<Q_c}{\Phi_{n+1}}$} after integration, and should thus contain all-order factors identical to the latter (i.e.\ dynamic arguments for one power of $\alpha_s$, and only one Sudakov factor). Finally, the unitarization of $\obs{n+2}$ terms should result in an appropriate Sudakov factor to make the $\obs{n+1}$ contribution exclusive. Thus, any all-order weighting of the tree-level two-jet contribution is seriously constrained, leading to eqs.\ \ref{eq:two-parton-O2} and \ref{eq:two-parton-O1}. Some more discussion can also be found in the following points.

%\subsubsection{Three-parton observables are LO+PS accurate}

\begin{center}
{\textit{Three-additional-parton observables are LO+PS accurate}}
\end{center}

Only the last term in eq.\ \ref{eq:nnnlops},
\begin{eqnarray}
\ss{n+3}{0}{Q_{n+3}>Q_c}{\Phi_{n+3}} &\otimes&
\Delta_{n} (t_+,t_{n+1}) \Delta_{n+1} (t_{n+1},t_{n+2}) \Delta_{n+2} (t_{n+2},t_{n+3})\\
&\otimes& \w{n+1} \w{n+2}\w{n+3} ~\otimes~\f{\mathcal{F}}{n+3}{\infty}{\Phi_{n+3}, t_{n+3},t_-}~,\nonumber
\end{eqnarray}
contributes to $\obs{n+3}$ observables. The unitarity of the parton shower guarantees that for inclusive observables, this reduces to
\begin{eqnarray}
&&\ss{n+3}{0}{Q_{n+3}>Q_c}{\Phi_{n+3}} \otimes~
\Delta_{n} (t_+,t_{n+1}) \Delta_{n+1} (t_{n+1},t_{n+2}) \Delta_{n+2} (t_{n+2},t_{n+3})\nonumber\\
&&\qquad\qquad\qquad\qquad~~~~~~\, \otimes~ \w{n+1} \w{n+2}\w{n+3}\nonumber\\
&&\rightarrow
\ss{n+3}{0}{Q_{n+3}>Q_c}{\Phi_{n+3}} ~\times~\left(1+\mathcal{O}(\alpha_s)\right)~.
\end{eqnarray}
For exclusive $n+3$-parton observables (i.e.\ when vetoing $n+4$-parton states with separation larger than $Q_c$), the action of the parton shower eq.\ \ref{eq:ps1} attaches an additional Sudakov factor $\Delta_{n+3} (t_{n+3},t_c))$,
\begin{eqnarray}
&&\ss{n+3}{0}{Q_{n+3}>Q_c}{\Phi_{n+3}} \otimes~
\Delta_{n} (t_+,t_{n+1}) \Delta_{n+1} (t_{n+1},t_{n+2}) \Delta_{n+2} (t_{n+2},t_{n+3})\nonumber\\
&&\qquad\qquad\qquad\qquad~~~~~~\, \otimes~ \w{n+1} \w{n+2}\w{n+3} ~\otimes~\f{\mathcal{F}}{n+3}{\infty}{\Phi_{n+3}, t_{n+3},t_-}\nonumber\\
&&\rightarrow
\ss{n+3}{0}{Q_{n+3}>Q_c}{\Phi_{n+3}} \\
&&\qquad\qquad\otimes~
\Delta_{n} (t_+,t_{n+1}) \Delta_{n+1} (t_{n+1},t_{n+2}) \Delta_{n+2} (t_{n+2},t_{n+3}) \Delta_{n+3} (t_{n+3},t_c))\nonumber\\
&&\qquad\qquad\otimes~
\w{n+1} \w{n+2}\w{n+3}
~,
\end{eqnarray}
and thus reproduces the correct parton-shower resummation of the jet veto. Hence, three-parton states are described at LO+PS accuracy.

%\subsubsection{Two-parton observables are NLO+PS accurate} 

\begin{center}
{\textit{Two-additional-parton observables are NLO+PS accurate}}
\end{center}

The description of two-parton states in \nnnlops{} is very similar to the \unnlops{} scheme. 
For inclusive $n+2$-parton observables, the terms in \raisebox{1pt}{\tcboxmath[colback=blue!10,boxsep=1.5pt]{\cdot}} in eq.\ \ref{eq:nnnlops} cancel by construction, resulting in
a $\obs{n+2}$ contribution
\begin{eqnarray}
&&
~\ss{n+2}{0}{Q_{n+2}>Q_c}{\Phi_{n+2}} 
\nonumber\\
&&\qquad\otimes~
\Delta_{n} (t_+,t_{n+1})\Delta_{n+1} (t_{n+1},t_{n+2}) \f{w}{n+1}{\infty}{\Phi_{n+1}}  \f{w}{n+2}{\infty}{\Phi_{n+2}}
\nonumber\\
&&\qquad\otimes~
\Big[ 1 - \f{w}{n+1}{1}{\Phi_{n+1}} - \f{w}{n+2}{1}{\Phi_{n+2}} - \f{\Delta}{n}{1}{t_+,t_{n+1}} - \f{\Delta}{n+1}{1}{t_{n+1},t_{n+2}} \Big]
\nonumber\\
&&+~ \ss{n+2}{1}{Q_{n+2}>Q_c \land Q_{n+3}<Q_c}{\Phi_{n+2}} \nonumber\\
&&\qquad\otimes~\Delta_{n} (t_+,t_{n+1})\Delta_{n+1} (t_{n+1},t_{n+2}) \f{w}{n+1}{\infty}{\Phi_{n+1}}\f{w}{n+2}{\infty}{\Phi_{n+2}} \nonumber\\
&&+~ \intOne \ss{n+3}{0}{Q_{n+3}>Q_c}{\Phi_{n+3}}\nonumber\\
&&\qquad\otimes~
\Delta_{n} (t_+,t_{n+1}) \Delta_{n+1} (t_{n+1},t_{n+2}) \f{w}{n+1}{\infty}{\Phi_{n+1}}\f{w}{n+2}{\infty}{\Phi_{n+2}}\one{n+3}{n+2}~.
\label{eq:O2inc}
\end{eqnarray}
After expanding all shower factors, this gives
\begin{eqnarray}
&&~\ss{n+2}{0}{Q_{n+2}>Q_c}{\Phi_{n+2}} \cdot\left(1+\mathcal{O}(\alpha_s^2)\right)
~+~
   \ss{n+2}{1}{Q_{n+2}>Q_c \land Q_{n+3}<Q_c}{\Phi_{n+2}}  \cdot\left(1+\mathcal{O}(\alpha_s)\right)~
\nonumber\\
&&+~ \int \ss{n+3}{0}{Q_{n+3}>Q_c}{\Phi_{n+3}}  \cdot\left(1+\mathcal{O}(\alpha_s)\right)~\nonumber\\
&&= \ss{n+2}{0+1}{Q_{n+2}>Q_c}{\Phi_{n+2}}~,
\end{eqnarray}
such that the \nlo{} cross section is recovered. Note that the factor $\one{n+3}{n+2}$ ensures that the $(n+3)$-parton contribution combines with the $(n+2)$-parton contributions
in the correct manner, to produce an unbiased $\upint$-integration. In order to assess how eq.\ \ref{eq:O2inc} influences the all-order description (for one parton becoming unresolved, or both partons becoming unresolved in a strongly ordered manner), it is useful to reorder the terms into a ``parton-shower prediction" and a ``remnant term"~\cite{Hoeche:2014aia},
\begin{eqnarray}
\label{eq:two-parton-rem1}
&&\textnormal{Eq.\ \ref{eq:O2inc}} =\\
&&
\Delta_{n} (t_+,t_{n+1})\Delta_{n+1} (t_{n+1},t_{n+2}) \f{w}{n+1}{\infty}{\Phi_{n+1}} \f{w}{n+2}{\infty}{\Phi_{n+2}}
\left[ \ss{n+2}{0}{Q_{n+2}>Q_c}{\Phi_{n+2}} ~+~ \ss{n+2}{1}{\mathrm{REM}}{\Phi_{n+2}} \right]\nonumber
\end{eqnarray}
where
\begin{eqnarray}
\label{eq:two-parton-rem2}
&&
\ss{n+2}{1}{\mathrm{REM}}{\Phi_{n+2}} =
\ss{n+2}{1}{Q_{n+2}>Q_c}{\Phi_{n+2}} \\
&&\quad +~ \ss{n+2}{0}{Q_{n+2}>Q_c}{\Phi_{n+2}}
\textnormal{\smaller$\left( - \f{w}{n+1}{1}{\Phi_{n+1}} - \f{w}{n+2}{1}{\Phi_{n+2}} - \f{\Delta}{n}{1}{t_+,t_{n+1}} - \f{\Delta}{n+1}{1}{t_{n+1},t_{n+2}} \right)$} \nonumber
\end{eqnarray}
The first term in eq.\ \ref{eq:two-parton-rem1} is the desired parton-shower result. The remnant term in \ref{eq:two-parton-rem1} consists of an all-order factor multiplying an $\mathcal{O}(\alpha_s^3)$ correction (eq.\ \ref{eq:two-parton-rem2}). This correction contains all non-universal \nlo{} corrections, since all universal parts (as defined by the shower approximation) are removed from the complete \nlo{} corrections. Thus, the ``remnant" does not impair the shower accuracy. Instead, it can be considered favorable, since it allows to view the cross section \ref{eq:two-parton-rem1} as separated into a ``hard" coefficient (in brackets), dressed with identical all-order factors encapsulating the evolution of the state.

For exclusive $(n+2)$-parton observables, the terms in \raisebox{1pt}{\tcboxmath[colback=blue!10,boxsep=1.5pt]{\cdot}} in eq.\ \ref{eq:nnnlops} do no longer cancel since
$(n+3)$-parton states with separation larger than $Q_c$ are vetoed. The $(n+3)$-parton contribution can be written as
\begin{eqnarray}
&&\intOne \ss{n+3}{0}{Q_{n+3}>Q_c}{\Phi_{n+3}} \Delta_{n} (t_+,t_{n+1}) \Delta_{n+1} (t_{n+1},t_{n+2}) \f{w}{n+1}{\infty}{\Phi_{n+1}}\f{w}{n+2}{\infty}{\Phi_{n+2}}\nonumber\\
&&\qquad \Big[ ~
 \left(\one{n+3}{n+2}- 1\right) ~+~ \left(1 - \Delta_{n+2} (t_{n+2},t_{n+3})\w{n+1} \w{n+2}\right)~\Big]\\
&\approx&
\intOne \ss{n+3}{0}{Q_{n+3}>Q_c}{\Phi_{n+3}} \Delta_{n} (t_+,t_{n+1}) \Delta_{n+1} (t_{n+1},t_{n+2}) \f{w}{n+1}{\infty}{\Phi_{n+1}}\f{w}{n+2}{\infty}{\Phi_{n+2}}\nonumber\\
&&\qquad \Big[ ~ 1 - \tcboxmath[colback=blue!10]{~\Delta_{n+2} (t_{n+2},t_{n+3})\w{n+1} \w{n+2}~}~\Big]~.
\label{eq:two-parton-exc2}
\end{eqnarray}
The expansion \raisebox{1pt}{\tcboxmath[colback=blue!10,boxsep=1.5pt]{\cdot}} term removes hard $(n+3)$-parton events at lowest order. 
At all orders, it combines (by virtue of eq.\ \ref{eq:ps1}) with the other contributions to form a Sudakov factor reproducing the parton-shower resummation of the jet veto. 
As argued in Appendix \ref{app:integrals} the $\left(\one{n+3}{n+2}- 1\right)$ term does not impair the accuracy  according to the \emph{balanced} parton shower accuracy criterion, and has thus been dropped in eq.\ \ref{eq:two-parton-exc2}. In conclusion, two-parton states are described at \nlopps{} accuracy.

%\subsubsection{One-parton observables are NNLO+PS accurate}

\begin{center}
{\textit{One-additional-parton observables are NNLO+PS accurate}}
\end{center}

The description of $n+1$-parton observables $\obs{n+1}$ contains some of the most interesting features of the \nnnlops{} method. In inclusive $n+1$-parton observables, the construction \ref{eq:nnnlops} enforces a cancellation of the terms in \raisebox{1pt}{\tcboxmath[colback=green!10,boxsep=1.5pt]{\cdot}}, \raisebox{1pt}{\tcboxmath[colback=red!10,boxsep=1.5pt]{\cdot}}, \raisebox{1pt}{\tcboxmath[colback=orange!10,boxsep=1.5pt]{\cdot}}, as well as in \raisebox{1pt}{\tcboxmath[colback=blue!10,boxsep=1.5pt]{\cdot}}. After these simplifications, the $\obs{n+1}$ prediction is
\begin{eqnarray}
&&
\phantom{+~}
\Bigg(
\ss{n+1}{2}{Q_{n+2}<Q_c \land Q_{n+3}<Q_c}{\Phi_{n+1}} ~+~ \ss{n+1}{1}{Q_{n+2}<Q_c}{\Phi_{n+1}} ~ \Big( 1 - \f{w}{n+1}{1}{\Phi_{n+1}} - \f{\Delta}{n}{1}{t_+,t_{n+1}}\Big)    \nonumber\\
&&\quad
+~\s{n+1}{0}{\Phi_{n+1}}
\Bigg[                  
1 - \f{w}{n+1}{1}{\Phi_{n+1}} - \f{w}{n+1}{2}{\Phi_{n+1}}  - \f{\Delta}{n}{1}{t_+,t_{n+1}} -  \f{\Delta}{n}{2}{t_+,t_{n+1}}
\nonumber\\
&&\qquad\qquad + \left[\f{\Delta}{n}{1}{t_+,t_{n+1}}\right]^2 + \left[\f{w}{n+1}{1}{\Phi_{n+1}}\right]^2 + \f{w}{n+1}{1}{\Phi_{n+1}} \f{\Delta}{n}{1}{t_+,t_{n+1}} \Bigg]~
\Bigg)~\Delta_{n} (t_+,t_{n+1}) \f{w}{n+1}{\infty}{\Phi_{n+1}}
\nonumber\\
&&+~
\Bigg(
\intOne\ss{n+2}{1}{Q_{n+2}>Q_c \land Q_{n+3}<Q_c }{\Phi_{n+2}}\one{n+2}{n+1} 
~+~\intTwo\ss{n+3}{0}{Q_{n+3}>Q_c}{\Phi_{n+3}} \one{n+3}{n+1}  \nonumber\\
&&\quad
+~\intOne\ss{n+2}{0}{Q_{n+2}>Q_c}{\Phi_{n+2}} ~\Big[  1 - \f{w}{n+1}{1}{\Phi_{n+1}} - \f{\Delta}{n}{1}{t_+,t_{n+1}}\Big] \one{n+2}{n+1} 
~ \Bigg)~\Delta_{n} (t_+,t_{n+1}) \f{w}{n+1}{\infty}{\Phi_{n+1}}
\label{eq:O1inc}\\
&&\approx
\phantom{\Bigg(}
\ss{n+1}{2}{Q_{n+2}<Q_c \land Q_{n+3}<Q_c}{\Phi_{n+1}}\cdot\left(1+\mathcal{O}(\alpha_s)\right) ~+~ \ss{n+1}{1}{Q_{n+2}<Q_c}{\Phi_{n+1}} \cdot\left(1+\mathcal{O}(\alpha_s^2)\right) \nonumber\\
\phantom{\Bigg(}
&&\quad
+~\s{n+1}{0}{\Phi_{n+1}} \cdot\left(1+\mathcal{O}(\alpha_s^3)\right)
~+~
\int \ss{n+2}{1}{\mathrm{INC}}{\Phi_{n+2}} \cdot\left(1+\mathcal{O}(\alpha_s)\right)
~+~\int \ss{n+2}{0}{Q_{n+2}>Q_c}{\Phi_{n+2}} \cdot\left(1+\mathcal{O}(\alpha_s^2)\right) \nonumber\\
\phantom{\Bigg(}
&&=\ss{n+1}{0+1+2}{\mathrm{INC}}{\Phi_{n+1}}\cdot\left(1+\mathcal{O}(\alpha_s)\right)~.
\end{eqnarray}
where the $\one{n+2}{n+1}$ and $\one{n+3}{n+1}$ correction factors enforce an unbiased $\upint$-integrations.
The inclusive \nnlo{} cross section is thus reproduced correctly. The influence
of the matching on the all-order prediction -- if the parton becomes unresolved -- is best
discussed by rewriting eq.\ \ref{eq:O1inc} in terms of a parton-shower prediction and remnant term,
\begin{eqnarray}
\label{eq:one-parton-rem1}
&&\textnormal{Eq.\ \ref{eq:O1inc}} =
\Bigg(
\s{n+1}{0}{\Phi_{n+1}}
+ 
\ss{n+1}{1}{\mathrm{REM}}{\Phi_{n+1}}
+ 
\ss{n+1}{2}{\mathrm{REM}}{\Phi_{n+1}}
\Bigg)
\Delta_{n} (t_+,t_{n+1}) \f{w}{n+1}{\infty}{\Phi_{n+1}}~,\nonumber
\end{eqnarray}
where
\begin{eqnarray}
\label{eq:one-parton-rem2-1}
\ss{n+1}{1}{\mathrm{REM}}{\Phi_{n+1}}
&=&
\ss{n+1}{1}{Q_{n+2}<Q_c}{\Phi_{n+1}}
+ \upint \ss{n+2}{0}{Q_{n+2}>Q_c}{\Phi_{n+2}}\nonumber\\
&&+~ \s{n+1}{0}{\Phi_{n+1}} \left( - \f{w}{n+1}{1}{\Phi_{n+1}} - \f{\Delta}{n}{1}{t_+,t_{n+1}} \right)~,
\end{eqnarray}
and with the second-order remnant term defined in eq.\ \ref{eq:one-parton-rem2-2}. This again permits interpreting the matched result as a hard production coefficient (in brackets in eq.\ \ref{eq:one-parton-rem1}), dressed with the effect of soft- and collinear radiation. In the remnant term \ref{eq:one-parton-rem2-1}, all universal corrections are again subtracted, such that only non-universal corrections remain. An identical term appears in the \unnlops\ prescription. Thus, this term does not threaten the parton-shower accuracy of the matched result.

The second-order remnant is given by
\begin{eqnarray}
\label{eq:one-parton-rem2-2}
&&\ss{n+1}{2}{\mathrm{REM}}{\Phi_{n+1}}
=
\ss{n+1}{2}{Q_{n+2}<Q_c \land Q_{n+3}<Q_c}{\Phi_{n+1}} + \upint\ss{n+2}{1}{Q_{n+2}>Q_c}{\Phi_{n+2}} + \upiint\ss{n+3}{0}{Q_{n+3}>Q_c}{\Phi_{n+3}}\nonumber\\
&&\quad+~ \upint\ss{n+2}{0}{Q_{n+2}>Q_c}{\Phi_{n+2}} \left( - \f{w}{n+1}{1}{\Phi_{n+1}} - \f{\Delta}{n}{1}{t_+,t_{n+1}}\right)\nonumber\\
&&\quad+~ \ss{n+1}{1}{Q_{n+2}<Q_c}{\Phi_{n+1}} \left( - \f{w}{n+1}{1}{\Phi_{n+1}} - \f{\Delta}{n}{1}{t_+,t_{n+1}}\right) \nonumber\\
&&\quad+~ \s{n+1}{0}{\Phi_{n+1}} \Big( - \f{w}{n+1}{2}{\Phi_{n+1}}  -  \f{\Delta}{n}{2}{t_+,t_{n+1}}
  + \left[\f{\Delta}{n}{1}{t_+,t_{n+1}}\right]^2 + \left[\f{w}{n+1}{1}{\Phi_{n+1}}\right]^2 \nonumber\\
&&\quad\qquad\qquad\qquad\quad+~  \f{w}{n+1}{1}{\Phi_{n+1}} \f{\Delta}{n}{1}{t_+,t_{n+1}} \Big)~.
\end{eqnarray}
These contributions are all interconnected, such that the impact of the second-order remnant on the all-order result is more subtle than for the first-order remnant. However, using eqs.\ \ref{eq:delta1} and \ref{eq:2nd-order-exp} as well as the identifications
\begin{eqnarray*}
\ss{n+1}{2}{\mathrm{INC}}{\Phi_{n+2}} &=& \ss{n+1}{2}{Q_{n+2}<Q_c \land Q_{n+3}<Q_c}{\Phi_{n+1}} + \upint\ss{n+2}{1}{Q_{n+2}>Q_c}{\Phi_{n+2}} + \upiint\ss{n+3}{0}{Q_{n+3}>Q_c}{\Phi_{n+3}}\\
\ss{n+1}{1}{\mathrm{INC}}{\Phi_{n+2}} &=& \ss{n+1}{1}{Q_{n+2}<Q_c}{\Phi_{n+1}} + \upint\ss{n+2}{0}{Q_{n+2}>Q_c}{\Phi_{n+2}}
\end{eqnarray*}
the second-order remnant may suggestively be written as 
\begin{eqnarray}
\ss{n+1}{2}{\mathrm{REM}}{\Phi_{n+1}}
&=&
\ss{n+1}{2}{\mathrm{INC}}{\Phi_{n+2}}\\
&-& \frac{\alpha_s(\mu)}{2\pi}\frac{\beta_0}{2}\ln\left(\frac{\mu}{t_{n+1}}\right) \left[\ss{n+1}{1}{\mathrm{INC}}{\Phi_{n+2}} + \s{n+1}{0}{\Phi_{n+1}} \upint\limits^{t_+}_{t_{n+1}} \!\frac{dt}{t} 
dz d\phi \frac{\alpha_s(\mu)}{2\pi} P(t,z,\phi)\right]\nonumber\\
&+& \upint\limits^{t_+}_{t_{n+1}} \frac{dt}{t} dz d\phi \frac{\alpha_s(\mu)}{2\pi} P^{(0)}(t,z,\phi) \left[\ss{n+1}{1}{\mathrm{INC}}{\Phi_{n+2}} 
+
\s{n+1}{0}{\Phi_{n+1}}
\frac{\alpha_s(\mu)}{2\pi}
\left( \frac{\beta_0}{2} \ln\left(\frac{\mu}{t}\right) + \frac{P^{(1)}}{P^{(0)}} \right)\right]
\nonumber\\
&+& \s{n+1}{0}{\Phi_{n+1}} \upint\limits^{t_+}_{t_{n+1}} \frac{dt}{t} dz d\phi \frac{\alpha_s(\mu)}{2\pi} P^{(0)}(t,z,\phi) \!\! \upint\limits^{t}_{t_{n+1}}\! \frac{d\bar t}{\bar t} d\bar z d\bar\phi \frac{\alpha_s(\mu)}{2\pi} P^{(0)}(\bar t,\bar z,\bar\phi)
\nonumber\\
&-&
\s{n+1}{0}{\Phi_{n+1}} \left(\frac{\alpha_s(\mu)}{2\pi}\right)^2\frac{\beta_1}{4}\ln\left(\frac{\mu}{t_{n+1}}\right)\nonumber
\end{eqnarray}
The \nnlo{} cross section {\smaller$\ss{n+1}{2}{\mathrm{INC}}{\Phi_{n+2}}$} contains all the other terms, which consequently act to remove universal higher orders from the \nnlo{} result. More specifically, in the bracket in the first line, only universal $\beta_0$-dependent terms remain. These, together with the prefactor, act to remove $\beta_0$-dependent terms from the \nnlo{} result. Similarly, the bracket in the second line does not contain universal \nlo{} corrections to soft gluon couplings\footnote{Provided that $ P^{(1)}/ P^{(0)}$ reproduces the two-loop cusp-anomalous dimension at the inclusive level, as is the case in most modern showers.}, so that only Sudakov-like effects remain. Multiplied with the prefactor, this factor acts to remove further universal terms. The remaining terms again remove universal contributions to the \nnlo{} cross section. Thus, in total, {\smaller$\ss{n+1}{2}{\mathrm{REM}}{\Phi_{n+1}}$} contains -- from the viewpoint of the parton shower -- only non-universal terms, meaning that including {\smaller$\ss{n+1}{2}{\mathrm{REM}}{\Phi_{n+1}}$} as prefactor to all-order terms does not impair the desired parton-shower accuracy. 

Finally, for exclusive $n+1$-parton observables (i.e.\ sensitive to exactly one additional parton), several all-order terms in eq.\ \ref{eq:nnnlops} do no longer cancel since $n+2$-parton states with separation larger than $Q_c$ are vetoed. These uncanceled terms mirror, by design, the inclusive $n+2$-parton prediction exactly. The latter provide an \nlopps{} accurate description when one of the two additional partons becomes much softer (or more collinear) than the other. Thus, the uncanceled all-order terms in exclusive $n+1$-parton observables provide a parton-shower accurate resummation of the jet veto.

In conclusion, $n+1$-parton observables are described with \nnlopps{} accuracy. Incidentally, this also means that omitting the $n$-parton contribution to \nnnlops{} yields an improved \nnlopps{} matching for $n+1$ parton processes.

%\subsubsection{Zero-parton observables are N3LO+PS accurate}

\begin{center}
{\textit{Zero-additional-parton observables are N3LO+PS accurate}}
\end{center}

In the prediction of inclusive $n$-parton observables $\obs{n}$, the cancellation between higher-multiplicity predictions and all-order subtractions take maximal effect. After replacing all observables in eq.\ \ref{eq:nnnlops} with $\obs{n}$ (since inclusive $n$-parton observables are by definition insensitive to additional partons), the matching formula reduces to
\begin{eqnarray}
&&\ff{\mathcal{F}}{n}{\infty}{\mathrm{\nnnlops}\!\!}{\Phi_n, t_+,t_-} \nonumber\\
&&= \obs{n}~\Bigg\{ 
\ss{n}{0+1+2+3}{\mathrm{EXC}}{\Phi_{n}}
+~
\ss{n+1}{0+1+2}{Q_{n+2}<Q_c \land Q_{n+3}<Q_c}{\Phi_{n+1}} 
+~\ss{n+2}{0}{Q_{n+2}>Q_c}{\Phi_{n+2}}~\nonumber\\
&&
+~\ss{n+2}{1}{Q_{n+2}>Q_c \land Q_{n+3}<Q_c }{\Phi_{n+2}}
+~\ss{n+3}{0}{Q_{n+3}>Q_c}{\Phi_{n+3}}
~\Bigg\}
~=~
\obs{n}~
\ss{n}{0+1+2+3}{\mathrm{INC}}{\Phi_{n}}~.
\end{eqnarray}
Thus, the inclusive \nnnlo{} cross section is recovered without any higher-order corrections. For exclusive $n$-parton observables, the all-order subtraction to balance the $n+1$-parton prediction observables remains uncanceled, and thus produces a resummation of the jet veto. The inclusive $n+1$-parton spectrum recovers the parton-shower resummation when the additional parton becomes soft or collinear. By virtue of the mechanism in eq.\ \ref{eq:ps2}, this means that the all-order subtractions correctly reproduce the jet veto resummation for exclusive $n$-parton observables.
To summarize, $n$-parton observables are described with \nnnlopps{}  accuracy.

\section{Matching inclusive N3LO calculations to parton showers}
\label{app:inclusive-n3lops}

\noindent
The \nnnlops{} matching method developed in the main text relies on exclusive fixed-order calculations. The rationale behind this assumption is that exclusive fixed-order calculations might be easier to calculate or approximate. The \unnlops{} method~\cite{Hoeche:2014aia,Hoche:2014dla} for example employed well-known results from analytic resummation to construct exclusive cross sections. 
It is however also straight-forward to rearrange the \nnnlops{} method to rely on (a mix of exclusive and) inclusive cross sections. This opens many avenues to re-use known results to construct a matched calculation. Using an inclusive \nnnlo{} calculation may be the most relevant re-arrangement. In this case, the \nnnlops{} formula reads
\begin{eqnarray}
&&\ff{\mathcal{F}}{n}{\infty}{\mathrm{i\nnnlops}\!\!}{\Phi_n, t_+,t_-} 
%\nonumber\\
%&&
\coloneqq \obs{n}~\Bigg\{ 
\ss{n}{0+1+2+3}{\mathrm{INC}}{\Phi_{n}}\nonumber\\
&&
-~
\intOne\ss{n+1}{2}{Q_{n+2}<Q_c \land Q_{n+3}<Q_c}{\Phi_{n+1}}~
~\tcboxmath[colback=black!00,colframe=red,boxrule=0.5pt,boxsep=1pt,hypertarget=cfredB-i,hyperlink=cfredA-i]{\!\!\phantom{\int}
\Delta_{n} (t_+,t_{n+1}) \f{w}{n+1}{\infty}{\Phi_{n+1}}  ~}
\nonumber\\
&&
-~
\intOne\s{n+1}{0}{\Phi_{n+1}}~
\tcboxmath[colback=black!00,colframe=myolive,boxrule=0.5pt,boxsep=1pt,hypertarget=cfoliveB-i,hyperlink=cfoliveA-i]{
\begin{aligned}
&~\Delta_{n} (t_+,t_{n+1}) \f{w}{n+1}{\infty}{\Phi_{n+1}} ~\nonumber\\
&\quad\cdot\Big(                  
1 - \f{w}{n+1}{1}{\Phi_{n+1}} - \f{w}{n+1}{2}{\Phi_{n+1}}  - \f{\Delta}{n}{1}{t_+,t_{n+1}} -  \f{\Delta}{n}{2}{t_+,t_{n+1}}
\nonumber\\
&\quad~ + \left[\f{\Delta}{n}{1}{t_+,t_{n+1}}\right]^2 + \left[\f{w}{n+1}{1}{\Phi_{n+1}}\right]^2 + \f{w}{n+1}{1}{\Phi_{n+1}} \f{\Delta}{n}{1}{t_+,t_{n+1}} \Big)
\end{aligned}
}
\nonumber\\
&&
-~\intOne\ss{n+1}{1}{Q_{n+2}<Q_c}{\Phi_{n+1}}
~\tcboxmath[colback=black!00,colframe=mypurple,boxrule=0.5pt,boxsep=1pt,hypertarget=cfpurpleB-i,hyperlink=cfpurpleA-i]{\!\!\phantom{\int}
\Delta_{n} (t_+,t_{n+1})  \f{w}{n+1}{\infty}{\Phi_{n+1}} ~\Big( 1 - \f{w}{n+1}{1}{\Phi_{n+1}} - \f{\Delta}{n}{1}{t_+,t_{n+1}}\Big) ~}\nonumber\\
&&
-~\intTwo\ss{n+2}{0}{Q_{n+2}>Q_c}{\Phi_{n+2}}~
~\tcboxmath[colback=mybrown!20,hypertarget=brownB-i,hyperlink=brownA-i]{\!\!\phantom{\int}
\Delta_{n} (t_+,t_{n+1}) \f{w}{n+1}{\infty}{\Phi_{n+1}} ~\Big(  1 - \f{w}{n+1}{1}{\Phi_{n+1}} - \f{\Delta}{n}{1}{t_+,t_{n+1}}\Big) \one{n+2}{n+1} ~}
\nonumber\\
&&
-~\intTwo\ss{n+2}{1}{Q_{n+2}>Q_c \land Q_{n+3}<Q_c }{\Phi_{n+2}}
~\tcboxmath[colback=mypurple!20,hypertarget=purpleB-i,hyperlink=purpleA-i]{\!\!\phantom{\int}
~ \Delta_{n} (t_+,t_{n+1}) \f{w}{n+1}{\infty}{\Phi_{n+1}} \one{n+2}{n+1} ~}\nonumber\\
&&
-~\intThree\ss{n+3}{0}{Q_{n+3}>Q_c}{\Phi_{n+3}}
~\tcboxmath[colback=mymidgrey!20,hypertarget=greyB-i,hyperlink=greyA-i]{~ \Delta_{n} (t_+,t_{n+1}) \f{w}{n+1}{\infty}{\Phi_{n+1}} \one{n+3}{n+1} ~} ~\Bigg\}\nonumber\\
&&+~\obs{n+1}~\Bigg\{ 
\ss{n+1}{2}{Q_{n+2}<Q_c \land Q_{n+3}<Q_c}{\Phi_{n+1}}~\tcboxmath[colback=black!00,colframe=red,boxrule=0.5pt,boxsep=1pt,hypertarget=cfredA-i,hyperlink=cfredB-i]{\!\!\phantom{\int}
\Delta_{n} (t_+,t_{n+1}) \f{w}{n+1}{\infty}{\Phi_{n+1}} ~}\nonumber\\
&&+~\s{n+1}{0}{\Phi_{n+1}}~\otimes~
\tcboxmath[colback=black!00,colframe=myolive,boxrule=0.5pt,boxsep=1pt,hypertarget=cfoliveA-i,hyperlink=cfoliveB-i]{
\begin{aligned}
&~\Delta_{n} (t_+,t_{n+1}) \f{w}{n+1}{\infty}{\Phi_{n+1}} ~\nonumber\\
&\quad\cdot\Big(                  
1 - \f{w}{n+1}{1}{\Phi_{n+1}} - \f{w}{n+1}{2}{\Phi_{n+1}}  - \f{\Delta}{n}{1}{t_+,t_{n+1}} -  \f{\Delta}{n}{2}{t_+,t_{n+1}}
\nonumber\\
&\quad~ + \left[\f{\Delta}{n}{1}{t_+,t_{n+1}}\right]^2 + \left[\f{w}{n+1}{1}{\Phi_{n+1}}\right]^2 + \f{w}{n+1}{1}{\Phi_{n+1}} \f{\Delta}{n}{1}{t_+,t_{n+1}} \Big)
\end{aligned}
~}
\nonumber\\
&&+~ \ss{n+1}{1}{Q_{n+2}<Q_c}{\Phi_{n+1}} \nonumber\\
&&\qquad\otimes~
\tcboxmath[colback=black!00,colframe=mypurple,boxrule=0.5pt,boxsep=1pt,hypertarget=cfpurpleA-i,hyperlink=cfpurpleB-i]{\!\!\phantom{\int}
\Big[ 1 - \f{w}{n+1}{1}{\Phi_{n+1}} - \f{\Delta}{n}{1}{t_+,t_{n+1}}\Big] \Delta_{n} (t_+,t_{n+1})\f{w}{n+1}{\infty}{\Phi_{n+1}} ~}  \nonumber\\
&&+~\intOne\ss{n+2}{0}{Q_{n+2}>Q_c}{\Phi_{n+2}} 
%\nonumber\\
%&&\qquad
\otimes~ \Delta_{n} (t_+,t_{n+1}) \f{w}{n+1}{\infty}{\Phi_{n+1}}
\nonumber\\
&&\qquad\otimes~\!\Big[~\tcboxmath[colback=mybrown!20,hypertarget=brownA-i,hyperlink=brownB-i]{\!\!\phantom{\int} \Big(  1 - \f{w}{n+1}{1}{\Phi_{n+1}} - \f{\Delta}{n}{1}{t_+,t_{n+1}}\Big) \one{n+2}{n+1} ~}
\nonumber\\
&&\qquad~
-
~
\Delta_{n+1} (t_{n+1},t_{n+2}) \f{w}{n+2}{\infty}{\Phi_{n+2}}\nonumber\\
&&\qquad\qquad\otimes
~\tcboxmath[colback=green!10,hypertarget=greenB-i,hyperlink=greenA-i]{\!\!\phantom{\int}
\Big( 1 - \f{w}{n+1}{1}{\Phi_{n+1}} - \f{w}{n+2}{1}{\Phi_{n+2}} - \f{\Delta}{n}{1}{t_+,t_{n+1}} - \f{\Delta}{n+1}{1}{t_{n+1},t_{n+2}} \Big)
~}~\Big]
\nonumber\\
&&+~\intOne\ss{n+2}{1}{Q_{n+2}>Q_c \land Q_{n+3}<Q_c }{\Phi_{n+2}} \nonumber\\
&&\qquad\otimes~\Big[
~\tcboxmath[colback=mypurple!20,hypertarget=purpleA-i,hyperlink=purpleB-i]{\!\!\phantom{\int}  \Delta_{n} (t_+,t_{n+1}) \f{w}{n+1}{\infty}{\Phi_{n+1}} \one{n+2}{n+1} ~  }
%\nonumber\\
%&&\qquad~
- ~\tcboxmath[colback=red!10,hypertarget=redB-i,hyperlink=redA-i]{\!\!\phantom{\int}  \Delta_{n} (t_+,t_{n+1}) \Delta_{n+1} (t_{n+1},t_{n+2}) \f{w}{n+1}{\infty}{\Phi_{n+1}} \f{w}{n+2}{\infty}{\Phi_{n+2}} ~}~\Big] \nonumber\\
&&+~\intTwo \ss{n+3}{0}{Q_{n+3}>Q_c}{\Phi_{n+3}}\nonumber\\
&&\qquad\otimes~ \Big[
~\tcboxmath[colback=mymidgrey!20,hypertarget=greyA-i,hyperlink=greyB-i]{\!\!\phantom{\int} \Delta_{n} (t_+,t_{n+1}) \f{w}{n+1}{\infty}{\Phi_{n+1}} \one{n+3}{n+1}~}
%\nonumber\\
%&&\qquad~
- ~\tcboxmath[colback=orange!10,hypertarget=orangeB-i,hyperlink=orangeA-i]{\!\!\phantom{\int}
\Delta_{n} (t_+,t_{n+1}) \Delta_{n+1} (t_{n+1},t_{n+2}) \f{w}{n+1}{\infty}{\Phi_{n+1}} \f{w}{n+2}{\infty}{\Phi_{n+2}} \one{n+3}{n+2} ~}~\Big]
\Bigg\}
\nonumber\\
&&+~
\obs{n+2} ~\Bigg\{
~\ss{n+2}{0}{Q_{n+2}>Q_c}{\Phi_{n+2}} 
\nonumber\\
&&\qquad\otimes~
\Delta_{n} (t_+,t_{n+1})\Delta_{n+1} (t_{n+1},t_{n+2}) \f{w}{n+1}{\infty}{\Phi_{n+1}}  \f{w}{n+2}{\infty}{\Phi_{n+2}}
\nonumber\\
&&\qquad\otimes~
~\tcboxmath[colback=green!10,hypertarget=greenA-i,hyperlink=greenB-i]{\!\!\phantom{\int}\Big[ 1 - \f{w}{n+1}{1}{\Phi_{n+1}} - \f{w}{n+2}{1}{\Phi_{n+2}} - \f{\Delta}{n}{1}{t_+,t_{n+1}} - \f{\Delta}{n+1}{1}{t_{n+1},t_{n+2}} \Big]~}
\nonumber\\
&&+~ \ss{n+2}{1}{Q_{n+2}>Q_c \land Q_{n+3}<Q_c}{\Phi_{n+2}} \nonumber\\
&&\qquad\otimes~\tcboxmath[colback=red!10,hypertarget=redA-i,hyperlink=redB-i]{\!\!\phantom{\int}
\Delta_{n} (t_+,t_{n+1})\Delta_{n+1} (t_{n+1},t_{n+2}) \f{w}{n+1}{\infty}{\Phi_{n+1}}\f{w}{n+2}{\infty}{\Phi_{n+2}} ~} \nonumber\\
&&+~ \intOne \ss{n+3}{0}{Q_{n+3}>Q_c}{\Phi_{n+3}} \Big[ ~\tcboxmath[colback=orange!10,hypertarget=orangeA-i,hyperlink=orangeB-i]{\!\!\phantom{\int}
\Delta_{n} (t_+,t_{n+1}) \Delta_{n+1} (t_{n+1},t_{n+2}) \f{w}{n+1}{\infty}{\Phi_{n+1}}\f{w}{n+2}{\infty}{\Phi_{n+2}} \one{n+3}{n+2}~}
\nonumber\\
&&-~\tcboxmath[colback=blue!10,hypertarget=blueB-i,hyperlink=blueA-i]{\!\!\phantom{\int}
\Delta_{n} (t_+,t_{n+1}) \Delta_{n+1} (t_{n+1},t_{n+2}) \Delta_{n+2} (t_{n+2},t_{n+3})\w{n+1} \w{n+2} \w{n+3}~}~\Big] 
\Bigg\}\nonumber\\
&&+~ \ss{n+3}{0}{Q_{n+3}>Q_c}{\Phi_{n+3}} ~\tcboxmath[colback=blue!10,hypertarget=blueA-i,hyperlink=blueB-i]{\!\!\phantom{\int} 
\Delta_{n} (t_+,t_{n+1}) \Delta_{n+1} (t_{n+1},t_{n+2}) \Delta_{n+2} (t_{n+2},t_{n+3})  \w{n+1} \w{n+2}\w{n+3}  ~}\nonumber\\
&&\quad\otimes~\f{\mathcal{F}}{n+3}{\infty}{\Phi_{n+3}, t_{n+3},t_-}~.
\label{eq:nnnlops-inc-n3lo}
\end{eqnarray}
This is trivially obtained from eq.\ \ref{eq:nnnlops} by setting $\one{n+1}{n} = \one{n+2}{n} = \one{n+3}{n} = 0$. Other combinations of inclusive with exclusive calculations may be obtained in a similar manner. This simple change does, however, not guarantee that exclusive fixed-order cross sections are recovered exactly, since the unitarity subtractions might introduce biases. This was the reason why the use of exclusive cross-sections was prioritized in the main text. It would be valuable to assess and address the bias when using inclusive calculations in the future.

\bibliography{ref}{}
\bibliographystyle{h-physrev}

\end{document}